\documentclass[apj]{emulateapj}
\usepackage{apjfonts}
\newcommand       \IRAS         {{\it IRAS}}
\newcommand       \ISO          {{\it ISO}}
\newcommand       \Spitzer      {{\it Spitzer}}

\newcommand       \IRS          {{\rm IRS}}
\newcommand       \IRAC         {{\rm IRAC}}
\newcommand       \MIPS         {{\rm MIPS}}

\newcommand       \LongISO      {{\it Infrared Space Observatory}}
\newcommand       \LongSpitzer  {{\it Spitzer Space Telescope}}

\newcommand       \LongIRS      {{\rm Infrared Spectrograph}}

\newcommand       \Angstrom     {\,\textnormal{\AA}}
\newcommand       \um           {\,{\micron}}

\newcommand       \K            {\,{\rm K}}
\newcommand       \pc           {\,{\rm pc}}

\newcommand       \LSun         {\,\ensuremath{L_\sun}}
\newcommand       \MSun         {\,\ensuremath{M_\sun}}
\newcommand       \LIR          {\ensuremath{L_{\rm IR}}}
\newcommand       \nH           {\ensuremath{n_{\rm H}}}    
\newcommand       \NH           {\ensuremath{N_{\rm H}}}    
\newcommand       \Min          {{\rm min}}
\newcommand       \Max          {{\rm max}}
\newcommand       \Cold         {{\it cold}}
\newcommand       \Cool         {{\it cool}}
\newcommand       \Warm         {{\it warm}}
\newcommand       \Hot          {{\it hot}}
\newcommand       \ISRF         {{\it ISRF}}
\newcommand       \AGN          {{\it AGN}}
\newcommand       \Starburst    {{\it SB}}
\newcommand       \Lines        {{\it lines}}
\newcommand       \PAHs         {{\it PAHs}}
\newcommand       \cold         {{\rm cold}}
\newcommand       \cool         {{\rm cool}}
\newcommand       \warm         {{\rm warm}}
\newcommand       \hot          {{\rm hot}}
\newcommand       \isrf         {{\rm ISRF}}
\newcommand       \agn          {{\rm AGN}}
\newcommand       \starburst    {{\rm SB}}
\newcommand       \lines        {{\rm lines}}
\newcommand       \pahs         {{\rm PAHs}}
\newcommand       \WD           {{\rm WD01}}
\newcommand       \data         {{\rm data}}
\newcommand       \cont         {{\rm cont}}
\newcommand       \phot         {{\rm phot}}
\newcommand       \dust         {{\rm dust}}
\newcommand       \gas          {{\rm gas}}
\newcommand       \source       {{\rm source}}

\newcommand       \ice          {{\rm ice}}
\newcommand       \HAC          {{\rm HAC}}
\newcommand       \abs          {{\rm abs}}
\newcommand       \scat         {{\rm scat}}
\newcommand       \ext          {{\rm ext}}
\newcommand       \gra          {{\rm gra}}
\newcommand       \sil          {{\rm sil}}
\newcommand       \MMP          {{\rm MMP}}
\newcommand       \Tavg         {{\bar{T}}}
\shorttitle{Decomposing Dusty Galaxies. I.}
\shortauthors{Marshall et al.}
\begin{document}
\title{Decomposing Dusty Galaxies. I.
  Multi-Component Spectral Energy Distribution Fitting}

\author{
  J.~A. Marshall,\altaffilmark{1,2}
  T.~L. Herter,\altaffilmark{1}
  L. Armus,\altaffilmark{3}
  V. Charmandaris,\altaffilmark{4,5}
  H.~W.~W Spoon,\altaffilmark{1}
  J. Bernard-Salas,\altaffilmark{1} and
  J.~R. Houck\altaffilmark{1}
}

\email{jam258@caltech.edu}

\altaffiltext{1}{Cornell University, Department of Astronomy,
  Ithaca, NY 14853}
\altaffiltext{2}{Caltech/JPL, 
  Pasadena, CA 91125}
\altaffiltext{3}{Spitzer Science Center, MS 220-6, Caltech,
  Pasadena, CA 91125}
\altaffiltext{4}{University of Crete, Department of Physics,
  P.O. Box 2208 GR-71003, Heraklion, Greece}
\altaffiltext{5}{IESL/FORTH, P.O. Box 1527, GR-71110, Heraklion, Greece, and
  Chercheur Associ\'{e}, Observatoire de Paris, F-75014, Paris, France}


\begin{abstract}
  We present a new multi-component spectral energy distribution (SED)
  decomposition method and use it to analyze the ultraviolet to millimeter
  wavelength SEDs of a sample of dusty infrared-luminous galaxies.  SEDs are
  constructed from spectroscopic and photometric data obtained with the
  \LongSpitzer\ in conjunction with photometry from the literature. Each SED
  is decomposed into emission from populations of stars, an AGN accretion
  disk, PAHs, atomic and molecular lines, and distributions of graphite and
  silicate grains. Decompositions of the SEDs of the template starburst
  galaxies NGC\,7714 and NGC\,2623 and the template AGNs PG\,0804+761 and
  Mrk\,463 provide baseline properties to aid in quantifying the strength of
  star-formation and accretion in the composite systems NGC\,6240 and
  Mrk\,1014.  We find that obscured radiation from stars is capable of
  powering the total dust emission from NGC\,6240. The presence of a small
  quantity of $1260\K$ dust in this source suggests a $\sim$2\% AGN
  contribution, although we cannot rule out a larger contribution from a
  deeply embedded AGN visible only in X-rays.  The decomposition of
  Mrk\,1014 is consistent with $\sim$65\% of its power emerging from an AGN
  and $\sim$35\% from star-formation.  We suggest that many of the
  variations in our template starburst SEDs may be explained in terms of the
  different mean optical depths through the clouds of dust surrounding the
  young stars within each galaxy. Prompted by the divergent far-IR
  properties of our template AGNs, we suggest that variations in the
  relative orientation of their AGN accretion disks with respect to the
  disks of the galaxies hosting them may result in different amounts of
  AGN-heated cold dust emission emerging from their host galaxies. We
  estimate that 30--50\% of the far-IR and PAH emission from Mrk\,1014 may
  originate from such AGN-heated material in its host galaxy disk. In a
  subsequent paper, this decomposition method will be applied to a large
  sample of dusty galaxy SEDs to further test the validity of these
  suggestions.
\end{abstract}


\section{Introduction}
\label{sec:Introduction}

Optical and near-IR imaging has revealed that most Luminous and
Ultra-Luminous Infrared Galaxies
%
\citep[LIRGs and ULIRGs---see][]{1996ARA&A..34..749S},
as well as many lower luminosity dusty galaxies (e.g. starbursts and AGNs),
are interacting systems which have recently merged or are in the process of
merging
%
\citep{1987AJ.....94..831A, 1988ApJ...328L..35S, 1996AJ....111.1025M}.
Large quantities of gas and dust driven into the nascent nuclei of these
systems catalyze massive bursts of star formation and/or fuel accretion onto
supermassive black holes
%
\citep{1996ApJ...464..641M, 1999PhR...321....1S}.
Ultraviolet and optical emission from many of these dusty galaxies is
significantly obscured by the same gas and dust providing their infrared
power---a fact which greatly complicates the process of diagnosing the
star-formation and accretion contributions to their bolometric luminosities.
LIRGs and ULIRGs are believed to be the largest contributors to the far-IR
background and star-formation energy density at $z \approx 1$--$3$
%
\citep{2002PhR...369..111B, 2003Sci...300..270E},
and local ULIRGs are frequently used as analogues for the luminous sub-mm
galaxy population. The development of a proper understanding of the
histories of star-formation and galaxy evolution in the universe therefore
requires the construction of robust and reliable methods for use in
diagnosing the dominant luminosity source in obscured dusty galactic nuclei.


Great progress towards this goal was made using data from the \LongISO\
(\ISO).  Diagnostic diagrams based upon high-ionization mid-IR lines (e.g.
[\ion{O}{4}] and [\ion{Ne}{5}]), the slope of the mid-IR continuum, as well
as the strengths of the $6.2$ and $7.7\um$ PAH emission features were
constructed to quantify the relative contributions from accretion and
starburst activity in many sources
%
\citep[e.g.][]{1998ApJ...498..579G, 1998A&A...333L..75L,
  2000A&A...359..887L, 2001ApJ...552..527T}.
These observations provided evidence that ULIRGs are powered primarily by
star-formation, with the fraction of AGN activity increasing with bolometric
luminosity. However, due to the limited wavelength coverage and sensitivity
of the \ISO\ spectrometers, this analysis was limited to a relatively small
number of bright nearby sources. These observational challenges have largely
been abated with the improved wavelength baseline and sensitivity of the
\LongIRS \footnote{The IRS was a collaborative venture between Cornell
  University and Ball Aerospace Corporation funded by NASA through the Jet
  Propulsion Laboratory and the Ames Research Center.}
%
\citep[\IRS;][]{2004ApJS..154...18H}
on the \LongSpitzer\
%
\citep{2004ApJS..154....1W}.
Large samples of starburst galaxies
%
\citep{2006ApJ...653.1129B},
AGNs
%
\citep{2005ApJ...625L..75H, 2005ApJ...633..706W, 2006ApJ...649...79S},
local ULIRGs
%
\citep{2007ApJ...656..148A},
and their high-redshift counterparts
%
\citep[e.g.][]{2005ApJ...622L.105H, 2005ApJ...628..604Y,
  2005ApJ...625L..83L}
have been observed with the \IRS. As with the previous generation of studies
with \ISO, diagnostic diagrams have been constructed using the properties of
high-ionization lines (with much improved sensitivity to place strict limits
on [\ion{Ne}{5}] emission), mid-IR spectral slopes (with a much broader
mid-IR wavelength baseline), and PAH feature emission (now detected to $z
\ga 2$).


In addition to these observational diagnostics, many tools have been
developed to model the SEDs of emission from dusty galaxies. Models
containing a variety of grains-sizes and chemical compositions have been
used to calculate the spectrum of emission from astronomical dust embedded
in single
%
\citep[e.g.][]{1990A&A...237..215D, 2001ApJ...554..778L, Draine2007}
and multiple
%
\citep[e.g.][]{2001ApJ...549..215D, 2002ApJ...576..762L}
incident radiation fields.  Additionally, numerous radiative-transfer models
have been published, including several to calculate the emission from an AGN
and its obscuring torus
%
\citep[e.g.][]{2002ApJ...570L...9N, 2005A&A...436...47D}.
Finally, methods intended to decompose the SEDs of galaxies into a number of
components have also been developed.  For example,
%
\citet{2001A&A...379..823K}
present such a method and use it to decompose the far-IR SEDs of a sample of
ULIRGs into contributions from several dust-modified blackbodies at
different characteristic temperatures.


We present in this work a new method used to decompose the SEDs of galaxies
into emission from populations of stars, an AGN accretion disk, PAH
molecules, atomic and molecular lines, and distributions of thermally heated
graphite and silicate grains at different fitted characteristic
temperatures.  Given the many observational and theoretical techniques to
study dusty galactic nuclei described above, it is fair to question our
decision to introduce yet another. We are motivated to do so for two
principal reasons.  First, while the radiative-transfer methods
incorporating realistic dust models described above provide the most
physically complete method of calculating the SEDs of dusty galaxies, these
methods are also the most dependent on the assumed geometry of the source.
Unfortunately, detailed spatial information to constrain these assumptions
is available for only a handful of nearby galaxies, so the parameter space
in which to search for best-fitting models in agreement with observations is
both sizable and degenerate. Second, the methods for calculating the
emission from dusty galaxies described above are primarily intended to model
starburst galaxies {\it or} AGNs.  None of them, however, is particularly
well-suited to calculating the emission from sources containing a starburst
{\it and} an AGN, each of which may be extinguished by different levels of
obscuration (see \S\ref{sec:ComparisonWithOtherMethods} for an expanded
discussion).


Our goal for this paper is therefore to convince the reader that our
decomposition method is well-suited to analyzing the SEDs of composite
sources, and is motivated by the desire to address: (1) What energy source
is primarily responsible for powering the emission from dusty galaxies?  (2)
How homogeneous are the SEDs of dusty galaxies which have similar AGN and
starburst fractions? and (3) What geometries can explain any inhomogeneities
in the SEDs of similarly powered dusty galaxies?  To meet this goal we apply
our decomposition method to a small sample of well-studied galaxies,
including the relatively unextinguished starburst NGC\,7714, the obscured
starburst NGC\,2623, the quasar PG\,0804+761, and the Seyfert 2 Mrk\,463.
These fits provide templates by which to understand the properties of pure
starburst galaxies and AGNs---the constituents of composite sources such as
many ULIRGs.  Additionally, as examples of such composite sources, we
present fits to the LIRG NGC\,6240 (believed to be powered primarily by
star-formation) and the ULIRG Mrk\,1014 (believed to be powered primarily by
accretion).


We begin in \S\ref{sec:SpectralDecomposition} with an overview of the
spectral decomposition method. In \S\ref{sec:SourceEmission} we detail the
method used to calculate the spectrum of emission from the various source
components. In \S\ref{sec:DustModel} and \S\ref{sec:DustEmission} we
describe the adopted dust model and the method used to calculate the thermal
emission from dust. In \S\ref{sec:PAHEmission} and \S\ref{sec:LineEmission}
we describe PAH feature and atomic and molecular line emission,
respectively. In \S\ref{sec:ConstructingDustySEDs} we describe the sample of
galaxies analyzed in this work, in \S\ref{sec:DecomposingDustyGalaxies} we
present their decompositions, and in \S\ref{sec:Discussion} we present
detailed analysis of the results.  The breadth of our conclusions are
clearly limited by the size of our sample. In a subsequent paper our method
will be applied to a much larger sample of starburst galaxies, AGNs, LIRGs,
and ULIRGs obtained as part of the \Spitzer\ GTO program.

\section{Spectral Decomposition Method\protect\footnotemark}
\footnotetext{See
  http://isc.astro.cornell.edu/\textrm{\textasciitilde}jam258/cafe for more
  information about the method and the Continuum And Feature Extraction
  ({\sc Cafe}) software.}
\label{sec:SpectralDecomposition}

\subsection{Decomposition Components}
\label{sec:DecompositionComponents}

The principal assumption in our decomposition method is that the ultraviolet
to millimeter wavelength ($10\Angstrom$--$1000\um$) SED of a dusty galaxy
may be completely described in terms of emission from evolved stars in the
disk of the host galaxy (\ISRF\footnote{All names of decomposition
  components are printed in italic type throughout this work.}), populations
of young stars in starbursts (\Starburst), an AGN accretion disk (\AGN), PAH
molecules (\PAHs), atomic and molecular lines (\Lines), and several
distributions of thermally heated graphite and silicate grains (\Hot, \Warm,
\Cool, and \Cold). In this paper the \ISRF, \Starburst, and \AGN\ components
are collectively referred to as `source' emission components since they
represent the primary illuminating sources heating dust within a galaxy.
The \Hot, \Warm, \Cool, and \Cold\ components are collectively referred to
as `dust' emission components. Assuming these components form a complete
basis describing the emission from dusty galaxies, the observed flux density
may be expressed as\footnote{Note that $f_\lambda$ units may be substituted
  for $f_\nu$ units throughout this work and that variables accented with
  tildes are free parameters in the fitting.}
\begin{equation}
  \label{eq:TotalFlux}
    f_\nu^{\rm total} = 
    \frac{L_\source}{4 \pi D_{\rm L}^2}
    \sum_\source \tilde{\alpha}_i \hat{f}_\nu^i + 
    \frac{L_\dust}{4 \pi D_{\rm L}^2}
    \sum_\dust \tilde{\alpha}_j \hat{f}_\nu^j + 
    f_\nu^\pahs + f_\nu^\lines,
\end{equation}
where $L_\source$ and $L_\dust$ are the total apparent luminosities of the
source and dust components derived from the fit, $D_{\rm L}$ is the
luminosity distance to the galaxy, $\tilde{\alpha}_i \equiv L_i / L_\source$
and $\tilde{\alpha}_j \equiv L_j / L_\dust$ are the fitted contributions to
the total source and dust luminosities from each component, the $\hat{f}_\nu
\equiv f_\nu / \int f_\nu \, d\nu$ are the normalized flux densities of each
source or dust component, and $f_\nu^\pahs$ and $f_\nu^\lines$ are the
fitted PAH and emission line flux densities. We include a maximum of four
dust components which are typically heated to characteristic temperatures of
$\Tavg_\hot\approx1400\K$, $\Tavg_\warm\approx200\K$,
$\Tavg_\cool\approx80\K$, and $\Tavg_\cold\approx35\K$. The actual
temperature of each dust component is determined from the fitted magnitude
of its illuminating radiation field energy density.

%
\begin{figure}
  \epsscale{1.1}
  \plotone{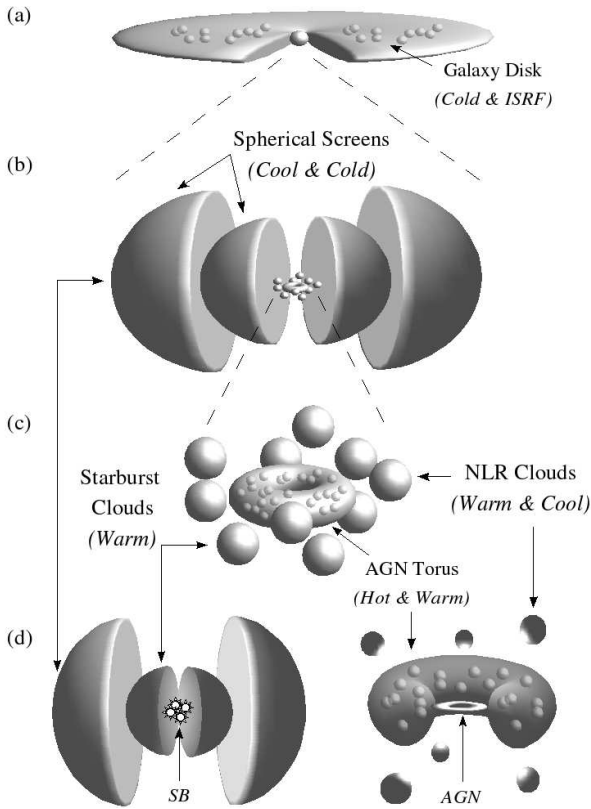}
  \caption{Example geometry of a dusty galaxy illustrating the potential
    roles played by each decomposition component. Panel (a) depicts the host
    galaxy disk containing \Cold\ dust and knots of \ISRF\ emission from
    evolved stars.  Panel (b) shows an enlarged view of the core of the
    galaxy containing a nuclear region embedded within obscuring screens of
    \Cool\ and \Cold\ dust.  Panels (c) and (d) depict the composite
    nucleus. The starburst contains \Starburst\ emission from young stars,
    \Warm\ dust in the surrounding shells, and possibly \Cool\ and/or \Cold\
    dust in spherical screens around each individual star-forming cloud. The
    AGN contains an \AGN\ accretion disk, \Hot\ and \Warm\ dust in the
    obscuring torus, and \Warm\ and \Cool\ dust in narrow-line-region (NLR)
    clouds.}
  \label{fig:GalaxyModel}
  \epsscale{1.0}
\end{figure}
%

Figure~\ref{fig:GalaxyModel} presents an illustration of one possible
geometric structure of a dusty galaxy within which to interpret the roles
played by each decomposition component in equation~(\ref{eq:TotalFlux}).
Panel (a) depicts the optically-thin (at mid-IR wavelengths) disk of the
host galaxy from which evolved photospheric (\ISRF) and \Cold\ dust emission
emerges. Panel (b) shows an enlarged view of the central region of the
galaxy consisting of an AGN and a starburst surrounded by screens of \Cool\
and \Cold\ dust.  Panel (c) depicts the composite nucleus of the dusty
galaxy, while the details of each nuclear component are shown in panel (d).
The starburst is composed of \Starburst\ emission from young stars, \Warm\
dust in the shells surrounding the young stars, and possibly \Cool\ and/or
\Cold\ dust in spherical screens around each individual star-forming cloud.
The AGN consists of emission from the \AGN\ accretion disk, \Hot\ and \Warm\
dust on the illuminated and shaded sides of the dust clouds creating the
obscuring torus (or dust in illuminated and shielded regions of a `classic'
torus with a smooth dust distribution---see Fig.~\ref{fig:SourceModel}), and
\Warm\ and \Cool\ dust in narrow-line-region clouds (again from dust in
illuminated and shaded regions).  We emphasize that our method does not
require the definition of {\it any} specific geometry (i.e.  we derive the
contributions from each emission component without the need to define its
proximity to any other component), and the actual geometry of a dusty galaxy
may look quite different from the one presented. The illustration in
Figure~\ref{fig:GalaxyModel} should therefore not be taken as a literal
description of the geometry of a particular galaxy (since our method
requires none), but instead is intended to provide a structure within which
to visualize the multiple roles potentially played by each decomposition
component.


We note that star-formation in the disk of the host galaxy and in starburst
regions are both explicitly accounted for via the \ISRF\ and \Starburst\
components. Furthermore, while Figure~\ref{fig:GalaxyModel} depicts the
specific case of a nuclear starburst, the actual location of the
star-formation (e.g. nuclear versus off-center) has no effect on the
decomposition (assuming that any off-center emission is contained within the
spectroscopic slits and photometric apertures used to construct the SEDs).
For example, in the decomposition of an off-center starburst, the \Cool\ and
\Cold\ decomposition components (or portions thereof) could be interpreted
as belonging to off-center obscuring clouds. Note, however, that if a galaxy
contains both a highly-enshrouded nuclear starburst and a less-enshrouded
off-center starburst, our decomposition method cannot explicitly fit both
components simultaneously.  Instead, a mean optical depth somewhere between
the optical depths of the two individual components is obtained. Thus, care
must be taken when interpreting the optical depths inferred from a globally
integrated SED (this admonishment is not unique to our analysis, but is a
general statement about the nature of spatially integrated data).

\subsection{Decomposition Algorithm}
\label{sec:DecompositionAlgorithm}

Once assembled, the ultraviolet to millimeter wavelength SEDs of the
galaxies in our sample are fit to the model of the total flux density from
equation~(\ref{eq:TotalFlux}). The best-fit parameters are determined using
a Levenberg-Marquardt least-squares routine to minimize
\begin{equation}
  \label{eq:Chi2}
  \chi^2 = \sum_k w_k
  \left[f_\nu^\data(\lambda_k) - f_\nu^{\rm total}(\lambda_k)\right]^2,
\end{equation}
where the $w_k$ are the weights applied at each wavelength $\lambda_k$ (see
\S\ref{sec:DataWeighting}), $f_\nu^\data$ is the observed flux density, and
$f_\nu^{\rm total}$ is the modeled flux density from
equation~(\ref{eq:TotalFlux}). For photometric data, we integrate the
modeled flux density over the filter transmission curve (if available) to
properly calculate the broadband flux density. This is important for
photometric bands containing emission and absorption features (e.g. L-band
photometry containing the $3.3\um$ PAH, $3.1\um$ water-ice, and $3.4\um$ HAC
features---see \S\ref{sec:IceOpacity} and \S\ref{sec:PAHEmission}), as well
as those bands covering spectral regions that change rapidly with wavelength
(e.g.  J-band photometry sampling extinguished photospheric emission---see
the near-IR fit to NGC\,6240 in \S\ref{sec:SpectralDecompositions}).


Experience has taught us that the greater the number of free parameters in a
fitting model, the less likely the resulting `best-fit' will converge to the
desired global minimum in the $\chi^2$ function. Correspondingly, it becomes
increasingly likely that the fit will converge to a local minimum, often
with very unsatisfactory results. This is a practical matter and not a
mathematical one since a fit with any number of parameters will reach the
global minimum given sufficiently well-chosen initial conditions.  In
practice, however, it is desirable to have a more robust fitting method such
that any {\it reasonably} chosen initial conditions will converge to the
global minimum.  To meet this goal, we have developed an algorithm which
separates the processes of fitting the continuum---defined as the sum of the
source and dust components---and the individual PAH features and atomic and
molecular emission lines. Such a division into two separate least-squares
processes greatly reduces the total number of free parameters in each fit.
Parameter errors are calculated after the first step of the fitting
procedure and propagated through to the following fit (see
\S\ref{sec:DataWeighting}), so that the net accumulated uncertainty is
reliably calculated.


After an SED has been fit, we perform a series of tests to ensure that all
components are well-constrained (i.e. that the parameter errors are smaller
than their values). First, if the \ISRF\ luminosity is $< 1$\% the sum of
the luminosities of all source components, then the fit is repeated without
the \ISRF\ component. Similarly, if any of the observed dust component
luminosities are $< 1$\% the sum of the apparent luminosities of all dust
components, the fit is repeated without those components. All three source
components and the \Hot\ dust component emit strongly at ultraviolet to
near-IR wavelengths. For some galaxies these overlapping emission components
are degenerate and result in unconstrained parameters. Our second check is
therefore to test if either the \Starburst\ or \AGN\ components are
unconstrained (i.e. if their luminosity uncertainties exceed their
luminosities). If the \Starburst\ component is unconstrained, we fix the
\Starburst-to-\PAHs\ luminosity ratio (i.e. the ratio of the unextinguished
\Starburst\ luminosity to the \PAHs\ luminosity---see
\S\ref{sec:DecompositionParametersAndConstraints}) to the value derived from
fits to the template starbursts NGC\,7714 and NGC\,2623, and restart the
fit. Similarly, if the \AGN\ component is unconstrained, we fix the
\AGN-to-\Hot\ dust luminosity ratio (i.e. the ratio of the unextinguished
\AGN\ luminosity to the unextinguished \Hot\ dust luminosity---see
\S\ref{sec:DecompositionParametersAndConstraints}) to the value derived from
fits to the template AGNs PG\,0804+761 and Mrk\,463, and restart the fit.
As our final test, if any of the fitted optical depths are unconstrained, we
fix the most unconstrained to zero and restart the fit.  This sequence
provides a well-defined method by which to sequentially impose constraints
until a successful fit is obtained.


The steps in the fitting method are therefore: (1) Input the observed SED
and the initial estimates of the \PAHs\ and \Lines\ components (see
\S\ref{sec:PAHEmission} and \S\ref{sec:LineEmission}); (2) Subtract the
initial \PAHs\ and \Lines\ components from the observed SED to create an
observed continuum emission spectrum; (3) Fit this spectrum to obtain the
model continuum; (4) Subtract the model continuum from the observed SED to
create an observed PAH feature and line emission spectrum; (5) Fit the
individual features in this spectrum; (6) Perform the sequence of tests to
ensure that the fit is well-constrained; and (7) If necessary, repeat the
fitting process with the constrained components.

\subsection{Data Weighting}
\label{sec:DataWeighting}

The success or failure of a fitting algorithm and the reliability of the
results is largely dependent upon the chosen method of weighting the data in
equation~(\ref{eq:Chi2}). We adopt the function
\begin{equation}
  \label{eq:Weights}
  w_k = \frac{\hat{\Lambda}_k}{\sigma^2(\lambda_k)},
\end{equation}
where $\sigma(\lambda_k)$ is the total $1$--$\sigma$ uncertainty in the flux
density at wavelength $\lambda_k$, and $\hat{\Lambda}_k$ is a term which
compensates for non-uniform wavelength sampling (i.e. ensuring that a region
of the SED does not dominate the fit simply because it is sampled more
finely---see Appendix~\ref{app:DataWeighting}). The total uncertainty in any
observed data is given by the quadratic sum of the statistical and
calibration (i.e.  systematic) uncertainties. For \IRS\ spectroscopic data,
we use the difference of two spectra obtained at different nod positions (as
produced during a typical \IRS\ observation) to calculate the statistical
uncertainties (see Appendix~\ref{app:IRSUncertainties} for details).
Carefully extracted low-resolution \IRS\ spectra may still exhibit some
residual calibration uncertainties (e.g. fringes in the long-low modules),
which we estimate to be $\sim$2\% of the flux density at any given
wavelength. For photometric data, the total uncertainty is taken to be equal
to the quadratic sum of the uncertainties provided in the literature and an
absolute calibration uncertainty (e.g. from scaling to the \IRS\ apertures)
taken to be 5\%. The actual systematic uncertainties in the photometric data
are difficult to estimate, although we note that the adopted uncertainties
yield reasonable reduced $\chi^2$ values, so are likely good estimates.


As described in \S\ref{sec:DecompositionAlgorithm}, each decomposition
consists of two separate least-squares fits. In the first (step 3 above),
the flux density to fit is obtained by subtracting the \PAHs\ and \Lines\
components from the observed SED, i.e.  $f_\nu = f_\nu^\data - f_\nu^\pahs -
f_\nu^\lines$. When a component is subtracted from observed data to create
an SED to be fitted, the uncertainty of the subtracted component must be
propagated through the analysis. Thus, the total uncertainty input into the
least-squares routine for this first fit is $\sigma^2 = \sigma^2_\data +
\sigma^2_\pahs + \sigma^2_\lines$.  Here, $\sigma_\data$ is the uncertainty
in the observed flux density, while $\sigma_\pahs$ and $\sigma_\lines$ are
the uncertainties in the estimated \PAHs\ and \Lines\ components,
respectively.  Since the estimated strengths of the \PAHs\ and \Lines\
components are derived from the observed spectrum, $\sigma_\pahs$ and
$\sigma_\lines$ are dominated by the \IRS\ uncertainties. We therefore adopt
$\sigma_\pahs = \sigma_\lines = 0$, since the uncertainties in the \IRS\
data on which they depend are already included through $\sigma_\data$. In
the second fit (step 5 above), the flux density to fit is obtained by
subtracting the fitted continuum from the observed SED, i.e.  $f_\nu =
f_\nu^\data - f_\nu^\cont$, so that the total uncertainty input into the
least-squares routine is $\sigma^2 = \sigma^2_\data + \sigma^2_\cont$. Here,
$\sigma_\cont$ is the formal uncertainty in the continuum calculated using
the full covariance matrix from the first fit.

\section{Source Emission}
\label{sec:SourceEmission}

\subsection{\ISRF\ Component Emission}
\label{sec:ISRFEmission}

%
\begin{figure}
  \epsscale{1.1}
  \plotone{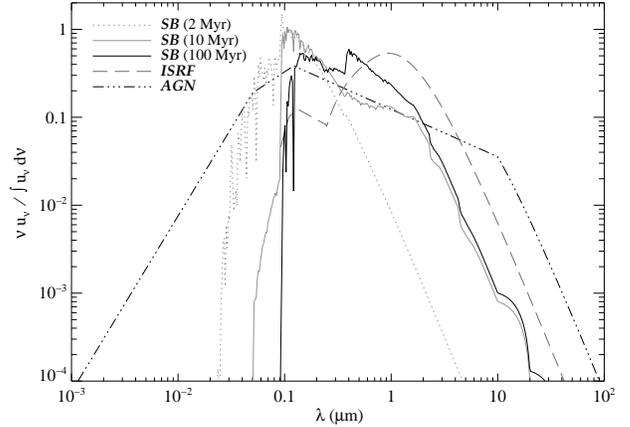}
  \caption{Normalized $10\Angstrom$--$1000\um$ energy densities of the
    principal luminosity sources in dusty galaxies including: emission from
    starbursts 2, 10, and 100~Myr after an instantaneous burst (\Starburst);
    evolved disk stars (\ISRF); and an AGN accretion disk (\AGN).}
  \label{fig:SourceSEDs}
  \epsscale{1.0}
\end{figure}
%

\begin{figure}
  \epsscale{1.1}
  \plotone{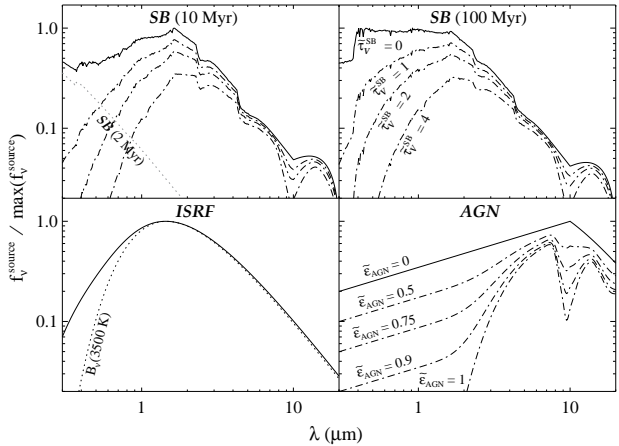}
  \caption{Example SEDs of emission from a starburst (\Starburst) 10 and
    100~Myr after an instantaneous burst, evolved disk stars (\ISRF), and an
    AGN accretion disk (\AGN). Also shown are the SEDs of the \Starburst\
    components extinguished by obscuring screens with the indicated optical
    depths, and the SEDs of the \AGN\ component behind an $A_V \approx 25$
    screen for the indicated covering factors ({\it dash-dotted lines}).
    Note that emission from the 2~Myr starburst population declines steeply
    at $\lambda > 3000\Angstrom$ and that the \ISRF\ component is very
    similar to a $3500\K$ blackbody for $\lambda > 1\um$ ({\it dotted
      lines}---see text).}
  \label{fig:SourceEmission}
  \epsscale{1.0}
\end{figure}

We adopt the model of the average interstellar radiation field presented in
%
\citet{1982A&A...105..372M}
and
%
\citet{2001ApJ...554..778L}
to represent the SED of emission from stars in the disk of a galaxy. In this
model, the mean energy density in the solar neighborhood, $c\,u_\nu^\isrf =
f_\nu^{\isrf,\odot} = 4 \pi J_\nu^{\isrf,\odot}$, is given by
\begin{equation}
  \label{eq:SourceSED-Phot}
  u_\nu^\isrf \propto
  \cases{0 & $\lambda < 912\Angstrom$, \cr
    \lambda^{5.4172} & $912\Angstrom < \lambda < 1100\Angstrom$, \cr
    \lambda^2 & $1100\Angstrom < \lambda < 1340\Angstrom$, \cr
    \lambda^{0.3322} & $1340\Angstrom < \lambda < 2460\Angstrom$, \cr
    \sum_n W_n B_\nu(T_n) & $\lambda > 2460\Angstrom$,}
\end{equation}
where $B_\nu$ is the Planck function per unit frequency and the $W_n =
[0.025$, $0.25$, $1]$ are weighting factors for each blackbody component of
temperature $T_n = [7500\K$, $4000\K$, $3000\K]$. This model spectrum is
shown in Figure~\ref{fig:SourceSEDs}, normalized to have unit integrated
energy density. Emission from the \ISRF\ decomposition component is assumed
to emerge nearly unobscured from a galaxy disk, so that the line-of-sight
extinction to the emitting stars is small in the infrared ($\lambda >
1\um$).  The flux density per unit frequency interval of emission from the
\ISRF\ component for use in equation~(\ref{eq:TotalFlux}) is therefore
\begin{equation}
  \label{eq:ISRFEmission}
  f_\nu^\isrf \propto u_\nu^\isrf.
\end{equation}
An example \ISRF\ component SED, $f_\nu^\isrf$, is shown in
Figure~\ref{fig:SourceEmission}. Displayed alongside the \ISRF\ component is
the emission from a $3500\K$ blackbody (characteristic of evolved stars)
scaled to match at $1.5\um$. The two curves are nearly identical at $\lambda
> 1\um$, so that fitting near-IR data with the \ISRF\ component is very
similar to fitting it with a $3500\K$ blackbody.

%
%

\subsection{\Starburst\ Component Emission}
\label{sec:StarburstEmission}

In contrast to the modest $\sim$1$\MSun \, {\rm yr}^{-1}$ star-formation
rate characteristic of emission from the \ISRF\ component, the rates in
starbursts are often much higher (e.g. $\ga 10$--$100\MSun \, {\rm
  yr}^{-1}$), resulting in significantly different stellar populations. In
order to properly characterize the spectral properties and bolometric
luminosities of these populations, we use Starburst99
%
\citep{1990ApJS...73....1L}
to generate SEDs of instantaneous bursts at different epochs.  We assume a
Kroupa IMF having a $M^{1.3}$ power-law in the $0.1$--$0.5\MSun$ range and a
$M^{2.3}$ power-law in the $0.5$--$100\MSun$ range. We further assume Padova
AGB tracks with solar metallicity.  Populations of stars in starbursts
evolve dramatically during the first 10~Myr after a burst. At 4~Myr the most
massive stars still exist on the main sequence, but by 10~Myr all of the
photoionizing stars have evolved into supergiants.  As shown in
Figure~\ref{fig:SourceSEDs}, the emission from a starburst 2~Myr after a
burst emerges almost exclusively at ultraviolet wavelengths.  Since we have
very little data in this wavelength range for the sources in our sample, we
are unable to explicitly constrain the presence or absence of such a
population.  In contrast to this, the 10 and 100~Myr populations emit a
large fraction of their luminosity at optical and longer wavelengths, where
our SEDs are better constrained. Therefore, in order to adequately sample
the evolutionary stages in a starburst that we are sensitive to, and to
smoothly transition into the evolved population modeled with the \ISRF\
component, we create a composite starburst spectrum containing equal
contributions from the 10 and 100~Myr populations---i.e. $u_\nu^\starburst =
(u_\nu^{\starburst10} + u_\nu^{\starburst100}) / 2$. We note, however, that
the omission of an ultraviolet-luminous young stellar population from
$u_\nu^\starburst$ may result in an underprediction of the total \Starburst\
luminosity for some sources, perhaps by up to $\sim$$1/3$ if the 2~Myr
population contributes at the same level as each of the older populations.


Unlike emission from the \ISRF\ component, we assume that \Starburst\
component emission emerges from regions that may be obscured by screens of
dust out to infrared wavelengths. If the obscuration to the \Starburst\
component is dominated by dust in the cocoons of material surrounding each
star-forming knot, and not by extinction between the various knots, then the
majority of the obscuration occurs via a screened geometry.
%
\citet{2007ApJ...654L..45L}
use radiative transfer models to show that such a screened geometry is
actually required to reproduce the deep $9.7\um$ silicate features observed
in many ULIRGs, suggesting that the assumption of a screened geometry is
reasonable for highly obscured sources. At lower optical depths (i.e. $\tau
< 1$), mixed and screened geometries both give rise to extinctions scaling
approximately as $f / f_0 \approx 1 - \tau$, so that the choice between the
form of extinction is less critical. We therefore model the flux density per
unit frequency interval of emission from the \Starburst\ component as
\begin{equation}
  \label{eq:StarburstEmission}
  f_\nu^\starburst \propto u_\nu^\starburst e^{-\tau_\starburst},
\end{equation}
where the optical depth through the obscuring screen to the \Starburst\
component is given by 
\begin{equation}
  \label{eq:StarburstTau}
  \tau_\starburst(\lambda) = \tilde{\tau}_{V}^\starburst
  \frac{\Sigma_\abs(\lambda)}{\Sigma_\abs(5500\Angstrom)}.
\end{equation}
Here, $\tilde{\tau}_{V}^\starburst$ is the screen optical depth to the
\Starburst\ component at $5500\Angstrom$, and $\Sigma_\abs = \Sigma_\ext -
\Sigma_\scat$ is the total dust absorption opacity obtained by subtracting
the scattering opacity from the total extinction opacity (see
\S\ref{sec:DustOpacity}). We use the absorption and not the extinction
opacity in equation~(\ref{eq:StarburstTau}) since scattering tends to cancel
itself out in sufficiently spherical geometries, such as depicted in
Figure~\ref{fig:GalaxyModel} (i.e. light scattered out of the beam along the
direct path to the observer is offset by light scattered into the beam along
another path).  Furthermore, the extinction and absorption opacities of dust
in our model have similar slopes between $1000\Angstrom$ and $3\um$, where
the \Starburst\ component emits strongly.  Since the derived optical depth
depends only on the opacity slope and not its absolute value, the omission
of scattering has very little effect. We note, however, that this omission
may introduce some uncertainties at $\lambda \la 3\um$, where the scattering
opacity is largest.  Example 10 and 100~Myr SEDs, $f_\nu^{\starburst10}$ and
$f_\nu^{\starburst100}$, used to construct the composite \Starburst\ model,
$f_\nu^\starburst$, are shown in Figure~\ref{fig:SourceEmission}.  Also
shown are example SEDs obscured by screens of dust with
$\tilde{\tau}_{V}^\starburst = 1$, $2$, and $4$.

\subsection{\AGN\ Component Emission}
\label{sec:AGNEmission}

The spectral properties of radiation emerging from an AGN accretion disk
depend sensitively upon the assumed geometry of the surrounding obscuring
structure---i.e. the putative torus of AGN unification models
%
\citep{1993ARA&A..31..473A}. Important geometrical factors include the
height of the inner edge of the torus, its proximity to the nucleus, the
clumpy versus smooth structure of the obscuring medium, and the orientation
of the torus with respect to the observer. Radiative-transfer calculations
incorporating these geometric properties have been used to construct models
of the emission from an AGN accretion disk and its obscuring torus
%
\citep[e.g.][]{1992ApJ...401...99P, 1993ApJ...402..441L,
  1994MNRAS.268..235G, 2002ApJ...570L...9N, 2005A&A...436...47D}.
Since our intent in this work is to characterize the general properties of
the emergent SEDs of large samples of galaxies, and not to discern the
detailed properties of the obscuring torus in individual sources, such
modeling is unnecessary for our purposes.


Instead, we adopt an empirical model of the emission from an AGN accretion
disk and assume that this SED may be extinguished by a foreground screen of
dust. We use the accretion disk model from
%
\citet{2005A&A...437..861S}
in which the emergent energy density, shown in Fig.~\ref{fig:SourceSEDs}, is
modeled by
\begin{equation}
  \label{eq:SourceSED-Disk}
  u_\nu^\agn \propto
  \cases{0 & $\lambda < 10\Angstrom$, \cr
    \lambda^3 & $10\Angstrom < \lambda < 500\Angstrom$, \cr
    \lambda^{1.8} & $500\Angstrom < \lambda < 1216\Angstrom$, \cr
    \lambda^{0.46} & $1216\Angstrom < \lambda < 10\um$, \cr
    B_\nu(1000\K) & $\lambda > 10\um$.}
\end{equation}
The flux density per unit frequency interval of emission from the \AGN\
component, including an obscuring screen with a non-uniform covering factor,
is modeled as
\begin{equation}
  \label{eq:AGNEmission}
  f_\nu^\agn \propto u_\nu^\agn 
  \left[\left(1 - \tilde{\epsilon}_\agn\right) + 
    \tilde{\epsilon}_\agn e^{-\tau_\agn}\right],
\end{equation}
where $0 \le \tilde{\epsilon}_\agn \le 1$ is the fraction of the accretion
disk obscured by a screen of optical depth $\tau_\agn$, with
\begin{equation}
  \label{eq:AGNTau}
  \tau_\agn(\lambda) = \tilde{\tau}_{V}^\agn
  \frac{\Sigma_\ext(\lambda)}{\Sigma_\ext(5500\Angstrom)}.
\end{equation}
Here, $\tilde{\tau}_{V}^\agn$ is the screen optical depth to the \AGN\
component at $5500\Angstrom$. An example unobscured \AGN\ component SED,
$f_\nu^\agn$, is shown in Figure~\ref{fig:SourceEmission}.  Also shown are
the SEDs of \AGN\ component emission obscured by a $\tilde{\tau}_{V}^\agn =
25$ screen, for several values of the screen covering factor
$\tilde{\epsilon}_\agn$. Note that significant ultraviolet and optical
emission may still emerge from a highly extinguished accretion disk if the
covering factor is less than unity.


In the bottom panel of Figure~\ref{fig:GalaxyModel}, the AGN accretion disk
is depicted as being surrounded by an obscuring structure composed of
discrete clouds and/or a smooth dusty component. The covering factor,
$\tilde{\epsilon}_\agn$, may therefore be interpreted as the fraction of the
accretion disk covered by this obscuring material (from the point-of-view of
the observer), which has an approximate visual band optical depth equal to
$\tilde{\tau}_{V}^\agn$. Note that the same qualifications regarding the
relative importance of scattering and the subsequent uncertainties in the
emergent flux density for $\lambda \la 3\um$ that were made for the
\Starburst\ component apply here as well. Since the clumpy torus is by
definition not spherically symmetric, we use the full extinction opacity
(including scattering) in equation~(\ref{eq:AGNTau}).  As described for the
\Starburst\ component, this choice has very little effect on the results
since the extinction and absorption opacities have similar slopes from the
ultraviolet to the near-IR, where the \AGN\ component emits strongly.

\section{Astronomical Dust Model}
\label{sec:DustModel}

\subsection{Grain Properties}
\label{sec:GrainProperties}

We adopt the grain-size distribution function from
%
\citet[][hereafter \WD]{2001ApJ...548..296W},
and the optical properties of graphitic carbon and smoothed astronomical
silicate from
\citet{1984ApJ...285...89D} and \citet{1993ApJ...402..441L},
with modifications from
\citet{2001ApJ...548..296W} and \citet{2001ApJ...554..778L}.
The \WD\ distribution function has been modified to include a small
grain-size exponential cutoff to model the deficit of small grains resulting
from sublimation in intense heating conditions (see
\S\ref{sec:GrainSublimation}). The modified distribution function for
graphite and silicate grains takes the form
\begin{equation}
  \label{eq:GrainSizeDF}
  \frac{1}{\nH} \frac{dn_i}{da} = 
  \left[\frac{1}{\nH} \frac{dn_i}{da}\right]_\WD
  \left(1 - \exp{\left[-\left(
          \frac{a}{a_{-}^i}\right)^{\beta_{-}}\right]}\right),
\end{equation}
where $[\nH^{-1} dn_i/da]_\WD$ is the \WD\ distribution function for grains
with radii satisfying $3.5\Angstrom \le a \la 1\um$, $a_{-} =
a_{-}(\tilde{U})$ is the minimum surviving grain-size in a distribution
embedded in a radiation field with an energy density of magnitude
$\tilde{U}$ (see \S\ref{sec:GrainSublimation} and
eq.~[\ref{eq:ThermalHeating}]), and $\beta_{-}$ determines the exponential
cutoff rate. In the limit of $\beta_{-} \rightarrow \infty$, the exponential
provides a sharp cutoff at $a = a_{-}$.  We choose $\beta_{-} = 3$, the same
value adopted for the large grain-size exponential cutoff in \WD.
Absorption and extinction efficiencies, $Q_\abs^i$ and $Q_\ext^i$, have been
tabulated for graphite and silicate spheres as described in
%
\citet{1993ApJ...402..441L}
for wavelengths $10\Angstrom < \lambda < 1000\um$. Absorption cross sections
are calculated using $C_\abs^\gra(a,\lambda) = \pi a^2
Q_\abs^\gra(a,\lambda)$ and $C_\abs^\sil(a,\lambda) = \pi a^2
Q_\abs^\sil(a,\lambda)$, with similar expressions for the extinction cross
sections $C_\ext^\gra(a,\lambda)$ and $C_\ext^\sil(a,\lambda)$.

\subsection{Grain Sublimation}
\label{sec:GrainSublimation}

Graphite and silicate grains sublimate when heated to temperatures above
$1750$ and $1400\K$, respectively
%
\citep{1993ApJ...402..441L}.
The minimum surviving grain-size in a distribution, $a_{-}$, is determined
by the size of the smallest grain which does not sublimate in the radiation
field in which the distribution is embedded. As the magnitude of the
radiation field energy density increases, the grain distribution becomes
increasingly depleted of small grains. A stronger radiation field therefore
changes the SED of emission from a distribution of grains both by increasing
the temperature of the grains, and by weighing the grain-size distribution
function to larger grains.  Additionally, due to their lower sublimation
temperatures, silicate grains become depleted before graphite grains.
Sublimation is one of many physical processes which can set the minimum
grain-size within a distribution. Shock waves
%
\citep{1994ApJ...433..797J, 1995Ap&SS.233..111D},
sputtering
%
\citep{1979ApJ...231...77D},
and grain-grain collisions
%
\citep{1994ApJ...431..321T}
can also influence the minimum grain-size. However, these other effects are
more difficult to model than sublimation, since they require a dynamic model
of the region where the grains are located. We do not consider them further.

\subsection{Thermal Heating}
\label{sec:ThermalHeating}

The equilibrium temperature attained by a grain is a function of its
composition, size, and the spectrum and strength of the radiation field in
which it is embedded. Our adopted dust model contains grains spanning
several orders of magnitude in size, so the temperatures of individual
grains within a distribution embedded in a given radiation field vary
greatly. For a graphite or silicate grain embedded in a radiation field with
normalized energy density $\hat{u}_\nu$ (see Fig.~\ref{fig:SourceSEDs})
which has a fitted magnitude $\tilde{U}$, energy conservation requires that
\begin{equation}
  \label{eq:ThermalHeating}
  \int C_\abs^i(a, \nu) \tilde{U} c \hat{u}_\nu \, d\nu =
  \int C_\abs^i(a, \nu) 4 \pi B_\nu[T_{\rm eq}^i(a)] \, d\nu.
\end{equation}
Here, the left and right sides represent the rates at which energy is
absorbed and emitted by a grain, respectively.
Equation~(\ref{eq:ThermalHeating}) is therefore an implicit equation for the
equilibrium temperatures, $T_{\rm eq}^i(a)$, of the graphite and silicate
grains in a distribution. These temperatures are used in
equation~(\ref{eq:DustEmissivity_i}) to properly calculate the spectrum of
emission from each decomposition component. We calculate the temperatures of
grains in the \Hot\ dust component using the normalized \AGN\ energy
density, $\hat{u}_\nu^\agn$, and the temperatures of grains in the \Warm,
\Cool, and \Cold\ dust components using the normalized \ISRF\ energy
density, $\hat{u}_\nu^\isrf$ (see Fig.~\ref{fig:SourceSEDs}). Note that
although the shape of the illuminating radiation field does affect the
temperatures of grains (since grains absorb and emit more efficiently at
certain wavelengths), similar grain temperatures may be obtained from two
different radiation fields by suitably adjusting the magnitude of the energy
density of each field. Thus, the specific choices about which illuminating
radiation fields we use to calculate our dust equilibrium temperatures do
not have a significant impact on our results. Note also that once the shape
of the normalized energy density is chosen, the temperature of a grain in
that field will scale with the fitted magnitude of the energy density,
$\tilde{U}$.

\subsection{Stochastic Heating}
\label{sec:StochasticHeating}

The heat capacities of very small grains are sufficiently small that they
undergo large temperature fluctuations upon absorption of individual
photons, and reradiate most of the deposited energy on timescales much
shorter than the typical time between photon strikes. To estimate the
importance of this stochastic heating for our dust components, we calculate
the threshold grain-size below which grains fall out of thermal equilibrium.
For graphite grains,
%
\citet{2001ApJ...551..807D}
calculate this to occur for
\begin{equation}\label{eq:aVSG}
  a_\Min \approx 100\Angstrom \left(\frac{u}{u_\MMP}\right)^{-0.2},
\end{equation}
where $u \equiv \int u_\nu d\nu$ is the integrated spectral energy density
of the illuminating radiation field, and $u_\MMP$ is the integrated spectral
energy density of the local interstellar radiation field from
%
\citet{1982A&A...105..372M}.
For our \Cold, \Cool, \Warm, and \Hot\ dust components, typical threshold
grain-sizes from equation~(\ref{eq:aVSG}) are $a_\Min \approx 53$, $14$,
$7.5$, and $3\Angstrom$.


In the model of
%
\citet{2001ApJ...554..778L},
stochastically heated grains exhibit strong spectral feature emission (i.e.
PAHs) but have very weak continua (due to the low continuum opacities of
small grains).  Most dusty galaxies contain a significant amount of
thermally heated $\bar{T} \ga 100\K$ dust which emits more continuum
radiation than these stochastically heated grains
%
\citep[see also][]{1993ApJ...402..441L}.
Thus, the continuum radiation from stochastically heated grains in our
\Cold\ component (i.e. grains with $a < 53\Angstrom$) is weak compared to
the thermal emission from warmer dust.  Additionally, since the threshold
grain-size decreases as the magnitude of the illuminating radiation field
energy density increases, stochastically heated continuum emission is even
less important for grains in our \Cool, \Warm, and \Hot\ components.  Since
the stochastic continuum emission of all components is weak compared to the
emission from thermally heated grains, and given that the stochastically
heated feature emission attributed to PAHs is included in our model through
the \PAHs\ component, we do not incorporate a detailed model of stochastic
heating.

\subsection{Dust Emissivity}
\label{sec:DustEmissivity}

The total emissivity per hydrogen nucleon of a distribution of grains (in
units of erg s$^{-1}$ Hz$^{-1}$ sr$^{-1}$ H$^{-1}$) is obtained by summing
over the emissivity of each grain species
\begin{equation}
  \label{eq:DustEmissivity}
  E_\nu = E_\nu^\gra + E_\nu^\sil.
\end{equation}
Here, the emissivities of graphite and silicate grains are
\begin{equation}
  \label{eq:DustEmissivity_i}
  E_\nu^i = \int_{50\Angstrom}^\infty
  \frac{1}{\nH} \frac{dn_i}{da} \ C_\abs^i(a, \lambda) 
  \ B_\nu[T_{\rm eq}^i(a,\tilde{\Tavg})] \, da,
\end{equation}
which is an average over the individual emissivities of the grains in the
distribution. Note that in contrast to models using a single effective
grain-size, the dust emissivity in our adopted model is {\it not} directly
proportional to the dust opacity (see \S\ref{sec:DustOpacity}), since the
equilibrium grain temperatures in equation~(\ref{eq:DustEmissivity_i})
depend upon grain-size and composition.

\subsection{Characteristic Distribution Temperature}
\label{sec:CharacteristicTemperature}

%
\begin{figure}
  \epsscale{1.1}
  \plotone{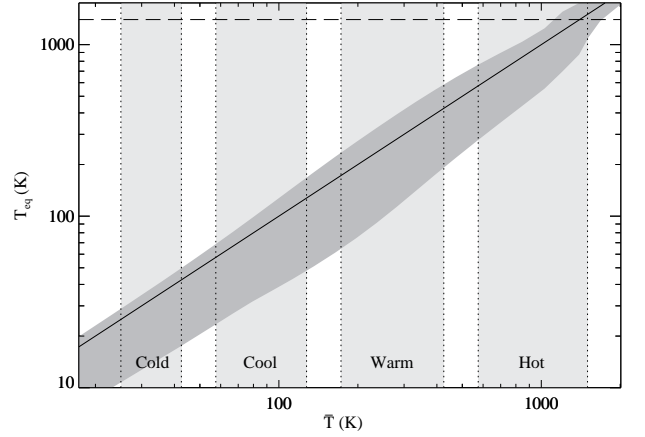}
  \caption{Range of equilibrium temperatures, $T_{\rm eq}$, of the grains
    within a distribution as a function of the characteristic distribution
    temperature $\Tavg$. The dark shaded region shows the range of
    equilibrium temperatures within a given distribution (i.e. at a fixed
    value of $\Tavg$), while the solid line shows their luminosity-weighted
    characteristic temperature. Light shaded regions indicate typical
    characteristic temperature ranges for the decomposition components.  The
    sublimation temperature of silicates is also labeled ({\it dashed
      line}).}
  \label{fig:DistributionTemperatures}
  \epsscale{1.0}
\end{figure}
%

As discussed in \S\ref{sec:ThermalHeating}, the many grains within a
distribution are brought to different equilibrium temperatures through
interactions with the radiation field in which they are embedded. In our
decomposition method this distribution of equilibrium temperatures is
determined from the fitted magnitude of the illuminating radiation field
energy density, $\tilde{U}$. A given dust component therefore has many
temperatures associated with it, none of which uniquely characterize the
properties of the distribution.  This is clearly not a desirable situation,
as we would like to associate a single characteristic temperature to each
value of $\tilde{U}$, preferably a temperature which communicates the
position of the spectral peak of the resulting emission from the
distribution. To meet this goal, we define the luminosity-weighted
characteristic distribution temperature
\begin{equation}
  \label{eq:CharacteristicTemperature}
  \Tavg \equiv \bar{T}(\tilde{U}) = \frac{
    \int_{50\Angstrom}^\infty 
    \sum_i f_L^i(a,\tilde{U}) T_{\rm eq}^i(a,\tilde{U}) \, da
  }{
    \int_{50\Angstrom}^\infty \sum_i f_L^i(a,\tilde{U}) \, da
  },
\end{equation}
where the sum is over the two grain compositions, and
\begin{equation}
  \label{eq:f_L}
  f_L^i(a,\tilde{U}) = \frac{1}{\nH} \frac{dn_i}{da} 
  \int C_\abs^i(a,\nu) B_\nu[T_{\rm eq}^i(a,\tilde{U})] \, d\nu
\end{equation}
is the grain-luminosity distribution function (which is essentially the
spectrally integrated emissivity of an individual grain---see
\S\ref{sec:DustEmissivity}). Note that the integration starts at $a =
50\Angstrom$, since smaller grains are not in thermal equilibrium.


Using this weighting function, $\Tavg$ is properly interpreted as the
composition-averaged temperature of the luminosity-dominating grain-size.
With this definition, the emission peak from a distribution of grains occurs
near the wavelength expected from the dust-modified Wien's law
\begin{equation}
  \label{eq:WiensLaw}
  \lambda_{\rm peak} \approx \left(\frac{300\K}{\Tavg}\right) 10\um,
\end{equation}
where the peak is defined to occur at the wavelength coinciding with the
maximum value of $E_\nu = \Sigma_\abs B_\nu(\Tavg)$ (i.e. the emissivity of
a distribution in the limit in which the temperatures of its constituent
grains are independent of their size and composition).
Figure~\ref{fig:DistributionTemperatures} shows the range of equilibrium
temperatures, $T_{\rm eq}$, of the grains within a distribution for a given
value of the characteristic distribution temperature, $\Tavg$. The dark
shaded region spans the range of equilibrium temperatures in a distribution,
while the solid line shows the luminosity-weighted average over that
distribution. Note that this average temperature is weighted towards the
warmer equilibrium temperatures since the grain-size distribution function
is weighted towards smaller (i.e. warmer) grains. The light shaded regions
indicate the typical ranges of the characteristic distribution temperatures
for the dust components in our decomposition (i.e. as determined by the
fitted values of $\tilde{U}$ for each component).

\subsection{Dust Opacity}
\label{sec:DustOpacity}

The total absorption opacity per hydrogen nucleon of a distribution of
grains (in units of cm$^2$ H$^{-1}$) is obtained by summing over the opacity
of each grain species
\begin{equation}
  \label{eq:DustOpacity}
  \Sigma_\abs(\lambda) = \Sigma_\abs^\gra(\lambda) + \Sigma_\abs^\sil(\lambda).
\end{equation}
Here, the opacities of graphite and silicate grains are
\begin{equation}
  \label{eq:DustOpacity_i}
  \Sigma_\abs^i(\lambda) = 
  \int_{50\Angstrom}^\infty \frac{1}{\nH} \frac{dn_i}{da} 
  C_\abs^i(a, \lambda) \, da,
\end{equation}
which is an average over the individual opacities of the grains in the
distribution.  The total extinction opacity per hydrogen nucleon,
$\Sigma_\ext$, is calculated in an analogous way with $C_\abs \rightarrow
C_\ext$. According to
%
\citet[][Table 6]{2001ApJ...554..778L},
$\Sigma_\abs \propto \lambda^{-2}$ over the range $20\um < \lambda < 700\um$
and $\Sigma_\abs \propto \lambda^{-1.68}$ over the range $700\um < \lambda <
10^4\um$ to within $\pm10$\%.

\subsection{Water-Ice and HAC Opacity}
\label{sec:IceOpacity}

%
\begin{figure}
  \epsscale{1.1}
  \plotone{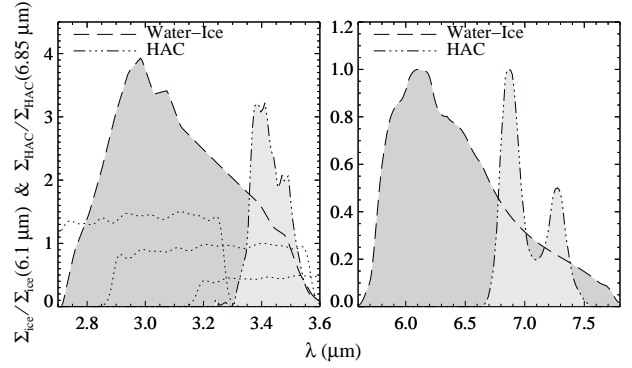}
  \caption{Water-ice and HAC opacity templates derived from L-band
    \citep{2006ApJ...637..114I} and \IRS\ \citep{2004ApJS..154..184S}
    spectra of ULIRGs (see text). Also shown in the left panel is the
    arbitrarily scaled \IRAC\ $3.6\um$ transmission curve for sources at $z
    = 0$, 0.1, and 0.2 (from right to left).}
  \label{fig:IceOpacity}
  \epsscale{1.0}
\end{figure}
%

Mid-IR spectra often exhibit 5.6--$7.8\um$ opacity produced from a
combination of water-ice absorption near $6.1\um$
%
\citep[see e.g.][]{2000ApJ...537..749C, 2000ApJ...536..347G}
and absorption from hydrogenated amorphous carbon (HAC) near $6.85\um$ and
$7.25\um$
%
\citep{1999ApJ...526..752F}.
Such water-ice and HAC absorption is particularly prominent in ULIRG spectra
%
\citep[e.g.][]{2002A&A...385.1022S}.
We derive 5.6--$7.8\um$ water-ice and HAC opacity templates from the \IRS\
spectrum of the heavily obscured ULIRG IRAS\,F00183-7111
%
\citep{2004ApJS..154..184S}.
This high signal-to-noise spectrum exhibits strong water-ice and HAC
absorption with very little PAH emission, thereby providing a clean spectrum
from which to extract profiles of the various opacity sources. We fit a
smooth spline to the observed spectrum between continuum points on either
side of the water-ice and HAC features, and use the ratio of the observed
spectrum to the estimated continuum to obtain the opacity templates
displayed in the right-panel of Figure~\ref{fig:IceOpacity}.


Additional opacity from $3.1\um$ water-ice and $3.4\um$ HAC features may
significant affect the broadband photometry of some sources (see
Fig.~\ref{fig:IceOpacity}). We therefore use the L-band spectra of ULIRGs
presented in
%
\citet{2006ApJ...637..114I}
to derive opacity templates for these features to use in our decompositions.
The opacities of the $3.1\um$ water-ice and $3.4\um$ HAC features are
derived from spectra of the deeply obscured ULIRGs IRAS\,00188-0856 and
IRAS\,08572+3915, respectively. Note that we do not derive the $6.1\um$
water-ice template from the \IRS\ spectrum of IRAS\,00188-0856 since it is
contaminated by $6.2\um$ PAH emission. As measured from their L-band and
\IRS\ spectra, the water-ice features of IRAS\,00188-0856 have apparent
$\tau_{3.1}^\ice / \tau_{6.1}^\ice \approx 3.9$ while the HAC features of
IRAS\,08572+3915 have apparent $\tau_{3.4}^\HAC / \tau_{6.85}^\HAC \approx
3.3$.  Our full opacity templates, shown in Figure~\ref{fig:IceOpacity}, are
constructed using these optical depth ratios to set the scalings between the
3.1--$3.4\um$ and 5.6--$7.8\um$ features.

\subsection{Total Optical Depth}
\label{sec:TotalOpticalDepth}

The total optical depth along a line-of-sight is the sum of the optical
depths from each opacity source. Thus, in our case
\begin{equation}
  \label{eq:OpticalDepth}
  \tau(\lambda) = \tau_\dust(\lambda) + \tau_\ice(\lambda) + \tau_\HAC(\lambda).
\end{equation}
The optical depth through a column of dust having total absorption or
extinction (depending on the situation) opacity $\Sigma(\lambda)$ is given
by
\begin{equation}
  \label{eq:DustOpticalDepth}
  \tau_\dust(\lambda) 
  = \NH \Sigma(\lambda)
  = \tilde{\tau}_{9.7}^\dust \frac{\Sigma(\lambda)}{\Sigma(9.7\um)},
\end{equation}
where $\NH$ and $\tilde{\tau}_{9.7}^\dust$ are the hydrogen nucleon column
density and $9.7\um$ optical depth through the dust, respectively. The
optical depth due to ice absorption is given by
\begin{equation}
  \label{eq:IceOpticalDepth}
  \tau_\ice(\lambda) = \tilde{\eta}_\ice \tilde{\tau}_{9.7}^\dust
  \frac{\Sigma_\ice(\lambda)}{\Sigma_\ice(6.1\um)},
\end{equation}
where $\Sigma_\ice$ is the ice opacity template from
Figure~\ref{fig:IceOpacity}, and $\tilde{\eta}_\ice \equiv \tau_{6.1}^\ice /
\tau_{9.7}^\dust$ determines the amount of ice opacity for a given dust
opacity. Similarly, the optical depth due to HAC absorption is given by
\begin{equation}
  \label{eq:HACOpticalDepth}
  \tau_\HAC(\lambda) = \tilde{\eta}_\HAC \tilde{\tau}_{9.7}^\dust
  \frac{\Sigma_\HAC(\lambda)}{\Sigma_\HAC(6.85\um)},
\end{equation}
where $\Sigma_\HAC$ is the HAC opacity template from
Figure~\ref{fig:IceOpacity}, and $\tilde{\eta}_\HAC \equiv \tau_{6.85}^\HAC
/ \tau_{9.7}^\dust$ determines the amount of HAC opacity for a given dust
opacity. The ratios of the apparent $6.1\um$ water-ice and $6.85\um$ HAC
optical depths to the apparent $9.7\um$ silicate optical depth are both
$\sim$0.1 for the various ULIRGs used to derive the opacity templates. We
therefore restrict the fitted values of these ratios to satisfy $0 \le
\tilde{\eta}_\ice \le 0.1$ and $0 \le \tilde{\eta}_\HAC \le 0.1$ in our
decomposition method.

\section{Dust Emission}
\label{sec:DustEmission}

\subsection{Optically-Thin Dust Emission}
\label{sec:OpticallyThinEmission}

%
\begin{figure}
  \epsscale{1.1}
  \plotone{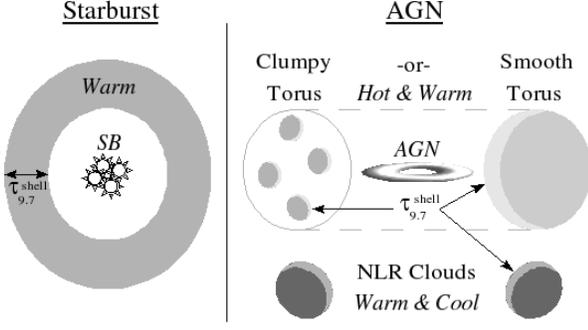}
  \caption{Cross-sectional views illustrating the optically-thin dust-shell
    approximation used to calculate the SEDs of dust components that are
    directly heated by ultraviolet and optical radiation from the
    \Starburst\ and \AGN\ source components (see text). The left-panel
    depicts a \Starburst\ source component surrounded by a shell of dust.
    The right-panel shows an \AGN\ accretion disk and torus (made either of
    discrete clouds as on the left {\it or} a smooth dust distribution as on
    the right), and more distant narrow-line region clouds.}
  \label{fig:SourceModel}
  \epsscale{1.0}
\end{figure}
%

Figure~\ref{fig:SourceModel} presents an illustration of the optically-thin
dust-shell approximation used to calculate the SEDs of emission from the
dust components in our decomposition that are heated directly by ultraviolet
and optical photons from the \Starburst\ and \AGN\ source components (see
also Fig.~\ref{fig:GalaxyModel}). The left-panel depicts a starburst
containing a \Starburst\ source component surrounded by a shell of dust.
The right-panel shows an AGN comprised of an \AGN\ accretion disk, a dusty
torus (either smooth or made of discrete clouds), and more distant
narrow-line region clouds.  In this geometry, ultraviolet and optical
photons emitted by the source components either escape freely or are
absorbed within shells of dust on the illuminated sides of clouds. By
definition, these shells have an optical depth of order unity at the
wavelength of peak energy absorption
\begin{equation}
  \label{eq:PeakEnergyAbsWave}
  \lambda_{\rm peak} = \frac{
    \int \Sigma_\abs(\lambda) u_\lambda \lambda \, d\lambda
  }{
    \int \Sigma_\abs(\lambda) u_\lambda \, d\lambda
  }.
\end{equation}
Here, the absorption-weighted distribution function is obtained by
integrating the energy absorbed by a single grain (i.e.  the left-side of
eq.~[\ref{eq:ThermalHeating}]) over the distribution. As described in
\S\ref{sec:ThermalHeating}, we assume that grains in the \Hot\ component are
heated by the \AGN\ radiation field (Fig.~\ref{fig:SourceSEDs}), while
grains in the other dust components are heated by the \ISRF\ field. The peak
absorption wavelengths from equation~(\ref{eq:PeakEnergyAbsWave}) for these
fields are $1470$ and $4550\Angstrom$, respectively. At these wavelengths,
the dust absorption opacity is approximately 20 and 5 times greater than its
value at $9.7\um$, so that the $9.7\um$ optical depths of the thin dust
shells illuminated by the \AGN\ and \ISRF\ radiation fields are
$\tau_{9.7}^{\rm shell} \approx 0.05$ and $0.2$, respectively.


All emission from these shells therefore emerges from regions which are
optically-thin at all infrared wavelengths (i.e. $\lambda \ga 1\um$).  The
emerging intensity from these shells, calculated from the radiative transfer
equation, is therefore
\begin{equation}
  \label{eq:DustIntensity}
  I_\nu = 
  \int_0^{\tau_{\rm shell}} S_\nu(\tau^\prime) 
  e^{-[\tau_{\rm shell} - \tau^\prime]} \, d\tau^\prime = 
  \left[1 - e^{-\tau_{\rm shell}}\right] S_\nu.
\end{equation}
Here, the source function $S_\nu \equiv E_\nu / \Sigma_\abs$, is assumed to
be constant (i.e. the magnitude of the radiation field energy density---and
thus the dust temperature distribution---is assumed to be uniform).  As
shown above, $\tau_{\rm shell} < 1$ for all infrared wavelengths, so that
equation~(\ref{eq:DustIntensity}) simplifies to
\begin{equation}
  \label{eq:DustIntensityApprox}
  I_\nu \approx \NH^{\rm shell} \, E_\nu,
\end{equation}
where $\NH^{\rm shell}$ is the hydrogen nucleon column density through the
dust shell (i.e. $\tau_{\rm shell} = \NH^{\rm shell} \Sigma_\abs$).

\subsection{Hot and Warm Component Emission}
\label{sec:HotWarmEmission}

%
\begin{figure}
  \epsscale{1.1}
  \plotone{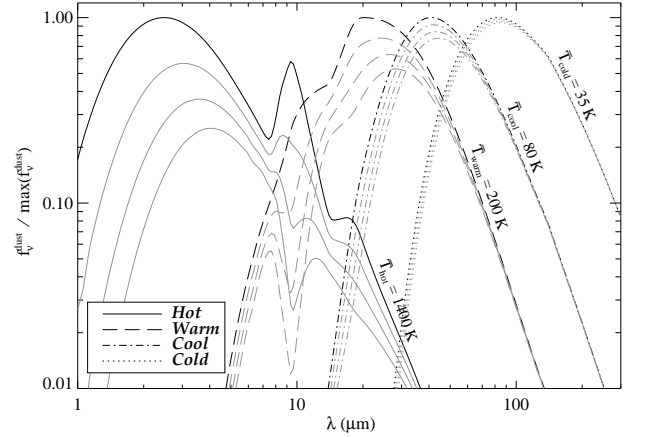}
  \caption{Example SEDs of emission from the \Hot, \Warm, \Cool, and \Cold\
    dust components for typical characteristic temperatures of $\Tavg =
    1400$, $200$, $80$, and $35\K$ ({\it black lines}).  Also shown are the
    SEDs obscured by screens ({\it gray lines}) having $\tau_{9.7} = 1$, 2,
    and 3 (from top to bottom).}
  \label{fig:DustEmission}
  \epsscale{1.0}
\end{figure}
%

Using equation~(\ref{eq:DustIntensityApprox}), the flux densities per unit
frequency interval of emission from the \Hot\ and \Warm\ components are
\begin{equation}
  \label{eq:HotWarmEmission}
  f_\nu^\hot \propto E_\nu(\Tavg_\hot) e^{-\tau_\hot} 
  \textrm{ and }
  f_\nu^\warm \propto E_\nu(\Tavg_\warm) e^{-\tau_\warm},
\end{equation}
where $\Tavg_\hot = \bar{T}(\tilde{U}_\hot)$ and $\Tavg_\warm =
\bar{T}(\tilde{U}_\warm)$ are the characteristic component temperatures
determined from the fitted values of $\tilde{U}_\hot$ and $\tilde{U}_\warm$
(see \S\ref{sec:CharacteristicTemperature}), and
\begin{equation}
  \label{eq:HotWarmTau}
  \tau_\hot = \tilde{\tau}_{9.7}^\hot 
  \frac{\Sigma_\abs(\lambda)}{\Sigma_\abs(9.7\um)}
  \textrm{ and }
  \tau_\warm = \tilde{\tau}_{9.7}^\warm
  \frac{\Sigma_\abs(\lambda)}{\Sigma_\abs(9.7\um)}
\end{equation}
are the optical depths through the screens of dust obscuring the \Hot\ and
\Warm\ components (e.g. provided by the \Cool\ and \Cold\ components in the
geometry of Fig.~\ref{fig:GalaxyModel}).  Here, $\tilde{\tau}_{9.7}^\hot$
and $\tilde{\tau}_{9.7}^\warm$ are the screen optical depths to the \Hot\
and \Warm\ components at $9.7\um$. Note that we use the absorption and not
the extinction opacity since scattering is negligible for $\lambda \ga
3\um$, where essentially all dust emission is radiated. Example \Hot\ and
\Warm\ component SEDs, $f_\nu^\hot$ and $f_\nu^\warm$, are shown in
Figure~\ref{fig:DustEmission} for the characteristic temperatures
$\Tavg_\hot = 1400\K$ and $\Tavg_\warm = 200\K$. The fitted \Hot\ component
characteristic temperature is constrained to satisfy $500\K < \Tavg_\hot <
1500\K$ and the fitted \Warm\ component characteristic temperature typically
satisfies $150\K < \Tavg_\warm < 500\K$. Also shown in
Figure~\ref{fig:DustEmission} are example \Hot\ and \Warm\ component SEDs
obscured by screens having $\tau_{9.7} = 1$, 2, and 3.  Note that departures
from a smooth continuum near $9.7$ and $18\um$ are caused by emission and
absorption from silicates.

\subsection{Cool and Cold Component Emission}
\label{sec:CoolColdEmission}

In the geometry of Figure~\ref{fig:GalaxyModel}, dust in the \Cool\ and
\Cold\ components may serve as sources of opacity to obscure the emission
from other components. Given this geometry, and the observation that the
mid-IR obscuration in dusty galaxies is often quite high
%
\citep[e.g. $\tau_{9.7} \approx 3$--$5$ for many of the local ULIRGs
in][]{2006ApJ...640..204A},
the optically-thin approximation used to obtain
equation~(\ref{eq:DustIntensityApprox}) is not necessarily valid for the
\Cool\ and \Cold\ components. However, as a result of their relatively cool
characteristic temperatures and $40$--$100\um$ SED peaks, the optical depth
through these components at the wavelengths over which they emit strongly is
a factor $\sim$10--$30$ times smaller than at $9.7\um$.  Therefore, despite
potentially large $9.7\um$ optical depths, both components are actually
optically-thin at the wavelengths over which the majority of their emission
is radiated. Thus, the conditions required for the validity of
equation~(\ref{eq:DustIntensityApprox}) are in fact still satisfied so that
the flux densities per unit frequency interval of emission from the \Cool\
and \Cold\ dust components are
\begin{equation}
  \label{eq:CoolColdEmission}
  f_\nu^\cool \propto E_\nu(\Tavg_\cool)
  \textrm{ and }
  f_\nu^\cold \propto E_\nu^\cold(\Tavg_\cold),
\end{equation}
where $\Tavg_\cool = \bar{T}(\tilde{U}_\cool)$ and $\Tavg_\cold =
\bar{T}(\tilde{U}_\cold)$ are the characteristic component temperatures
determined from the fitted values of $\tilde{U}_\cool$ and
$\tilde{U}_\cold$. We assume that emission from both the \Cool\ and \Cold\
components emerges unextinguished since the optical depths required to
achieve significant extinctions at $\lambda \ga 50\um$ where they emit
strongly would imply implausibly high optical depths at shorter wavelengths.
Example \Cool\ and \Cold\ component SEDs, $f_\nu^\cool$ and $f_\nu^\cold$,
are shown in Figure~\ref{fig:DustEmission} for the characteristic
temperatures $\Tavg_\cool = 80\K$ and $\Tavg_\cold = 35\K$.  The fitted
\Cool\ and \Cold\ component characteristic temperatures typically satisfy
$50\K < \Tavg_\cool < 100\K$ and $25\K < \Tavg_\cold < 40\K$, respectively.
Also shown in Figure~\ref{fig:DustEmission} are the \Cool\ and \Cold\ SEDs
obscured by screens having $\tau_{9.7} = 1$, 2, and 3, demonstrating that
the obscured \Cool\ and \Cold\ component SEDs are changed very little.

\subsection{Comparison with Other Methods}
\label{sec:ComparisonWithOtherMethods}

The emergent flux density from dust embedded in a distribution of radiation
fields with energy densities $U$ is
\begin{equation}
  \label{eq:IntegratedFlux}
  f_\nu^{\rm dust} \propto \int 
  \frac{dM_\dust}{dU} E_\nu[\bar{T}(U)] \, dU.
\end{equation}
Here, $M_\dust(U)$ is the mass of dust heated to the characteristic
temperature $\bar{T}(U)$ by the field with strength $U$. Our four-component
dust model characterizing the emission from a galaxy is derived from
equation~(\ref{eq:IntegratedFlux}) by taking
\begin{equation}
  \label{eq:dMdU_DDG}
  \frac{dM_\dust}{dU} = \sum_{i=1}^4 M_\dust^i \delta(U - U_i),
\end{equation}
where the sum is over four discrete values of $U_i$ illuminating a mass
$M_\dust^i$ of dust each. Clearly, equation~(\ref{eq:dMdU_DDG}) is an
oversimplification of the true distribution of mass in a galaxy.
%
\citet{2001ApJ...549..215D}
and
%
\citet{2002ApJ...576..762L}
both calculate the integrated emission from dust in galaxies by assuming a
power-law distribution of dust mass over heating intensity
\begin{equation}
  \label{eq:dMdU_Dale}
  \frac{dM_\dust}{dU} \propto U^{-\alpha},
\end{equation}
where $\alpha$ determines the relative contributions from each value of the
radiation field energy density. Given a sufficient number of discrete
temperature components, the model in equation~(\ref{eq:dMdU_DDG}) can be
used to approximate the emission from the power-law model to arbitrary
precision (with the appropriate weighting of each component through the
$M_\dust^i$).  The fact that both our multi-component model and these
power-law models produce excellent fits to the integrated spectra of
galaxies implies that the required `sufficient number' of discrete
components postulated above must be approximately four.


The real power of our multi-component approach stems from the fact that it
is capable not only of approximating the emission from the mass distribution
of star-forming galaxies in equation~(\ref{eq:dMdU_Dale}), but also the
emission from mass distributions which are {\it not} characterized by a
power-law (since we do not make the geometrical assumptions concerning the
distribution of dust within a galaxy implied by eq.~[\ref{eq:dMdU_Dale}]).
For example, the composite emission from a galaxy containing both a
starburst and an AGN may have a total mass distribution function dominated
by a star-formation driven power-law at low energy densities superposed with
an AGN driven power-law (having a different exponent) at high energy
densities.  Since the sum of two power-laws with different exponents is not
necessarily itself a power-law, a model such as
equation~(\ref{eq:dMdU_Dale}) cannot properly model this scenario.  This,
coupled with the fact that our method allows us to fit for different levels
of obscuration towards each component (as expected for physically distinct
AGN and starburst regions), drives our decision to utilize the
multi-component discrete temperature approach to describe the composite SEDs
of dusty galaxies.

\subsection{Dust Emission in Previous Publications}
\label{sec:PreviouslyPublished}

The decompositions of IRAS\,10214+4724 in
%
\citet{2006ApJ...638L...1T},
NGC\,6240 in
%
\citet{2006ApJ...640..204A},
and the BGS sample of ULIRGs in
\citet{2007ApJ...656..148A}
were performed using an earlier version of the method described in this work
which utilized a different model to calculate the emission from each dust
component. In these previous publications, dust emission is calculated from
an optically thin shell of constant density material surrounding the
illuminating source. As in equation~(\ref{eq:dMdU_Dale}), the dust in this
shell is exposed to different radiation fields (depending on its distance
from the central source), and is therefore brought to different equilibrium
temperatures. In these previous works, we assume that the dust shells had a
thickness $r_{\rm out} / r_{\rm in} = 10$, which fixes the range of
radiation fields and therefore dust temperatures, once the temperature at
the minimum radius is defined (taken to be the sublimation temperature).
Since each of these dust components contains a range of temperatures (from
dust at different radii), only three components were needed to fit the
sources instead of the four used in the current work. We choose to adopt the
four-component method for this work since it improves the fits at long
wavelengths and simplifies the interpretation.

\section{PAH Emission}
\label{sec:PAHEmission}

%
\begin{figure}
  \epsscale{1.1}
  \plotone{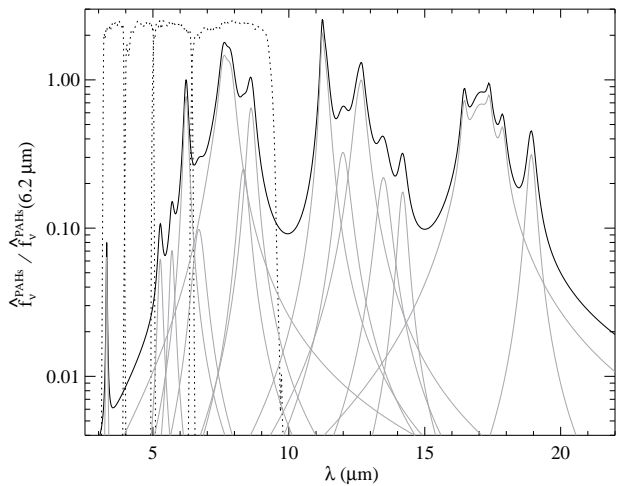}
  \caption{PAH emission template ({\it black line}) derived from the
    spectrum of the mean starburst galaxy from \citet{2006ApJ...653.1129B}.
    The individual complexes ({\it gray lines}) are modeled with Drude
    profiles \citep[see][]{2007ApJ...656..770S}. Note the significant
    `continuum' emission created by the addition of flux in the wings of the
    broad features. Also shown for reference are the four arbitrarily scaled
    \IRAC\ transmission curves.}
  \label{fig:PAHTemplate}
  \epsscale{1.0}
\end{figure}
%

Observations with \ISO\ and \Spitzer\ have shown that the presence of PAHs
%
\citep{1984A&A...137L...5L, 1985ApJ...290L..25A}
can be used as an indicator of star-formation
%
\citep[e.g.][]{2004A&A...419..501F, 2004ApJ...613..986P}.
We note, however, that 
%
\citet{2005ApJ...633..871C}
find that PAH emission does not correlate well with other star-formation
tracers in M\,51. Additionally, 
%
\citet{2005ApJ...628L..29E}
and
%
\citet{2006ApJ...639..157W}
find that PAH emission varies with metallicity, so may therefore not be a
useful tracer of star-formation in low-metallicity systems. With these
cautionary notes in mind, the presence or absence of PAH emission
nevertheless provides an important clue to help distinguish between
star-formation and accretion dominated ULIRGs
%
\citep[e.g.][]{1998ApJ...498..579G},
thereby necessitating the development of a method to carefully decompose the
contributions from PAHs to the total emission from dusty galaxies.


This process is complicated by the fact that it is difficult to establish
the level of the continuum beneath the PAH features, especially in the
presence of $9.7\um$ silicate absorption.  The method presented here
provides a systematic means of estimating this continuum---and therefore the
strength of the PAH features---without the need for the observer to define
the continuum level by hand. We model the flux density per unit frequency
interval of the \PAHs\ component as
\begin{equation}
  \label{eq:PAHFlux}
  f_\nu^\pahs = f_0^\pahs \frac{\hat{f}_\nu^\pahs}{\hat{f}_\nu^\pahs(6.2\um)},
\end{equation}
where $f_0^\pahs \equiv f_\nu^\data(6.2\um) - f_\nu^{\rm cont}(6.2\um)$ is
the estimated peak flux density of the $6.2\um$ PAH feature obtained from
the local continuum-subtracted observed spectrum, and
\begin{equation}
  \label{eq:PAHTemplateFlux}
  \hat{f}_\nu^\pahs = \sum_i \frac{\gamma_i^2 f_{\nu,0}^i}
  {\left(\lambda / \lambda_0^i - \lambda_0^i / \lambda \right)^2 + \gamma_i^2}
\end{equation}
is the PAH emission template shown in Figure~\ref{fig:PAHTemplate}
constructed by summing over a series of Drude profiles with the parameters
presented in Table~\ref{tab:PAHTemplate}. As suggested by the definition of
$f_0^\pahs$, the overall strength of the PAH template is held fixed during
the continuum fit, but the strengths of the individual features are allowed
to vary in the subsequent PAH fit (see \S\ref{sec:DecompositionAlgorithm}
and the following paragraph).

%
\begin{deluxetable}{ccccccc}
\tabletypesize{\scriptsize}
\tablewidth{0pc}

\tablecaption{PAH Template Parameters\label{tab:PAHTemplate}}

\tablehead{
  \colhead{Feature} & 
  \colhead{$\lambda_0^i$} &
  \colhead{$\gamma_i$} &
  \colhead{$f_{\nu,0}^i$} &
  \colhead{$L_j / \sum_j L_j\tablenotemark{a,b}$} &
  \colhead{Complex\tablenotemark{c}} \\
  \colhead{$i$} & 
  \colhead{$(\um)$} &
  \colhead{$(10^{-2})$} &
  \colhead{(Jy)} &
  \colhead{$(10^{-2})$} &
  \colhead{$j$}
}

\startdata
 1 &  3.30 &  1.2 & 0.07 &  0.6 &  1 \\
 2 &  5.27 &  3.4 & 0.06 &  1.0 &  2 \\
 3 &  5.70 &  3.5 & 0.07 &  1.1 &  3 \\
 4 &  6.22 &  3.0 & 0.77 &  9.7 &  4 \\
 5 &  6.69 &  7.0 & 0.10 &  2.7 &  5 \\
 6 &  7.42 & 12.6 & 0.24 & 39.3 &  6 \\
 7 &  7.60 &  4.4 & 0.89 &  --  & -- \\
 8 &  7.85 &  5.3 & 0.86 &  --  & -- \\
 9 &  8.33 &  5.0 & 0.25 &  3.9 &  7 \\
10 &  8.61 &  3.9 & 0.65 &  7.6 &  8 \\
11 & 11.23 &  1.2 & 1.29 & 11.3 &  9 \\
12 & 11.33 &  3.2 & 1.04 &  --  & -- \\
13 & 11.99 &  4.5 & 0.32 &  3.2 & 10 \\
14 & 12.62 &  4.2 & 0.85 &  7.9 & 11 \\
15 & 12.69 &  1.3 & 0.19 &  --  & -- \\
16 & 13.48 &  4.0 & 0.22 &  1.7 & 12 \\
17 & 14.19 &  2.5 & 0.18 &  0.8 & 13 \\
18 & 16.45 &  1.4 & 0.43 &  8.4 & 14 \\
19 & 17.04 &  6.5 & 0.64 &  --  & -- \\
20 & 17.38 &  1.2 & 0.29 &  --  & -- \\
21 & 17.87 &  1.6 & 0.26 &  --  & -- \\
22 & 18.92 &  1.9 & 0.31 &  0.8 & 15 
\enddata

\tablenotetext{a}{Fraction of total PAH luminosity emerging from each
  complex $j$.}

\tablenotetext{b}{$L_j = \sum_{i \in j} L_i$ is the total luminosity of a
  complex of Drude profiles, where $L_i = 4 \pi D_L^2 (\pi / 2) \lambda_0^i
  \gamma_i f_{\lambda,0}^i$ is the apparent total luminosity of a single
  profile, with $D_L$ the luminosity distance to the source.}

\tablenotetext{c}{As defined in \citet{2007ApJ...656..770S}, the $7.7$,
  $11.3$, $12.7$, and $17\um$ complexes are each composed of multiple Drude
  profiles.}

\end{deluxetable}
%

The central wavelengths and widths of our $\lambda_0 > 3.3\um$ features are
adapted from those used in the analysis of
%
\citet{2007ApJ...656..770S},
and their relative strengths are derived using {\sc Pahfit}\footnote{See
  http://turtle.as.arizona.edu/jdsmith/pahfit.php.} to fit the low
obscuration ($\tau_{9.7} \approx 0.24$) mean starburst galaxy spectrum
created from 13 \IRS\ spectra in
%
\citet{2006ApJ...653.1129B}.
The $3.3\um$ PAH feature central wavelength and width are obtained from
%
\citet{2001ApJ...554..778L},
and the peak flux density of this feature is derived from the ratio of the
strengths of the PAH features at $3.3$ and $6.2\um$ in the L-band and \IRS\
spectra of the star-forming ULIRG IRAS\,12112+0305 presented in
%
\citet{2006ApJ...637..114I}
and
%
\citet{2006ApJ...640..204A}.
We include this $3.3\um$ PAH feature so that its contribution to the
integrated L-band (or IRAC $3.6\um$ channel---see
Fig.~\ref{fig:PAHTemplate}) flux density is explicitly included in the
fitting method. As described in \S\ref{sec:DecompositionAlgorithm}, after
fitting the continuum, we fit the continuum-subtracted observed spectrum to
refine our \PAHs\ component (the properties of which are held fixed during
the continuum fit). In this fit, the central wavelengths and FWHM of the PAH
features are fixed to the values in Table~\ref{tab:PAHTemplate}, and the
peaks are allowed to vary freely. This fit provides accurate luminosities
for all PAH complexes, as well as error estimates derived from the full
covariance matrix.

\section{Atomic and Molecular Line Emission}
\label{sec:LineEmission}

%
\begin{deluxetable}{ccccccc}
\tabletypesize{\scriptsize}
\tablewidth{0pc}

\tablecaption{Atomic and Molecular Line Parameters\label{tab:Lines}}

\tablehead{
  \colhead{} &
  \colhead{} & 
  \colhead{$\lambda_0^i$} &
  \colhead{$\gamma_i^{z=0}$} &
  \colhead{$\gamma_i^{z=0.1}$} &
  \colhead{$\gamma_i^{z=0.25}$} &
  \colhead{$\gamma_i^{z=0.5}$} \\
  \colhead{i} & 
  \colhead{Line} &
  \colhead{$(\um)$} &
  \colhead{$(10^{-2})$} &
  \colhead{$(10^{-2})$} &
  \colhead{$(10^{-2})$} &
  \colhead{$(10^{-2})$}
}

\startdata
 1 &       H$_2$S(8) &  5.053 & \nodata &     6.0 &     6.0 &    12.1 \\
 2 &  [Fe\,{\sc ii}] &  5.340 &     6.0 &     6.0 &     6.0 &    12.1 \\
 3 &       H$_2$S(7) &  5.511 &     6.0 &     6.0 &     6.0 &    12.1 \\
 4 &  [Fe\,{\sc ii}] &  5.674 &     6.0 &     6.0 &     6.0 &    12.  \\
 5 &       H$_2$S(6) &  6.109 &     6.0 &     6.0 &    12.1 &    12.1 \\
 6 &  [Fe\,{\sc ii}] &  6.721 &     6.0 &     6.0 &    12.1 &    12.1 \\
 7 &       H$_2$S(5) &  6.910 &     6.0 &    12.1 &    12.1 &    12.1 \\
 8 &  [Ar\,{\sc ii}] &  6.985 &     6.0 &    12.1 &    12.1 &    12.1 \\
 9 &  [Ne\,{\sc vi}] &  7.652 &    12.1 &    12.1 &    12.1 &    12.1 \\
10 &       H$_2$S(4) &  8.025 &    12.1 &    12.1 &    12.1 &    12.1 \\
11 & [Ar\,{\sc iii}] &  8.991 &    12.1 &    12.1 &    12.1 &    12.1 \\
12 &       H$_2$S(3) &  9.665 &    12.1 &    12.1 &    12.1 &    16.9 \\
13 &   [S\,{\sc iv}] & 10.511 &    12.1 &    12.1 &    12.1 &    16.9 \\
14 &       H$_2$S(2) & 12.279 &    12.1 &    12.1 &    16.9 &    16.9 \\
15 &  [Ne\,{\sc ii}] & 12.814 &    12.1 &    12.1 &    16.9 &    16.9 \\
16 &   [Ne\,{\sc v}] & 14.322 &    16.9 &    16.9 &    16.9 &    33.9 \\
17 & [Ne\,{\sc iii}] & 15.555 &    16.9 &    16.9 &    22.3 &    33.9 \\
18 &       H$_2$S(1) & 17.035 &    16.9 &    16.9 &    33.9 &    33.9 \\
19 &  [S\,{\sc iii}] & 18.713 &    16.9 &    33.9 &    33.9 &    33.9 \\
20 &   [Ne\,{\sc v}] & 24.317 &    33.9 &    33.9 &    33.9 &    33.9 \\
21 &   [O\,{\sc iv}] & 25.890 &    33.9 &    33.9 &    33.9 & \nodata \\
22 &       H$_2$S(0) & 28.219 &    33.9 &    33.9 &    33.9 & \nodata \\
23 &  [S\,{\sc iii}] & 33.481 &    33.9 &    33.9 & \nodata & \nodata \\
24 &  [Si\,{\sc ii}] & 34.815 &    33.9 & \nodata & \nodata & \nodata \\
25 & [Ne\,{\sc iii}] & 36.014 &    33.9 & \nodata & \nodata & \nodata 
\enddata

\tablecomments{The FWHM = $\lambda_0^i \gamma_i$ of a particular atomic or
  molecular line is a function of the \IRS\ module in which it is detected,
  and therefore depends on the redshift of the source. We give example
  values of $\gamma_i$ for all lines at $z = 0$, 0.1, 0.25 and 0.5. Values
  of $\gamma_i$ are not given for lines that fall out of the \IRS\
  wavelength range at a given redshift.}

\end{deluxetable}
%

In order to accurately fit the continuum and PAH features in an \IRS\
spectrum, the contributions from atomic and molecular lines must be properly
fit as well. Of course, the strength of these lines is scientifically
interesting, but we caution against using our decomposition method to
measure them since the fitted continuum around each line is not necessarily
as accurate as one estimated by hand (a statement which cannot be made
concerning the underlying continuum beneath PAH features).  Therefore,
although we obtain integrated flux values for the lines we fit, we do not
recommend using these measurements for scientific purposes (at least not
without carefully checking the validity of the fitted local continuum around
each line). The one caveat to this recommendation where we do suggest using
the fitted integrated line flux is for the [\ion{Ne}{2}]~$12.81\um$ line,
which our decomposition method is capable of deblending from the $12.7\um$
PAH complex in low-resolution \IRS\ spectra.


We create our atomic and molecular emission line spectrum by estimating the
peak flux density of each line in Table~\ref{tab:Lines} from the local
continuum-subtracted \IRS\ spectrum. The flux density per unit frequency
interval of the \Lines\ component is modeled as a sum over unresolved
Gaussian profiles having central wavelengths $\lambda_0$, widths $\gamma_i$,
and peak flux densities $f_{\nu,0}$
\begin{equation}
  \label{eq:LineEmission}
  f_\nu^\lines = \sum_i f_{\nu,0}^i \exp\left[-\frac{1}{2} 
    \left(\frac{\lambda - \lambda_0^i}
      {\lambda_0^i \gamma_i / (2 \sqrt{2 \ln{2}})}\right)^2\right].
\end{equation}
Since the Gaussian profiles are unresolved, the width of each line is a
function of the \IRS\ module in which it is detected (see
Table~\ref{tab:Lines} for values of $\gamma$ as a function of source
redshift). As described in \S\ref{sec:PAHEmission} for the \PAHs\ component,
after fitting the continuum we perform a fit to the continuum-subtracted
observed spectrum in order to refine the \Lines\ component (the parameters
for which are held fixed in the continuum fit). In this fit, the central
wavelengths and FWHM of the atomic and molecular emission lines are fixed to
the values in Table~\ref{tab:Lines}, and their peak flux densities are
allowed to range freely. From this fit, we obtain line luminosities and
their uncertainties---although, as noted above, we recommend measuring all
unblended line fluxes by hand.

\section{Constructing Dusty Galaxy SED\lowercase{s}}
\label{sec:ConstructingDustySEDs}

\subsection{Sample Selection}
\label{sec:SampleSelection}

%
\begin{deluxetable}{ccccc}
\tabletypesize{\scriptsize}
\tablewidth{0pc}

\tablecaption{Dusty Galaxy Properties\label{tab:DustyGalaxyProperties}}

\tablehead{
  \colhead{} &
  \colhead{} &
  \colhead{$D_L$} &
  \colhead{} &
  \colhead{} 
  \\
  \colhead{Galaxy} &
  \colhead{$z$} &
  \colhead{(Mpc)} &
  \colhead{Class} &
  \colhead{Reference}
}

\startdata
NGC\,7714    & 0.009 &  39 &          Starburst &  1,2 \\
NGC\,2623    & 0.018 &  78 & Obscured Starburst &    2 \\
PG\,0804+761 & 0.130 & 610 &             Quasar &    5 \\
Mrk\,463     & 0.050 & 222 &          Seyfert 2 &    4 \\
NGC\,6240    & 0.024 & 105 &      Starburst+AGN &    3 \\
Mrk\,1014    & 0.163 & 781 &   Quasar+Starburst &    4
\enddata

\tablerefs{
  [1] \citet{2004ApJS..154..188B};
  [2] \citet{2006ApJ...653.1129B};
  [3] \citet{2006ApJ...640..204A};
  [4] \citet{2004ApJS..154..178A};
  [5] \citet{2005ApJ...625L..75H}.
}

\tablecomments{The \IRS\ spectra of the first three sources are from
  \Spitzer\ program 14, and the spectra of the last three are from \Spitzer\
  program 105. Luminosity distances are calculated assuming a flat
  $\Lambda$CDM cosmology with $H_0 = 70$~km~s$^{-1}$~Mpc$^{-1}$, $\Omega_M =
  0.3$, and $\Omega_\Lambda = 0.7$.}

\end{deluxetable}
%

Table~\ref{tab:DustyGalaxyProperties} presents a summary of the galaxies in
our sample. They have been selected with the goal of demonstrating the
capabilities of our decomposition method on a variety of dusty galaxy SEDs,
representative of the diverse properties of the group. We include examples
of unextinguished and obscured starbursts, a quasar, a Seyfert 2 galaxy, and
two composite systems powered by both AGN and starburst activity (see
\S\ref{sec:DetailedDescriptions} for details about each source). The first
four galaxies in this sample serve as templates to aid in understanding the
properties of the components from which composite systems are built. Our
sample includes star-forming galaxies covering a range of infrared
luminosities---from a minimum $\LIR \approx 5\times10^{10} \LSun$ for
NGC\,7714 to more than a factor $\sim$10 higher for NGC\,6240.  While it is
true that this sample does not include examples of `normal' star-forming
galaxies, our decomposition method is nonetheless valid for these sources as
well.  In fact, our method is applicable to {\it all} galaxies---including
AGNs, starbursts, normal, dwarfs, and ellipticals---so long as their SEDs
are composed of emission from stars, dust, PAHs, atomic and molecular lines,
and possibly an AGN accretion disk (i.e. our decomposition components).

\subsection{IRS Spectroscopy}
\label{sec:IRSSpectroscopy}

The \IRS\ spectra of the sources in our sample were first presented in the
papers referred to in Table~\ref{tab:DustyGalaxyProperties}. All spectra in
this paper have been extracted using the method described in
%
\citet{2006ApJ...640..204A}.
In brief, the \IRS\ pipeline at the Spitzer Science Center was used to
reduce the data, and one-dimensional spectra were extracted using the SMART
package
%
\citep{2004PASP..116..975H}.
After scaling and stitching the individual orders together, each spectrum
was scaled to match the \MIPS\ $24\um$ flux density if available (see
\S\ref{sec:IRACandMIPS}), or the \IRAS\ $25\um$ flux density otherwise.

\subsection{Supplementary Photometry}
\label{sec:SupplementaryPhotometry}

%
\begin{deluxetable}{cccccc}
\tabletypesize{\scriptsize}
\tablewidth{0pc}

\tablecaption{Supplementary Photometry\label{tab:SupplementaryPhotometry}}

\tablehead{
  \colhead{Galaxy} &
  \colhead{Ultraviolet} & 
  \colhead{Optical} &
  \colhead{Near-IR} &
  \colhead{Far-IR} &
  \colhead{Sub-mm}
}

\startdata
NGC\,7714    &  1 &   5 &     5 & 12,13 &   13,17 \\
NGC\,2623    &  2 &   6 &     9 &    12 &   17,18 \\
PG\,0804+761 &  3 &   3 & 10,11 & 14,15 & \nodata \\
Mrk\,463     &  1 &   7 &    10 & 14,16 & \nodata \\
NGC\,6240    &  4 &   8 &     9 & 12,16 &   16,18 \\
Mrk\,1014    &  3 &   7 &     9 & 14,15 & \nodata
\enddata

\tablerefs{
  [1] \citet{1993ApJS...86....5K};
  [2] GALEX (2006);
  [3] \citet{2005MNRAS.356.1029B};
  [4] \citet{1992ApJ...391L..81S};
  [5] \citet{2001ApJ...552..150L};
  [6] \citet{2005ApJ...630..784T};
  [7] \citet{2000AJ....120..604S};
  [8] \citet{1991trcb.book.....D};
  [9] \citet{2000AJ....119..991S};
  [10] 2MASS PSC;
  [11] \citet{1987ApJS...63..615N};
  [12] \citet{2003AJ....126.1607S};
  [13] \citet{1998A&A...331L...9K};
  [14] \citet{1990IRASF.C......0M};
  [15] \citet{2003A&A...402...87H};
  [16] \citet{2001A&A...379..823K};
  [17] \citet{2000MNRAS.315..115D};
  [18] \citet{1999PhDT.........5B}.
}

\end{deluxetable}
%

The spectral coverage of the \IRS\ low-resolution modules ends at an
observed-frame wavelength of $\sim$38$\um$. Emission from typical \Cool\
component dust ($\Tavg\sim80\K$) therefore only contributes to the last few
microns, while emission from typical \Cold\ component dust ($\Tavg\sim35\K$)
contributes negligibly at \IRS\ wavelengths.  To better constrain these
cooler components which dominate the dust mass and frequently provide
$\sim$50\% of the total dust luminosity, we supplement our \IRS\ spectra
with far-IR to millimeter wavelength photometry from the literature.
Furthermore, the spectral coverage of the \IRS\ modules begins at an
observed-frame wavelength of $\sim$5.2$\um$.  As seen in
Figure~\ref{fig:SourceSEDs}, our source components all radiate significantly
at shorter wavelengths. Thus, to better constrain these components, we
supplement our SEDs with ultraviolet to near-IR photometry from the
literature.  Table~\ref{tab:SupplementaryPhotometry} provides references to
the literature for all photometry used in our decompositions.


In each wavelength range, we construct our SEDs using data which most
closely samples the spatial region contained within the \IRS\ slits.  We
treat photometric points that sample larger spatial regions (i.e.  for
wavelength ranges where no data at a similar spatial resolution is available
in the literature) as upper-limits to the flux density of the area covered
by the \IRS\ spectrum.  To include such a limit as input to the $\chi^2$
fitting routine (which requires both the flux density of each data point and
its associated error), we assign the data to have an upper-limit flux
density $f_\nu^\downarrow = 0$ and an uncertainty $\sigma_\downarrow =
f_\nu^\data / 3$.  If in addition to an upper-limit to the flux density
contained in the \IRS\ slit, we also have an estimate of the lower-limit
(e.g. as estimated using a very small aperture measurement), we assign the
data point to have a flux density of $f_\nu^\updownarrow = (f_\nu^\downarrow
+ f_\nu^\uparrow) / 2$ with an associated uncertainty of
$\sigma_\updownarrow = (f_\nu^\uparrow - f_\nu^\updownarrow) / 3$, so that
the fit is constrained to fall between these two limiting values.

\subsection{IRAC and MIPS Photometry}
\label{sec:IRACandMIPS}

%
\begin{deluxetable}{cccccccc}
\tabletypesize{\scriptsize}
\tablewidth{0pc}

\tablecaption{\IRAC\ and \MIPS\ Nuclear Photometry\label{tab:IRACandMIPS}}

\tablehead{
  \colhead{} & \colhead{} &
  \multicolumn{2}{c}{\IRAC} & \colhead{} & \multicolumn{3}{c}{\MIPS} 
  \\
  \cline{3-4} \cline{6-8} 
  \\
  \colhead{} & \colhead{} &
  \colhead{$3.6$} & \colhead{$4.5$} & \colhead{} &
  \colhead{$24$} & \colhead{$70$} & \colhead{$160$}
  \\
  \colhead{Galaxy} &
  \colhead{PID} & 
  \colhead{(mJy)} & \colhead{(mJy)} & \colhead{} &
  \colhead{(Jy)} & \colhead{(Jy)} & \colhead{(Jy)}
}

\startdata
NGC\,7714    &      59 &    17.0 &    13.5 & &    1.96 &    6.67 &     4.64 \\
NGC\,2623    &      32 &    11.8 &    13.5 & &    1.33 &    16.4 &     7.56 \\
PG\,0804+761 &      49 & \nodata & \nodata & &    0.19 &    0.11 &    0.033 \\
Mrk\,463     &      32 &     115 &     156 & &    1.50 &    1.90 &     0.89 \\
NGC\,6240    & 32,3672 &    33.7 &    46.0 & &    2.71 &    14.7 &     9.08 \\
Mrk\,1014    &      32 &    19.2 &    23.2 & & \nodata & \nodata & \nodata
\enddata

\tablecomments{\IRAC\ flux density errors are estimated to be 5\% for
  Mrk\,1014 and Mrk\,463, and 10\% for the other three sources.  \MIPS\ flux
  density errors are driven by the post-BCD calibration uncertainties and
  are $\sim$10\% for the $24\um$ band and $\sim$20\% for the $70$ and
  $160\um$ bands. No color corrections are applied since they are much
  smaller than the uncertainties.  The PID are the \Spitzer\ program numbers
  for the \IRAC\ and \MIPS\ observations.}

\end{deluxetable}
%

All of the source components and the \Hot\ dust component radiate
significantly from $3$--$5\um$---i.e. between the JHK bands and the onset of
the \IRS\ spectral range (see Figs.~\ref{fig:SourceEmission} and
\ref{fig:DustEmission}). To better constrain the fits in this critical
region, we derive observed-frame $3.6$ and $4.5\um$ nuclear flux densities
of the central point-source for five of six galaxies in our sample using
archival \IRAC\ images.  This photometry is presented in
Table~\ref{tab:IRACandMIPS} and is derived using the post-BCD products
provided by the Spitzer Science Center. Flux densities for NGC\,7714,
NGC\,2623, and NGC\,6240 are calculated using a $5\arcsec$ radius circular
aperture to match both the near-IR literature photometry and the \IRS\
slit-width. We do not apply any aperture corrections for these sources since
they are all extended at these wavelengths. Mrk\,1014 and Mrk\,463 are both
point-like at these wavelengths, so we calculate their flux densities using
a $12.2\arcsec$ radius aperture, which requires no aperture correction to
recover the full point-source value. Uncertainties are estimated to be 5\%
for Mrk\,1014 and Mrk\,463, and 10\% for the other three sources. 


The \IRAS\ and \ISO\ beam-sizes are both much larger than the 5--$11\arcsec$
\IRS\ slit-widths, so that far-IR photometry of nearby extended sources
contains emission from regions outside the slits.  We therefore derive
mid-IR and far-IR nuclear flux densities of the central point-sources for
five of six galaxies in our sample using archival \MIPS\ images. This
photometry is presented in Table~\ref{tab:IRACandMIPS} and is derived using
the post-BCD products provided by the Spitzer Science Center. We use
Mrk\,463 to derive aperture corrections since it is well-detected in all
three bands and appears very point-like. To calculate the nuclear flux
densities, we measure the emission from each source within small apertures
($3\arcsec$, $8\arcsec$, and $20\arcsec$ radii at $24$, $70$, and $160\um$),
and scale these values by the appropriate aperture correction (3.37, 4.09,
and 2.95 for $24$, $70$, and $160\um$).  The \Spitzer\ beam-sizes at $24$,
$70$, and $160\um$ are $\sim$6$\arcsec$, $17\arcsec$, and $39\arcsec$. The
$24\um$ beam is therefore very similar to the $5$--$11\arcsec$ \IRS\ slits,
justifying our choice to scale the \IRS\ spectrum to this \MIPS\ point. The
beams at $70\um$ and $160\um$ are significantly smaller than the $>
100\arcsec$ \IRAS\ and \ISO\ beams, so that the \MIPS\ data provides a much
better estimate of the nuclear flux density. Flux density errors are
dominated by the post-BCD calibration uncertainties and are estimated to be
10\% at $24\um$ and 20\% at $70$ and $160\um$.  No color corrections are
applied to the \IRAC\ or \MIPS\ photometry since all corrections are much
smaller than the uncertainties.

\section{Decomposing Dusty Galaxies}
\label{sec:DecomposingDustyGalaxies}

\subsection{Spectral Decomposition Results}
\label{sec:SpectralDecompositions}

%
\begin{figure*}
  \epsscale{1.17}
  \plottwo{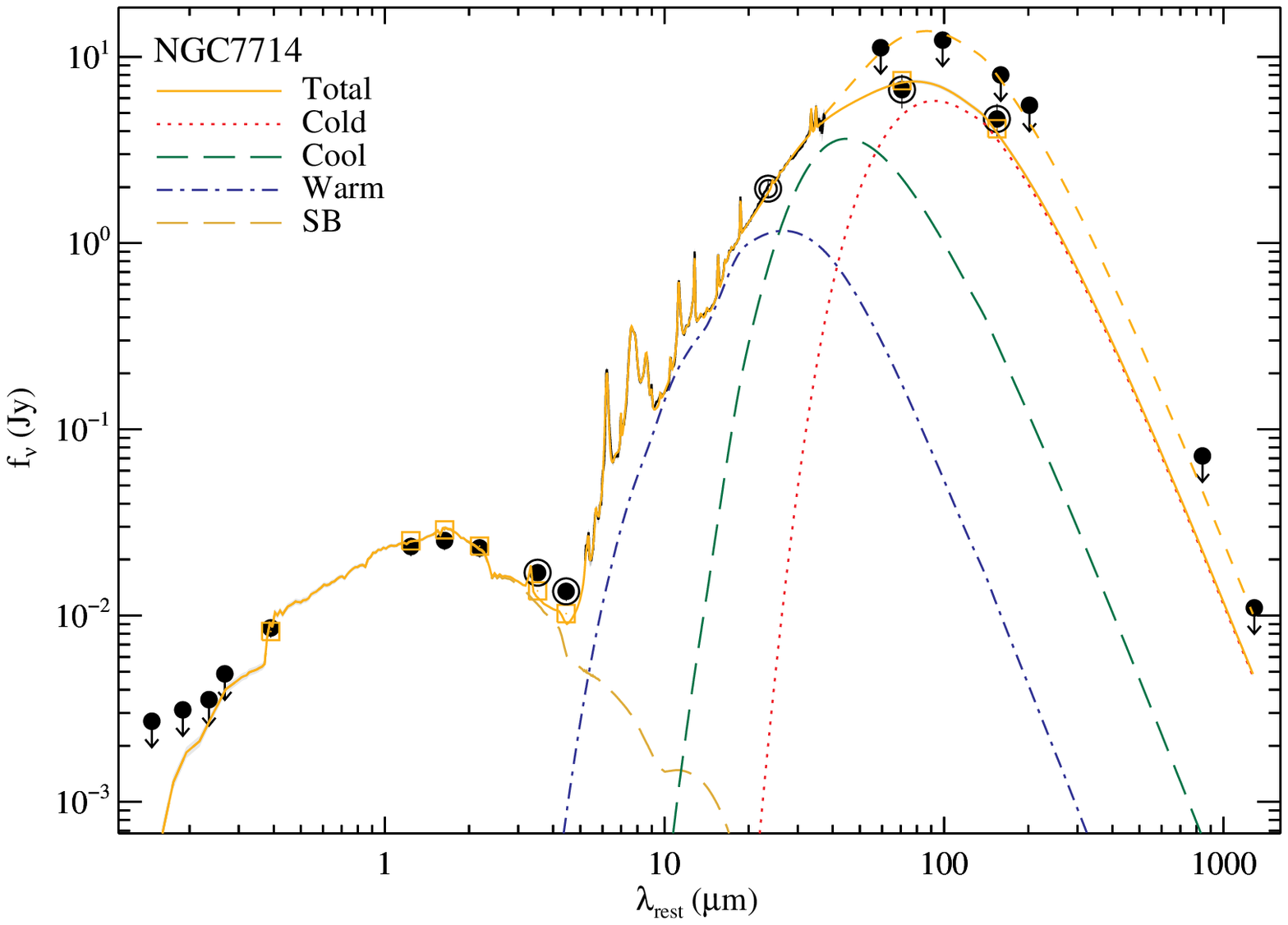}{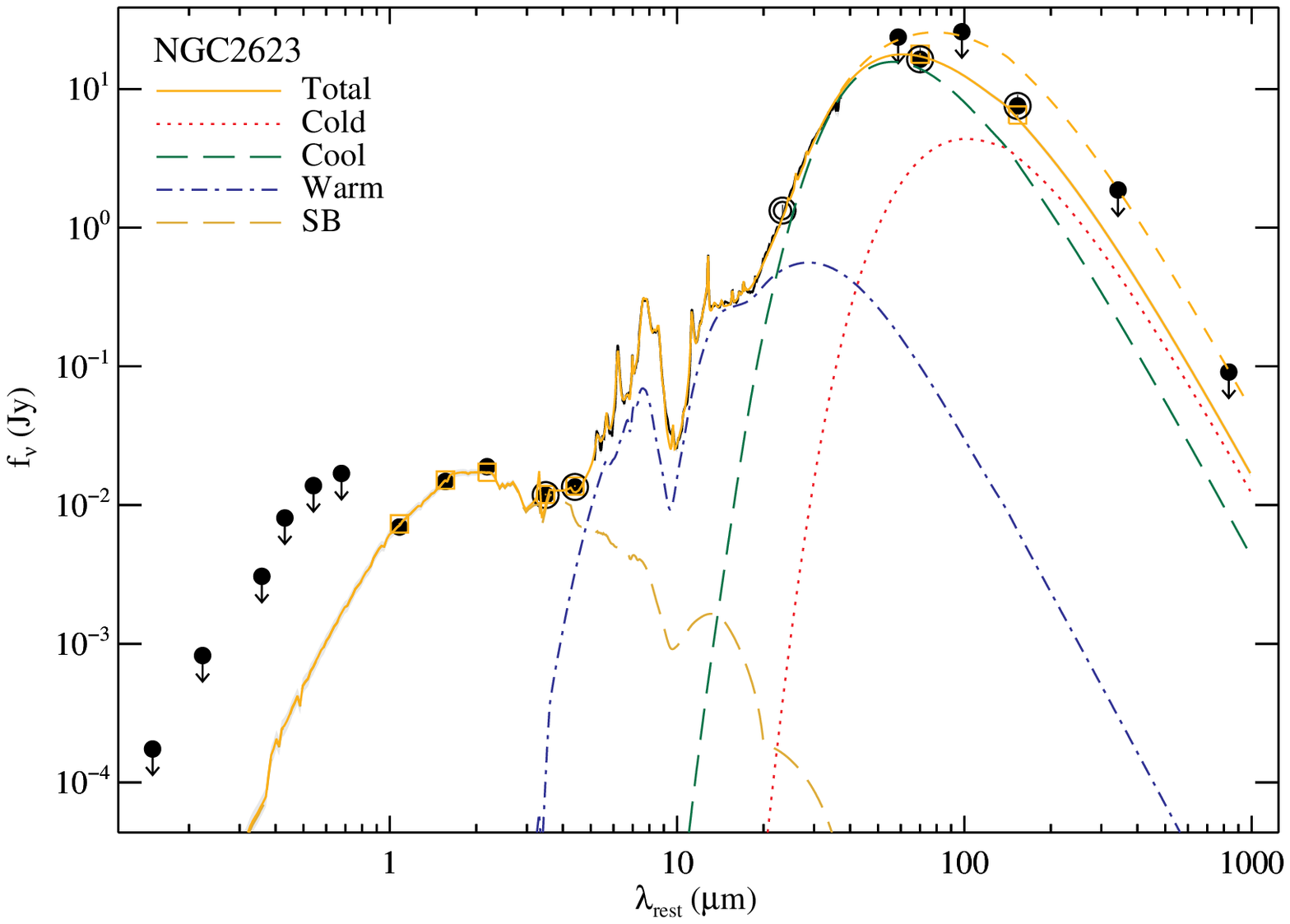}
  \plottwo{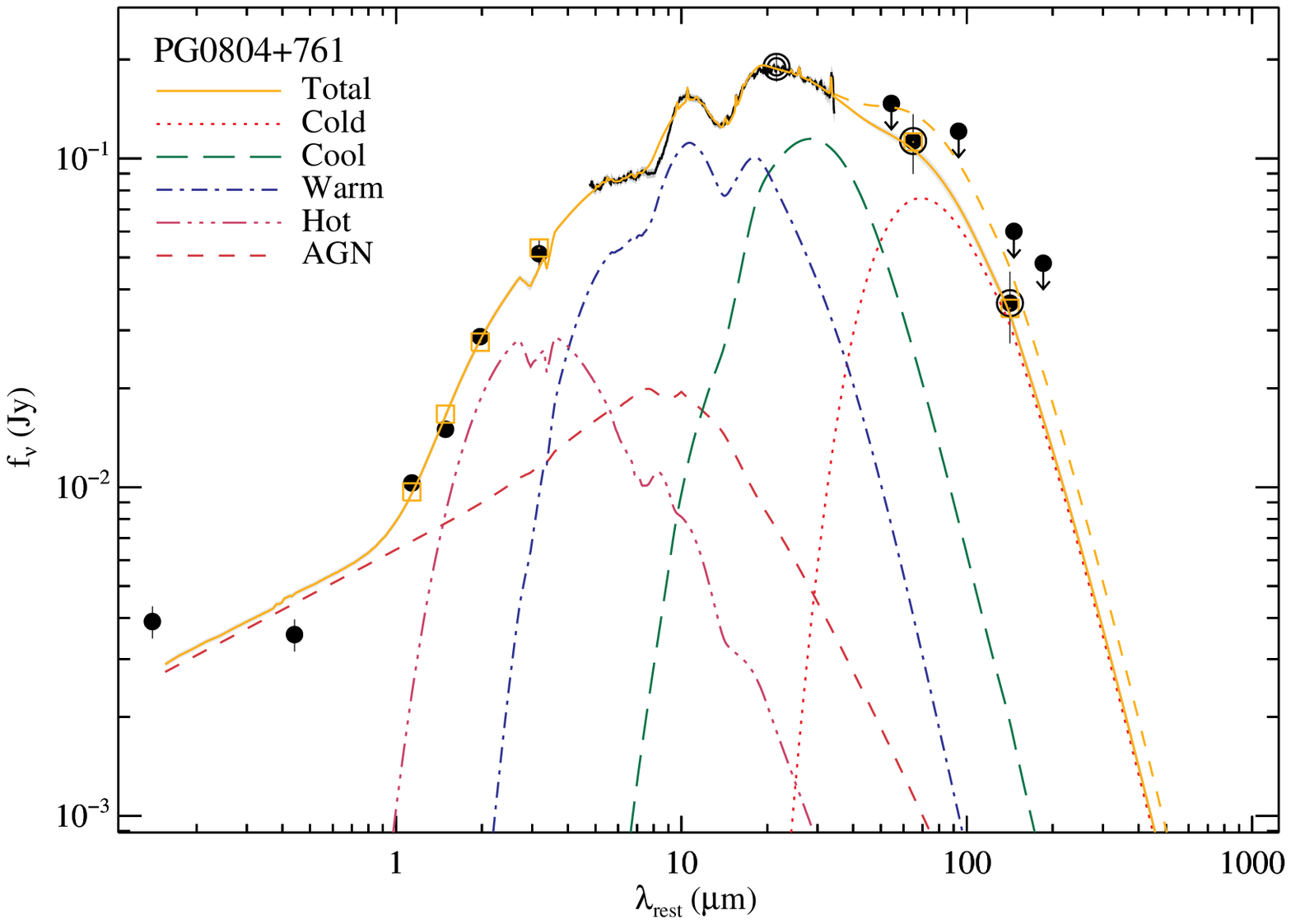}{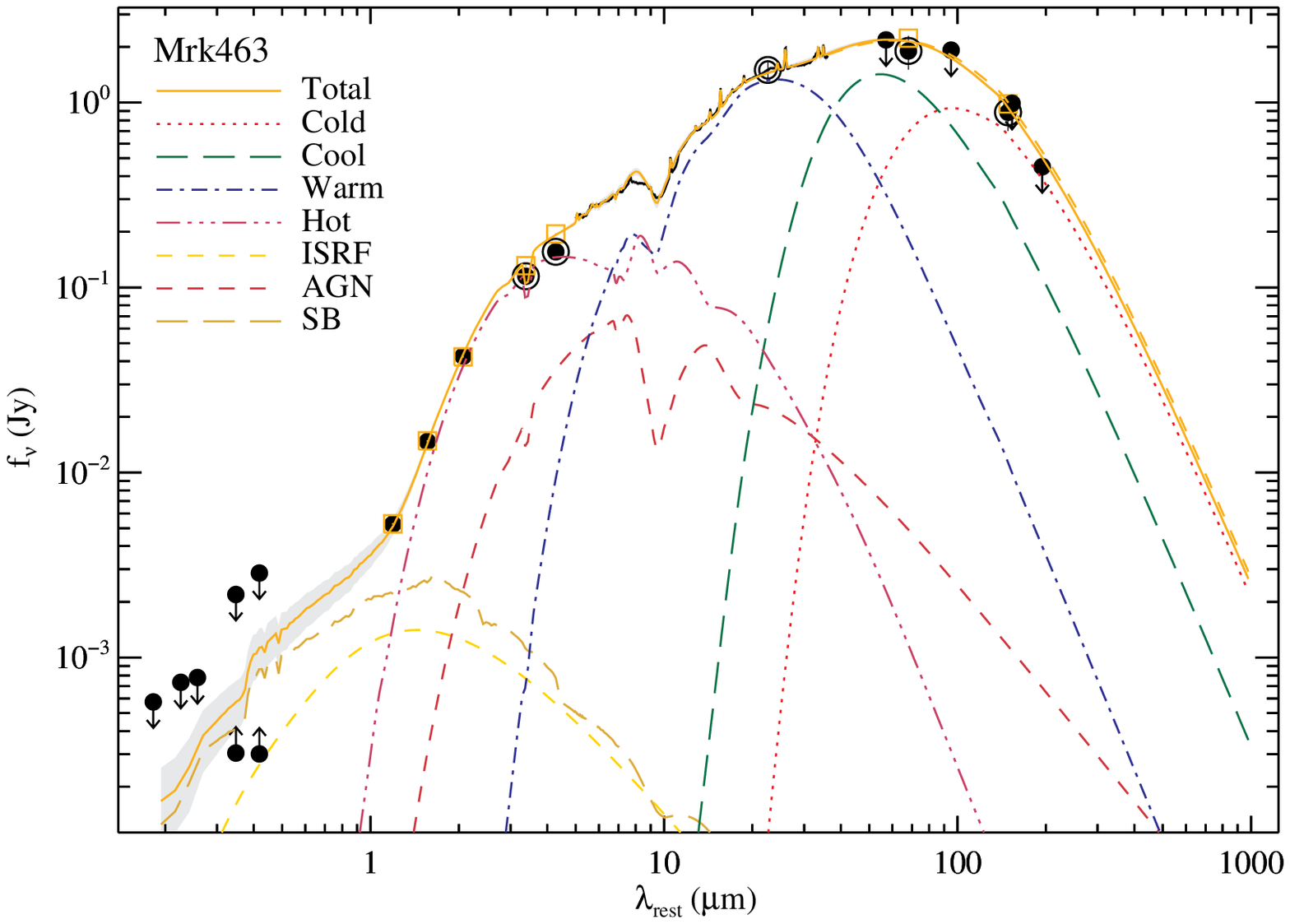}
  \plottwo{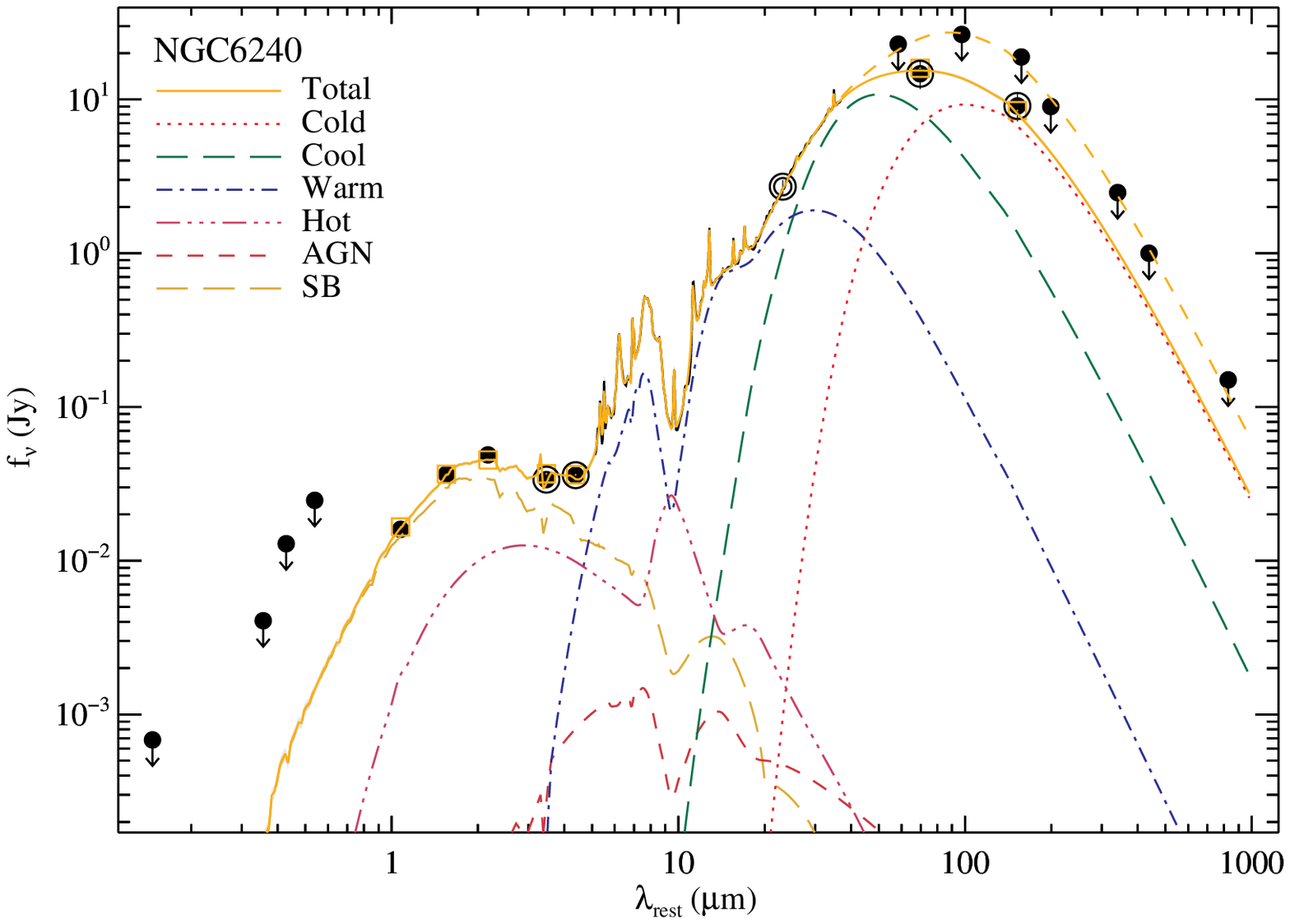}{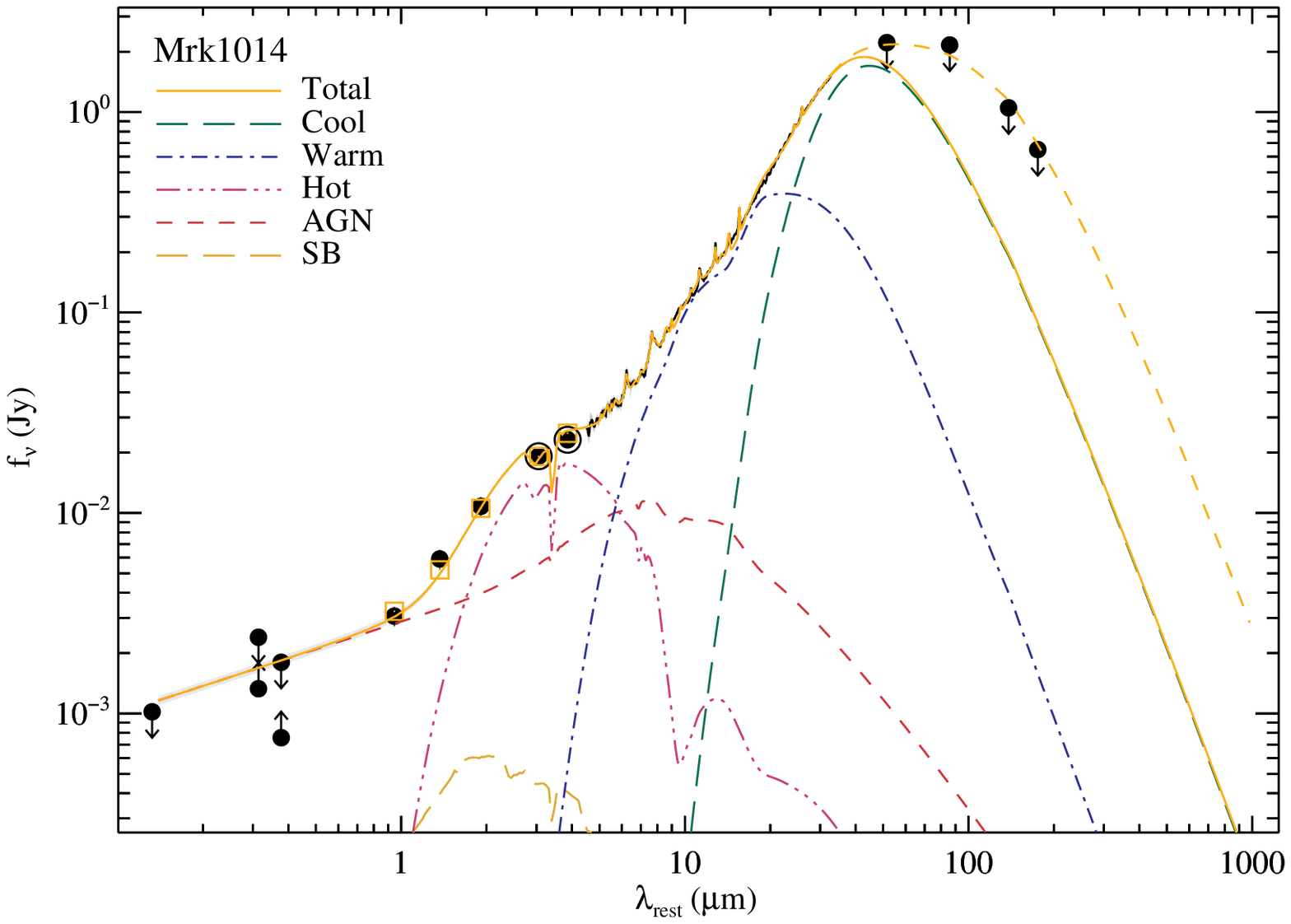}
  \caption{Spectral decompositions of the ultraviolet to millimeter
    wavelength SEDs ({\it black solid lines} and {\it filled circles}) of
    the nuclear regions of the dusty galaxies NGC\,7714, NGC\,2623,
    PG\,0804+761, Mrk\,463, NGC\,6240, and Mrk\,1014. Arrows indicate upper
    or lower-limits (see \S\ref{sec:SupplementaryPhotometry}). Unfilled
    circles indicate the position of the \MIPS\ $24\um$ point to which the
    \IRS\ spectra are scaled. \IRAC\ and \MIPS\ photometric points are
    circled. The various decomposition components are shown as indicated in
    the figure legend. Additionally, the total fitted SEDs obtained using
    the globally integrated far-IR data ({\it dashed `Total' line}) are
    displayed.}
  \label{fig:SEDdecompositions}
  \epsscale{1.0}
\end{figure*}
%
%
\begin{figure*}
  \epsscale{1.17}
  \plottwo{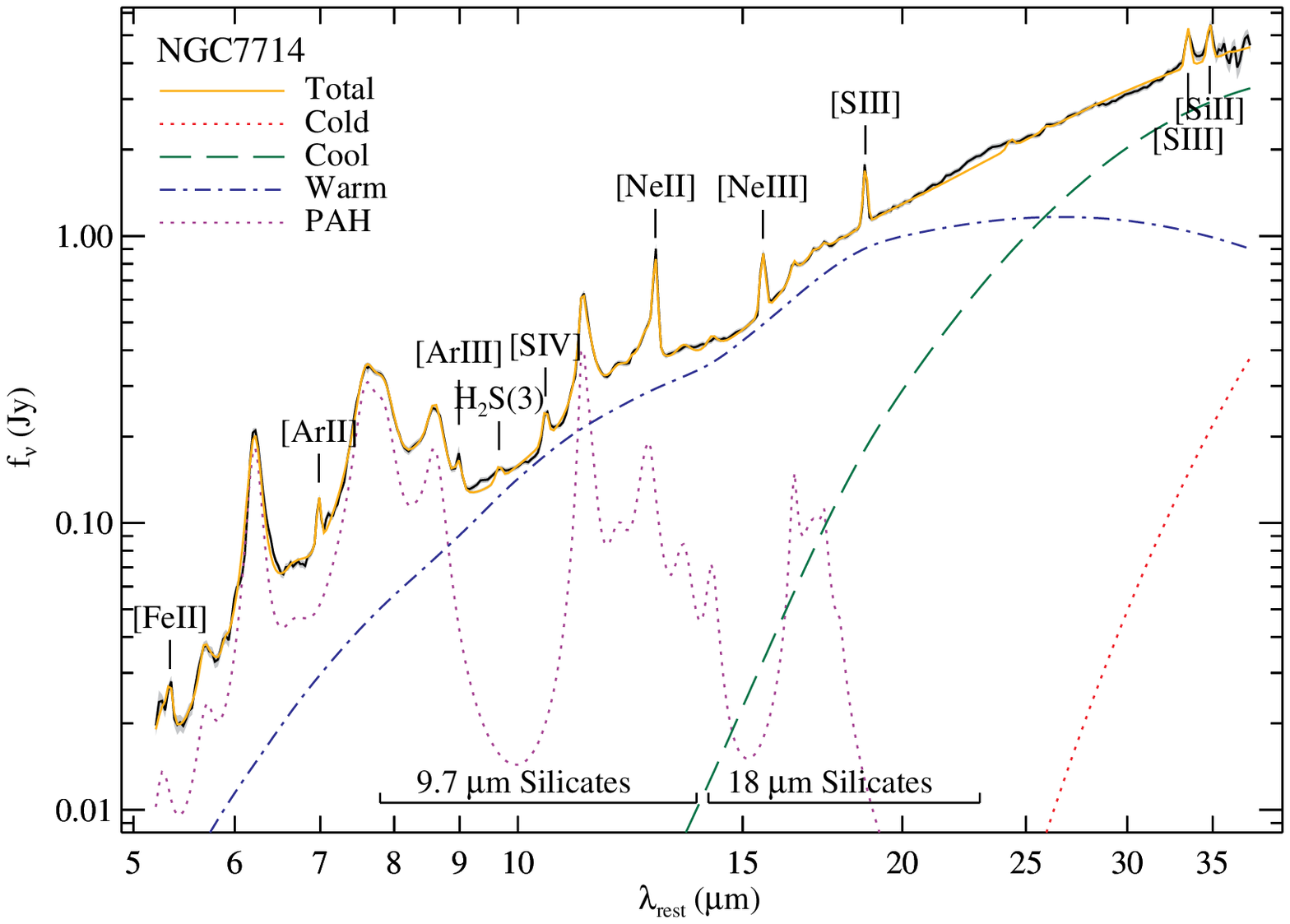}{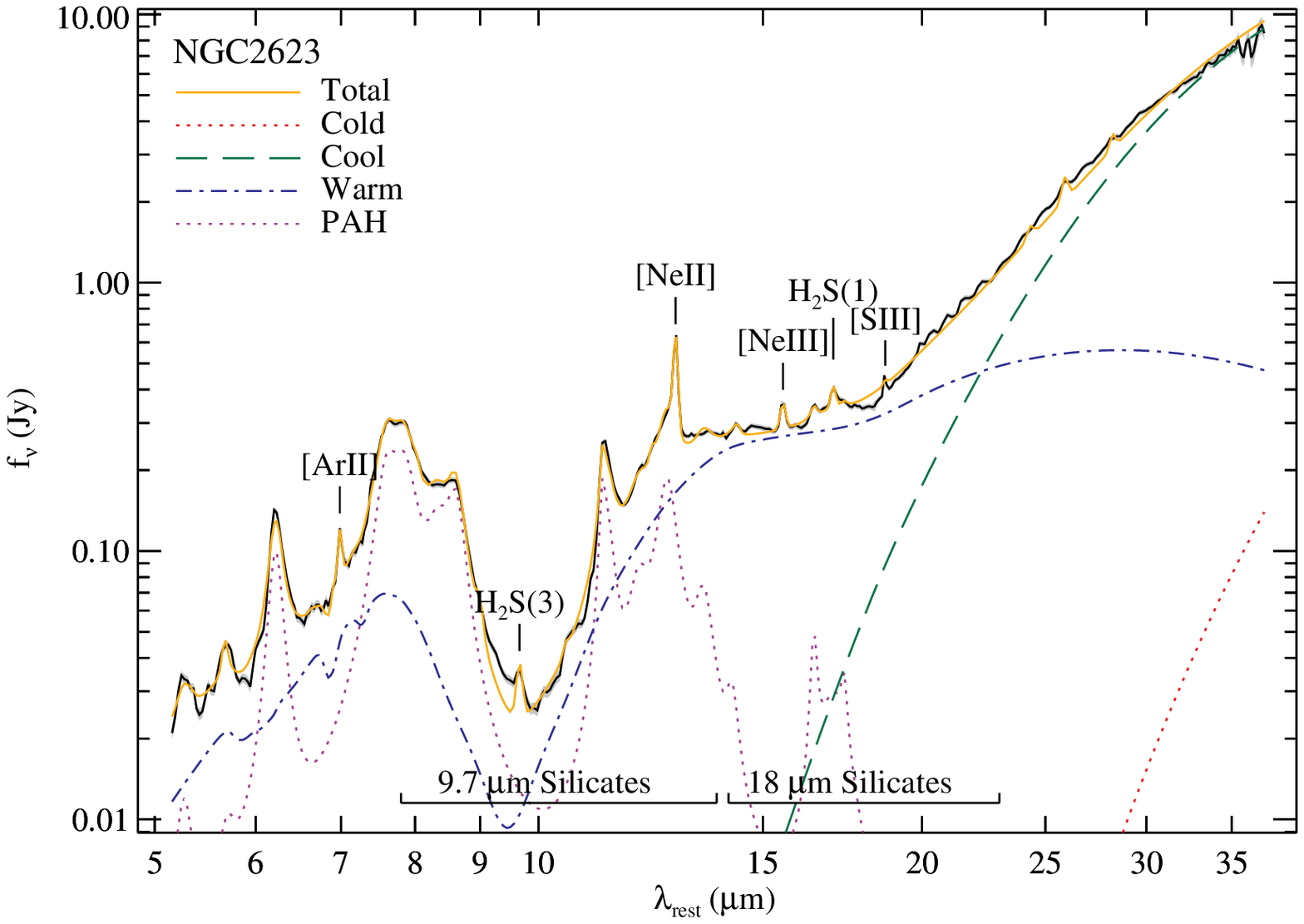}
  \plottwo{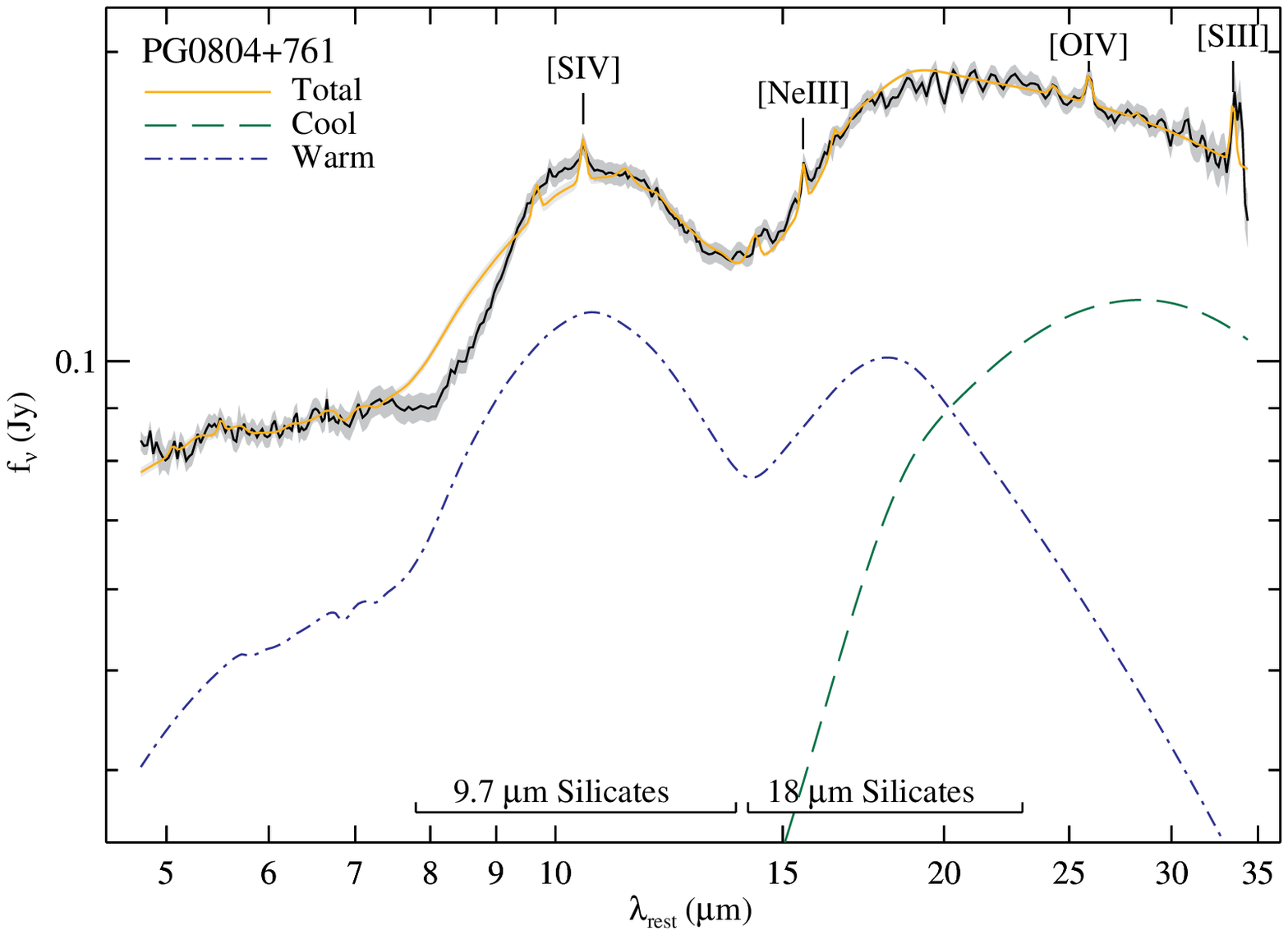}{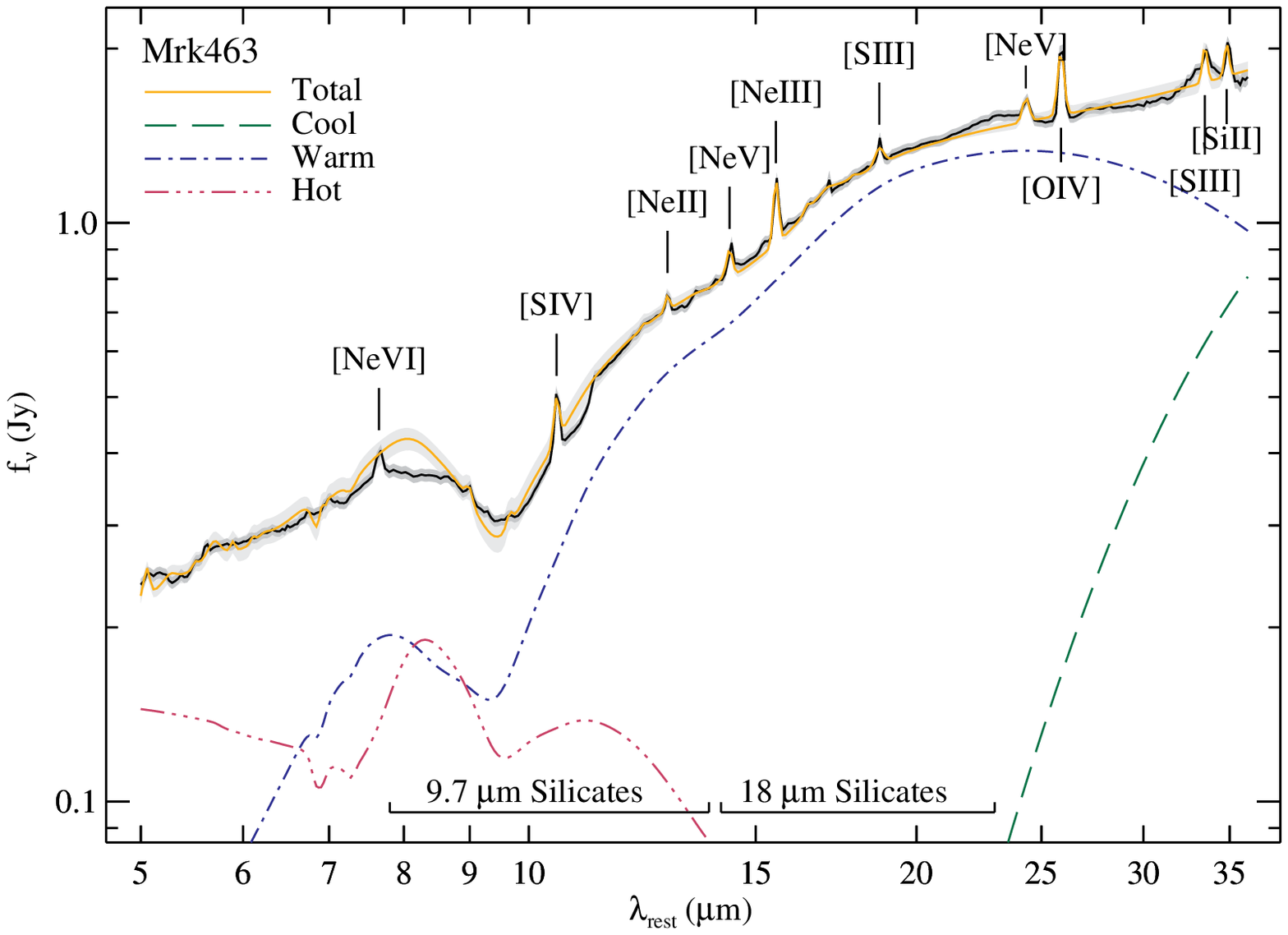}
  \plottwo{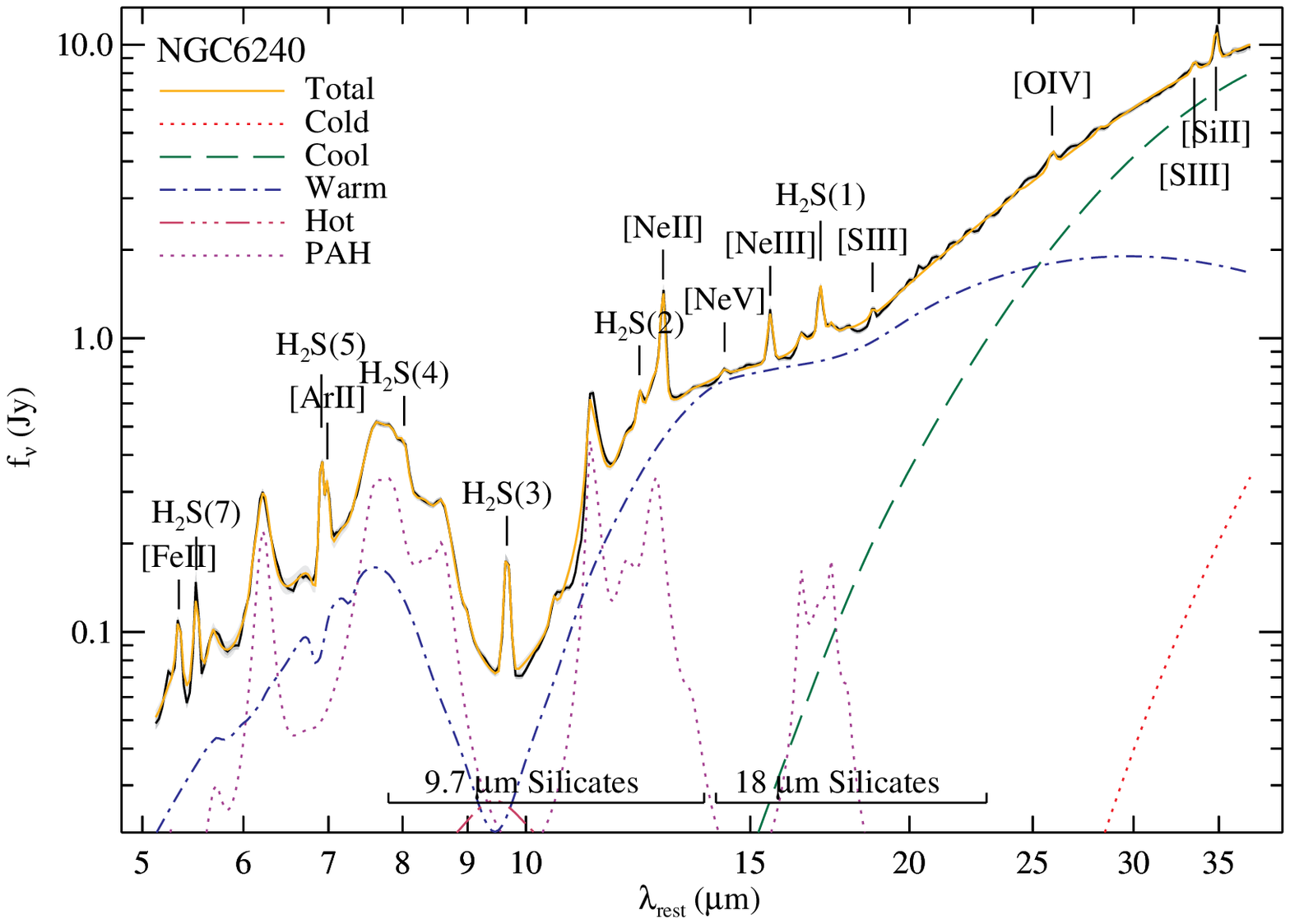}{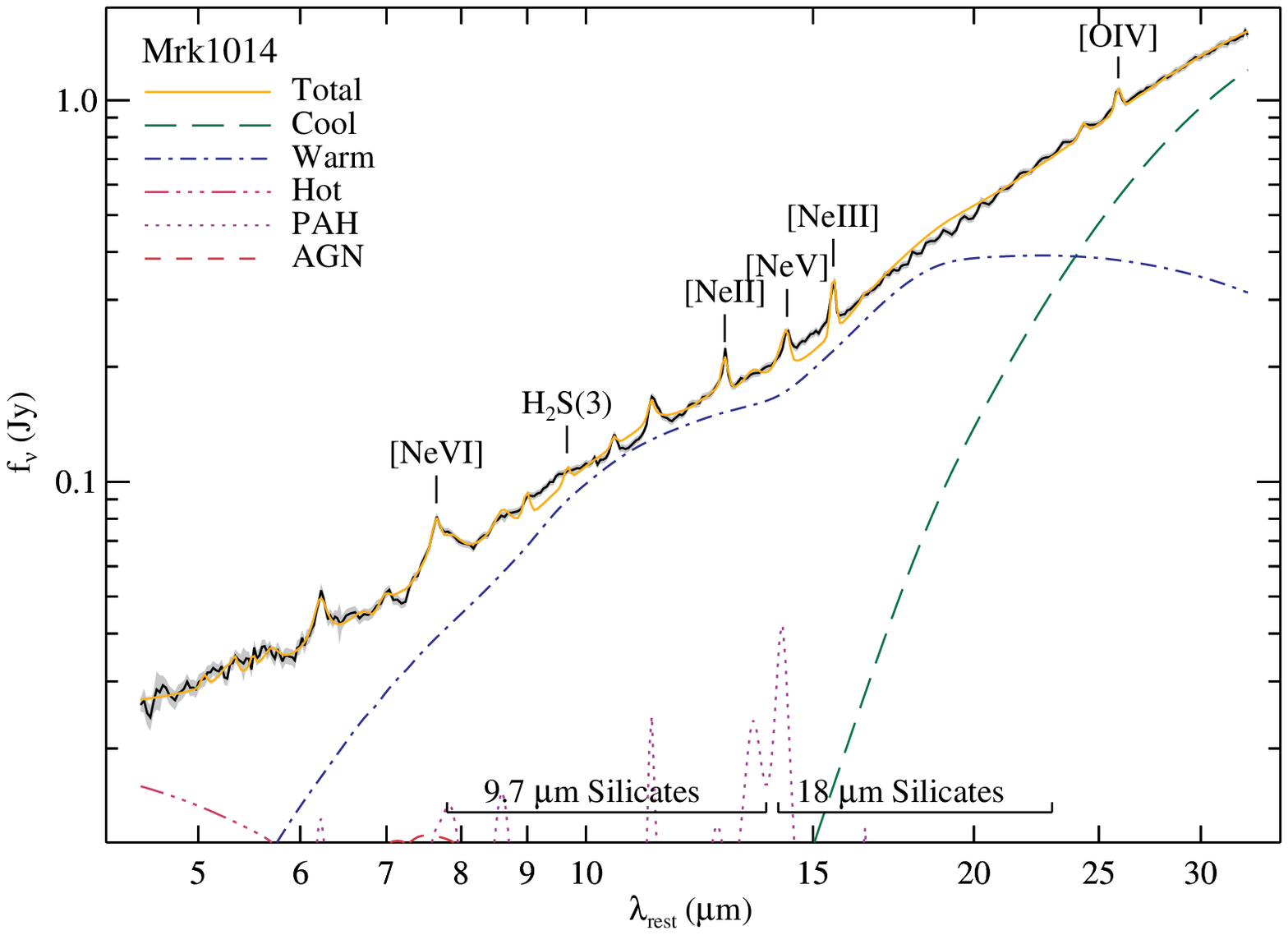}
  \caption{Closeup view of the \IRS\ wavelength ranges for the nuclear
    decompositions presented in Figure~\ref{fig:SEDdecompositions}.  The
    various decomposition components are shown as indicated in the figure
    legend.  Also shown are the $1$--$\sigma$ uncertainties in the \IRS\
    spectra ({\it dark shaded regions}) and the formal $1$--$\sigma$ error
    contours for the total fit ({\it light shaded regions}). Emission lines
    and the wavelength ranges over which silicate emission and absorption is
    observed are also labeled.}
  \label{fig:IRSdecompositions}
  \epsscale{1.0}
\end{figure*}
%

In Figures~\ref{fig:SEDdecompositions} and \ref{fig:IRSdecompositions} we
present our decompositions of the {\it nuclear} SEDs of the dusty galaxies
in our sample.  In these fits, the \MIPS\ photometric points are used to
constrain the nuclear far-IR SEDs while the globally integrated \IRAS, \ISO,
and sub-mm data provide upper-limits to this nuclear emission. The entire
ultraviolet to millimeter wavelength SEDs are presented in
Figure~\ref{fig:SEDdecompositions}, and enlarged views focusing on the \IRS\
regions are shown in Figure~\ref{fig:IRSdecompositions}. In
Figure~\ref{fig:SEDdecompositions}, we also show the {\it global} fitted
flux densities obtained using the full values of the large-beam \IRAS, \ISO,
and sub-mm data. The temperatures of the \Cold\ and \Cool\ dust components
obtained in the global fits are held fixed to these values in the nuclear
decompositions. Note that the \ISRF\ component contributes minimally (only
in Mrk\,463) since the small-aperture near-IR photometry used to match the
\IRS\ slits is dominated by emission from young stellar populations and not
evolved stars.

\subsection{Decomposition Parameters and Constraints}
\label{sec:DecompositionParametersAndConstraints}

%
\begin{deluxetable*}{ccccccccccccc}
\tabletypesize{\scriptsize}
\tablewidth{0pc}

\tablecaption{Nuclear SED Decomposition Parameters and Characteristic Dust
  Temperatures\label{tab:DecompositionParams}}

\tablehead{
  \colhead{} & 
  \colhead{} & 
  \colhead{} & 
  \colhead{$\Tavg_\cold$\tablenotemark{b}} &
  \colhead{$\Tavg_\cool$\tablenotemark{b}} &
  \colhead{$\Tavg_\warm$\tablenotemark{b}} &
  \colhead{$\Tavg_\hot$\tablenotemark{b}} &
  \colhead{} &
  \colhead{} &
  \colhead{} &
  \colhead{} &
  \colhead{} &
  \colhead{}
  \\
  \colhead{Galaxy} &
  \colhead{$\chi^2/{\rm dof}$\tablenotemark{a}} &
  \colhead{dof\tablenotemark{a}} &
  \colhead{(K)} &
  \colhead{(K)} &
  \colhead{(K)} &
  \colhead{(K)} &
  \colhead{$\tilde{\tau}_{9.7}^\warm$\tablenotemark{c}} &
  \colhead{$\tilde{\tau}_{9.7}^\hot$\tablenotemark{c}} &
  \colhead{$\tilde{\tau}_{V}^\starburst$\tablenotemark{d}} &
  \colhead{$\tilde{\epsilon}_\agn$\tablenotemark{e}} &
  \colhead{$\tilde{\eta}_\ice$\tablenotemark{f}} &
  \colhead{$\tilde{\eta}_\HAC$\tablenotemark{f}}
}

\startdata
   NGC\,7714 & $1.33$ & $382$ &    $31$ &    $71$ &   $165$ & \nodata &    $0.19$ &   \nodata &    $0.97$ &   \nodata &        $0$ &        $0$ \\
             &        &       &  $\pm1$ &  $\pm1$ &  $\pm1$ & \nodata & $\pm0.03$ &   \nodata & $\pm0.02$ &   \nodata &    \nodata &    \nodata \\
   NGC\,2623 & $2.46$ & $383$ &    $29$ &    $53$ &   $206$ & \nodata &    $3.52$ &   \nodata &    $4.81$ &   \nodata &      $0.1$ &      $0.1$ \\
             &        &       &  $\pm1$ &  $\pm1$ &  $\pm1$ & \nodata & $\pm0.06$ &   \nodata & $\pm0.11$ &   \nodata &    \nodata &    \nodata \\
PG\,0804+761 & $2.77$ & $363$ &    $42$ &   $150$ &   $400$ &  $1410$ &    $0.78$ &    $1.22$ &       $0$ &    $0.25$ &    $0.050$ &    $0.044$ \\
             &        &       &  $\pm2$ &  $\pm6$ &  $\pm3$ & $\pm10$ & $\pm0.01$ & $\pm0.03$ &   \nodata & $\pm0.01$ & $\pm0.012$ & $\pm0.035$ \\
    Mrk\,463 & $1.90$ & $364$ &    $30$ &    $55$ &   $211$ &  $1130$ &    $1.30$ &    $2.06$ &    $1.12$ &       $1$ &    $0.012$ &    $0.062$ \\
             &        &       &  $\pm2$ &  $\pm2$ &  $\pm1$ &  $\pm6$ & $\pm0.05$ & $\pm0.01$ & $\pm0.26$ &   \nodata & $\pm0.004$ & $\pm0.009$ \\
   NGC\,6240 & $1.77$ & $369$ &    $29$ &    $61$ &   $193$ &  $1260$ &    $3.64$ &       $0$ &    $4.68$ &       $1$ &    $0.067$ &      $0.1$ \\
             &        &       &  $\pm1$ &  $\pm1$ &  $\pm1$ & $\pm30$ & $\pm0.07$ &   \nodata & $\pm0.08$ &   \nodata & $\pm0.007$ &    \nodata \\
   Mrk\,1014 & $2.00$ & $362$ &    $31$ &    $71$ &   $195$ &  $1500$ &    $0.14$ &    $2.70$ &       $5$ &    $0.49$ &    $0.031$ &      $0.1$ \\
             &        &       &  $\pm1$ &  $\pm1$ &  $\pm1$ & \nodata & $\pm0.02$ & $\pm0.04$ &   \nodata & $\pm0.01$ & $\pm0.005$ &    \nodata
\enddata

\tablenotetext{a}{Total weighted reduced $\chi^2$ value (see
  eqs.~[\ref{eq:Chi2}] and [\ref{eq:Weights}]) for the indicated number of
  degrees of freedom (dof).}

\tablenotetext{b}{Characteristic dust temperature (see
  \S\ref{sec:ThermalHeating}) determined from the fitted value of the
  magnitude of the illuminating radiation field energy density,
  $\tilde{U}_i$.}

\tablenotetext{c}{$9.7\um$ optical depth through the screen obscuring the
  \Warm\ and \Hot\ dust components (see eqs.~[\ref{eq:HotWarmEmission}]).}

\tablenotetext{d}{$V$-band ($5500\Angstrom$) optical depth through the
  screen obscuring the \Starburst\ source component (see
  eqs.~[\ref{eq:StarburstEmission}] and [\ref{eq:StarburstTau}]).}

\tablenotetext{e}{Fraction of the \AGN\ source component covered by the
  obscuring screen (see eq.~[\ref{eq:AGNEmission}]).}

\tablenotetext{f}{Ratio of the $6.1\um$ water-ice and $6.85\um$ HAC optical
  depths to the $9.7\um$ optical depth of each dust component (see
  eqs.~[\ref{eq:IceOpticalDepth}] and [\ref{eq:HACOpticalDepth}]).}

\tablecomments{Formal statistical parameter uncertainties are given beneath
  each parameter value. If no uncertainty is given, the corresponding
  parameter was pegged at a limiting value. The $5500\Angstrom$ optical
  depth to the \AGN\ component is fixed to $\tilde{\tau}_{V}^\agn = 25$ for
  all fits.}

\end{deluxetable*}
%

The decomposition parameters obtained from the fits to the nuclear SEDs of
the dusty galaxies in our sample are presented in
Tables~\ref{tab:DecompositionParams} and
\ref{tab:SourceAndDustLuminosities}.  There are a total of 17 free
parameters. The \Starburst\ source component has two free parameters:
$\tilde{\alpha}_\starburst$ and $\tilde{\tau}_{V}^\starburst$ determine the
fractional \Starburst\ contribution to the total luminosity (see
eq.~[\ref{eq:TotalFlux}]) and the $5500\Angstrom$ optical depth (see
eq.~[\ref{eq:StarburstEmission}]). The \ISRF\ source component has only a
single free parameter: $\tilde{\alpha}_\isrf$ (see
eq.~[\ref{eq:ISRFEmission}]). The \AGN\ source component has a total of two
free parameters: $\tilde{\alpha}_\agn$ and $\tilde{\epsilon}_\agn$, where
the latter determines the fraction of the \AGN\ component covered by the
obscuring screen (see eq.~[\ref{eq:AGNEmission}]).  The optical depth
through the obscuring clouds to the \AGN\ component is fixed to
$\tilde{\tau}_{V}^\agn = 25$.  The \Hot\ and \Warm\ dust components each
have three free parameters: $\tilde{\alpha}_i$, $\tilde{\tau}_{9.7}^i$, and
$\tilde{U}_i$ determine the component luminosities, $9.7\um$ optical depths,
and dust temperatures via $\bar{T}_i \equiv T(\tilde{U}_i)$ (see
eq.~[\ref{eq:HotWarmEmission}]). The \Cool\ and \Cold\ components each have
two free parameters: $\tilde{\alpha}_i$ and $\tilde{U}_i$ (see
eq.~[\ref{eq:CoolColdEmission}]).  Finally, there are two additional free
parameters that control the water-ice and HAC contributions to the total
opacity: $\tilde{\eta}_\ice$ and $\tilde{\eta}_\HAC$ (see
eqs.~[\ref{eq:IceOpticalDepth}] and [\ref{eq:HACOpticalDepth}]).


We impose the following constraints to provide sensible limits over which
the free parameters may range: (1) The magnitude of the radiation field
energy density illuminating each dust component is constrained to give
characteristic dust temperatures satisfying $20\K < \Tavg_\cold < 50\K$,
$50\K < \Tavg_\cool < 150\K$, $150\K < \Tavg_\warm < 500\K$, and $500\K <
\Tavg_\hot < 1500\K$; (2) The maximum water-ice and HAC contributions to the
total opacity are constrained by their strengths in sources with very clean
absorption spectra (see \S\ref{sec:IceOpacity}); (3) The extinction to the
\AGN\ source component is fixed to $\tau_V = 25$; (4) The extinction to the
\Starburst\ source component must satisfy $\tau_V < 5$; (5) The optical
depths through the screens obscuring the \Hot\ and \Warm\ components must
not exceed the critical values beyond which their extinction-corrected
luminosities would exceed the total dust luminosity; (6) The ratio of the
unextinguished luminosity of the \Starburst\ component to the luminosity of
the \PAHs\ component must be between the values obtained for our unobscured
and obscured template starbursts NGC\,7714 and NGC\,2623; and (7) The \Hot\
dust covering fraction (i.e. the ratio of the unextinguished \Hot\ dust
luminosity to the unextinguished \AGN\ luminosity) must be $> 0.5$,
consistent with observations and corresponding to torus opening angles $>
30^\circ$.

\subsection{Derived Quantities}
\label{sec:DerivedQuantities}

\subsubsection{Source and Dust Component Luminosities}
\label{sec:SourceAndDustLuminosities}

%
\begin{deluxetable*}{ccccccccccccccc}
\tabletypesize{\scriptsize}
\tablewidth{0pc}

\tablecaption{Nuclear SED Decomposition Source and Dust Component
  Luminosities\label{tab:SourceAndDustLuminosities}}

\tablehead{
  \colhead{} &
  \multicolumn{6}{c}{Source} & \colhead{} &
  \multicolumn{7}{c}{Dust}
  \\
  \cline{2-7} \cline{9-15}
  \\
  \colhead{} &
  \colhead{$L_\source$} &
  \colhead{} & 
  \colhead{} & 
  \colhead{} & 
  \colhead{} &
  \colhead{} & 
  \colhead{} &
  \colhead{$L_\dust$} &
  \colhead{} & 
  \colhead{} & 
  \colhead{} & 
  \colhead{} & 
  \colhead{} &
  \colhead{}
  \\
  \colhead{Galaxy} &
  \colhead{($10^{10}\LSun$)} &
  \colhead{$\tilde{\alpha}_\isrf$} & 
  \colhead{$\tilde{\alpha}_\starburst$} & 
  \colhead{$\tilde{\alpha}_\starburst^\prime$} & 
  \colhead{$\tilde{\alpha}_\agn$} &
  \colhead{$\tilde{\alpha}_\agn^\prime$} & 
  \colhead{} &
  \colhead{($10^{10}\LSun$)} &
  \colhead{$\tilde{\alpha}_\cold$} & 
  \colhead{$\tilde{\alpha}_\cool$} & 
  \colhead{$\tilde{\alpha}_\warm$} & 
  \colhead{$\tilde{\alpha}_\warm^\prime$} & 
  \colhead{$\tilde{\alpha}_\hot$} &
  \colhead{$\tilde{\alpha}_\hot^\prime$}
}

\startdata
   NGC\,7714 &    $0.78$ &       $0$ &        $1$ &     $3.8$ &       $0$ &       $0$ &   &    $3.25$ &    $0.344$ &    $0.394$ &    $0.262$ &    $0.275$ &        $0$ &        $0$ \\
             & $\pm0.03$ &   \nodata &  $\pm0.05$ &  $\pm0.1$ &   \nodata &   \nodata &   & $\pm0.04$ & $\pm0.011$ & $\pm0.006$ & $\pm0.004$ & $\pm0.004$ &    \nodata &    \nodata \\
   NGC\,2623 &    $0.72$ &       $0$ &        $1$ &      $27$ &       $0$ &       $0$ &   &    $22.5$ &    $0.136$ &    $0.792$ &    $0.072$ &    $0.194$ &        $0$ &        $0$ \\
             & $\pm0.05$ &   \nodata &  $\pm0.09$ &    $\pm2$ &   \nodata &   \nodata &   &  $\pm0.2$ & $\pm0.010$ & $\pm0.010$ & $\pm0.001$ & $\pm0.004$ &    \nodata &    \nodata \\
PG\,0804+761 &      $99$ &       $0$ &        $0$ &       $0$ &       $1$ &    $1.27$ &   &    $62.5$ &    $0.036$ &    $0.146$ &    $0.457$ &    $0.639$ &    $0.361$ &        $1$ \\
             &    $\pm2$ &   \nodata &    \nodata &   \nodata & $\pm0.03$ & $\pm0.04$ &   &  $\pm1.1$ & $\pm0.002$ & $\pm0.003$ & $\pm0.011$ & $\pm0.013$ & $\pm0.014$ &  $\pm0.02$ \\
    Mrk\,463 &     $8.2$ &    $0.12$ &     $0.23$ &    $1.03$ &    $0.65$ &    $18.1$ &   &      $74$ &    $0.068$ &    $0.165$ &    $0.470$ &    $0.703$ &    $0.297$ &        $1$ \\
             &  $\pm0.6$ & $\pm0.03$ &  $\pm0.05$ & $\pm0.08$ & $\pm0.05$ &  $\pm1.4$ &   &    $\pm2$ & $\pm0.003$ & $\pm0.005$ & $\pm0.027$ & $\pm0.038$ & $\pm0.009$ &  $\pm0.03$ \\
   NGC\,6240 &    $2.61$ &       $0$ &    $>0.99$ &    $24.8$ &   $<0.01$ &    $0.29$ &   &    $45.2$ &    $0.257$ &    $0.528$ &    $0.198$ &    $0.514$ &    $0.017$ &    $0.017$ \\
             & $\pm0.07$ &   \nodata &  $\pm0.04$ &  $\pm0.7$ &   \nodata & $\pm0.01$ &   &  $\pm0.7$ & $\pm0.011$ & $\pm0.009$ & $\pm0.007$ & $\pm0.020$ & $\pm0.001$ & $\pm0.001$ \\
   Mrk\,1014 &     $120$ &       $0$ &    $0.018$ &    $0.49$ &    $0.98$ &    $1.86$ &   &     $337$ &    \nodata &    $0.583$ &    $0.336$ &    $0.352$ &    $0.081$ &     $0.66$ \\
             &    $\pm3$ &   \nodata & $\pm0.001$ & $\pm0.01$ & $\pm0.03$ & $\pm0.04$ &   &    $\pm2$ &    \nodata & $\pm0.005$ & $\pm0.003$ & $\pm0.003$ & $\pm0.002$ &  $\pm0.02$
\enddata

\tablecomments{Apparent and extinction-corrected luminosities of each source
  and dust component in the nuclear decompositions are given as fractions of
  $L_\source$ and $L_\dust$ (see eqs.~[\ref{eq:ApparentLuminosity}] and
  [\ref{eq:CorrectedLuminosity}]). Formal statistical uncertainties for the
  fitted luminosities are given beneath their values. The nuclear dust
  luminosity of Mrk\,1014 is uncertain by $\sim$20\% (i.e. the difference
  between its nuclear and global values) and its \Cold\ component luminosity
  is undetermined since no \MIPS\ data is available to constrain its far-IR
  SED.}

\end{deluxetable*}
%

Apparent and extinction-corrected source and dust component luminosities
from the nuclear decompositions are presented in
Table~\ref{tab:SourceAndDustLuminosities}.  The apparent luminosity of each
component is calculated from
\begin{equation}
  \label{eq:ApparentLuminosity}
  L_i = 4 \pi D_L^2 \int f_\nu^i \, d\nu \equiv
  \cases{\tilde{\alpha}_i L_\source & $L_i \in L_\source$, \cr
    \tilde{\alpha}_i L_\dust & $L_i \in L_\dust$,}
\end{equation}
where $D_L$ is the luminosity distance to the galaxy (see
Table~\ref{tab:DustyGalaxyProperties}), and $L_\source$ and $L_\dust$ are
the sums of the luminosities of all source and dust components,
respectively.  Extinction-corrected luminosities are calculated from
\begin{equation}
  \label{eq:CorrectedLuminosity}
  L_i^\prime = 4 \pi D_L^2 \int \frac{f_\nu^i}{\Upsilon_i(\nu)} \, d\nu \equiv
  \cases{\tilde{\alpha}_i^\prime L_\source & $L_i \in L_\source$, \cr
    \tilde{\alpha}_i^\prime L_\dust & $L_i \in L_\dust$,}
\end{equation}
where the $\Upsilon_i$ are the extinction terms for each component (i.e.
$\Upsilon_\agn = [1 - \tilde{\epsilon}_\agn] + \tilde{\epsilon}_\agn
\exp[-\tau_\agn]$ for the \AGN\ component and $\Upsilon_i = \exp[-\tau_i]$
for the \Warm, \Hot, and \Starburst\ components).

\subsubsection{Dust Component Masses}
\label{sec:DustComponentMasses}

%
\begin{deluxetable}{ccccc}
\tabletypesize{\scriptsize}
\tablewidth{0pc}

\tablecaption{Nuclear SED Decomposition Dust
  Masses\label{tab:NuclearDustMasses}}

\tablehead{
  \colhead{} &
  \colhead{$M_\cold$} &
  \colhead{$M_\cool$} &
  \colhead{$M_\warm$} &
  \colhead{$M_\hot$}
  \\
  \colhead{Galaxy} &
  \colhead{($10^{6}\MSun$)} &
  \colhead{($10^{4}\MSun$)} &
  \colhead{($10^{2}\MSun$)} &
  \colhead{($\MSun$)}
}

\startdata
   NGC\,7714 & $4.26\pm0.13$ & $4.90\pm0.04$ & $5.55\pm0.05$ &             $0$ \\
   NGC\,2623 &  $19.8\pm1.4$ &     $335\pm2$ &   $9.3\pm0.2$ &             $0$ \\
PG\,0804+761 & $1.53\pm0.07$ & $0.91\pm0.01$ & $2.95\pm0.04$ &   $1.75\pm0.09$ \\
    Mrk\,463 &  $23.7\pm0.9$ &     $185\pm3$ &    $101\pm13$ &     $7.9\pm0.2$ \\
   NGC\,6240 &  $67.3\pm2.6$ &     $201\pm1$ &      $69\pm6$ & $0.049\pm0.003$ \\
   Mrk\,1014 &       \nodata &     $736\pm5$ &     $333\pm3$ &   $1.13\pm0.03$
\enddata

\tablecomments{Formal statistical uncertainties for the fitted masses of the
  dust components in the nuclear decompositions are given along with their
  values. The mass of \Cold\ dust in the nucleus of Mrk\,1014 is
  undetermined since no \MIPS\ data is available to constrain its far-IR
  SED.}

\end{deluxetable}
%

The mass of dust in each of the \Cold, \Cool, \Warm, and \Hot\ nuclear
decomposition components are provided in Table~\ref{tab:NuclearDustMasses}.
We generalize the expression used to calculate dust masses for a single
grain-size and composition
%
\citep[e.g. see eq.~{[4]} of][]{2001A&A...379..823K}
to the case in which grains are distributed both in size and composition.
The mass of dust in each component is therefore given by
\begin{equation}
  \label{eq:DustMass}
  M_i = 
  D_L^2 \frac{f_\nu^i(\lambda_0)}{E_\nu^i(\lambda_0) e^{-\tau_i(\lambda_0)}} 
  \sum_j \int 
  \frac{4}{3} \pi a^3 \rho_j \frac{1}{\nH} \frac{dn_j}{da} \, da,
\end{equation}
which is the product of the number of hydrogen nucleons required to power
the fitted dust component and the total dust mass per hydrogen nucleon for
the adopted dust model (summed over the graphite and silicate compositions).
In this equation, $\lambda_0$ is an arbitrary wavelength at which the terms
are evaluated, and the densities of graphite and silicate are taken to be
$\rho_\gra = 2.24$ and $\rho_\sil = 3.50$ g$^{-1}$ cm$^{-3}$. For comparison
to the dust masses and luminosities obtained from the nuclear
decompositions, we present the corresponding properties for the global
decompositions in Table~\ref{tab:GlobalDustProperties}.

\subsubsection{Comparison With Previous Results}
\label{sec:ParameterComparison}

%
\begin{deluxetable*}{cccccc}
\tabletypesize{\scriptsize}
\tablewidth{0pc}

\tablecaption{Global SED Decomposition Dust
  Properties\label{tab:GlobalDustProperties}}

\tablehead{
  \colhead{} &
  \colhead{$L_\dust$} &
  \colhead{} &
  \colhead{} &
  \colhead{$M_\cold$} &
  \colhead{$M_\cool$}
  \\
  \colhead{Galaxy} &
  \colhead{($10^{10}\LSun$)} &
  \colhead{$\tilde{\alpha}_\cold$} &
  \colhead{$\tilde{\alpha}_\cool$} &
  \colhead{($10^6\MSun$)} &
  \colhead{($10^4\MSun$)}
}

\startdata
   NGC\,7714 & $4.50\pm0.05$ & $0.530\pm0.010$ & $0.280\pm0.006$ & $9.1\pm0.1$ &   $4.8\pm0.1$ \\
   NGC\,2623 &  $30.3\pm0.9$ & $0.365\pm0.036$ & $0.581\pm0.019$ &   $72\pm13$ &     $331\pm3$ \\
PG\,0804+761 &  $62.6\pm3.8$ & $0.050\pm0.004$ & $0.140\pm0.010$ & $2.2\pm0.1$ & $0.88\pm0.03$ \\
    Mrk\,463 &  $74.5\pm1.3$ & $0.076\pm0.014$ & $0.153\pm0.007$ &    $27\pm5$ &     $173\pm8$ \\
   NGC\,6240 &  $61.2\pm3.7$ & $0.461\pm0.039$ & $0.381\pm0.040$ &  $164\pm21$ &    $197\pm60$ \\
   Mrk\,1014 &    $414\pm23$ & $0.188\pm0.015$ & $0.469\pm0.054$ &  $303\pm18$ &   $730\pm280$
\enddata

\tablecomments{Global SED decomposition properties derived from fitting the
  total far-IR emission (see Fig.~\ref{fig:SEDdecompositions}).  All
  quantities are analogous to those in
  Tables~\ref{tab:SourceAndDustLuminosities} and
  \ref{tab:NuclearDustMasses}. The temperatures of the \Cold\ and \Cool\
  components are the same as given in Table~\ref{tab:DecompositionParams}.
  Formal statistical uncertainties for the fitted properties are given along
  with their values.}

\end{deluxetable*}
%

%
\citet{2001A&A...379..823K}
fit the far-IR to sub-mm wavelength SEDs of a sample of ULIRGs using
dust-modified blackbodies with temperatures typically ranging between 28 and
$50\K$. Included among their sources are two galaxies which are also in our
sample, Mrk\,463 and NGC\,6240, for which they fit single $\beta = 2$ dust
components with temperatures of $40$ and $33\K$, respectively. In a similar
study of quasars presented in
%
\citet{2003A&A...402...87H},
the far-IR SED of PG\,0804+761 is fitted with a $47\K$ dust component. The
temperatures we derive from our decompositions are all consistent with these
values. Our decomposition of Mrk\,463 includes contributions from $30\K$ and
$55\K$ dust components, while our fit to NGC\,6240 includes emission from
dust at $29\K$ and $61\K$. Finally, our decomposition of PG\,0804+761 is
dominated in the far-IR by emission from $42\K$ dust. For all three sources,
the temperatures of the single component fits in the literature fall between
the temperatures of the dust in our multi-component fits.
%
\citet{2001A&A...379..823K}
also fit the $\lambda > 60\um$ SED of NGC\,6240 with multiple $\beta = 2$
dust components and obtain temperatures of $31$ and $41\K$ (they also
include a $10\K$ component to fit to sub-mm upper-limits). Again, these
temperatures are quite comparable to ours, noting that our warmer component
is at a somewhat higher temperature in order to fit to the shorter
wavelength data included in our fits.


As illustrated by these decompositions, multi-component fits including dust
at different temperatures are rarely unique since several components can
usually be added together to approximate the SED of emission from a smaller
number of components. In fact, dust in galaxies is likely to be distributed
throughout many regions with various heating conditions, and is therefore
distributed over a continuum of temperatures.  Any multi-component
decomposition therefore serves only to characterize these distributions of
temperatures. Additionally, if either or both of the restrictions that dust
is optically-thin to its own emission and that its far-IR emissivity index
is characterized by $\beta = 2$ imposed above are lifted, yet different dust
temperatures may be obtained for the same SEDs.  For example,
%
\citet{2001A&A...379..823K}
perform fits in which the dust optical depth and emissivity are free
parameters, and obtain much higher temperatures---$52$ and $57\K$ for
Mrk\,463 and NGC\,6240, respectively. Given the multiple factors determining
the temperatures derived from spectral decompositions (e.g. fixed versus
variable $\beta$), we emphasize that caution must be taken when comparing
temperatures derived from different methods to ensure that they are in fact
comparable.


The total far-IR luminosity from dust inferred from a multi-component
decomposition is largely independent of the number of components and the
choice of their characteristic temperatures (since the luminosity is just
the integral under the SED). The total mass of dust, however, depends upon
these choices (since the mass is a strong function of temperature).  Since
the total mass of dust in galaxies is dominated by low-temperature material,
decompositions with multiple components (including very cold dust) produce
much higher dust masses. For example, \citet{2001A&A...379..823K} estimate
the mass of dust in Mrk\,463 to be greater than $5\times10^6\MSun$ and less
than $1.2\times10^7\MSun$ as derived from their single and multi-component
fits, respectively.  Similarly, they estimate that the mass of dust in
NGC\,6240 is between $3.6\times10^7\MSun$ and $5.8\times10^8\MSun$. Our
decompositions yield total globally integrated dust masses of
$2.8\times10^7\MSun$ and $1.6\times10^8\MSun$ for the two sources,
relatively consistent with the previously derived masses (within the
uncertainties of the methods).  \citet{2001A&A...379..823K} estimate the
${\rm H}_2$ mass of NGC\,6240 to be $3.7\times10^{10}\MSun$ (assuming
$M[{\rm H_2}] = 4.6 L_{\rm CO}$). This value, along with our estimate of the
total dust mass in NGC\,6240, gives a gas-to-dust ratio of $\sim$230, in
reasonable agreement with the canonical value of $\sim$165 from
%
\citet{Li2005}.
%

\subsubsection{Diagnostic Ratios}
\label{sec:DiagnosticRatios}

%
\begin{deluxetable*}{ccccccccccccc}
\tabletypesize{\scriptsize}
\tablewidth{0pc}

\tablecaption{Nuclear SED Decomposition Diagnostic
  Ratios\label{tab:DiagnosticRatios}}

\tablehead{
  \colhead{} &
  \multicolumn{3}{c}{Dust} & \colhead{} &
  \multicolumn{2}{c}{Starburst} & \colhead{} &
  \multicolumn{2}{c}{AGN} & \colhead{} &
  \multicolumn{2}{c}{PAHs}
  \\
  \cline{2-4} \cline{6-7} \cline{9-10} \cline{12-13}
  \\
  \colhead{} & 
  \colhead{} &
  \colhead{$L_\dust / M_\gas$\tablenotemark{a}} &
  \colhead{} & \colhead{} &
  \colhead{} & 
  \colhead{$L_\starburst^\prime / M_\gas$\tablenotemark{b}} &
  \colhead{} &
  \colhead{} &
  \colhead{$L_\agn^\prime / M_\gas$\tablenotemark{c}} &
  \colhead{} &
  \colhead{} &
  \colhead{$L_\pahs / M_\gas$\tablenotemark{d}}
  \\
  \colhead{Galaxy} & 
  \colhead{} &
  \colhead{($\LSun \MSun^{-1}$)} &
  \colhead{} & \colhead{} &
  \colhead{$L_\starburst^\prime / L_\dust$\tablenotemark{b}} &
  \colhead{($\LSun \MSun^{-1}$)} &
  \colhead{} &
  \colhead{$L_\agn^\prime / L_\dust$\tablenotemark{c}} &
  \colhead{($\LSun \MSun^{-1}$)} &
  \colhead{} &
  \colhead{$L_\pahs / L_\dust$\tablenotemark{d}} &
  \colhead{($\LSun \MSun^{-1}$)}
}

\startdata
   NGC\,7714 &   & $60.4\pm2.0$ &   &   & $0.912\pm0.018$ & $55.3\pm1.9$ &   &         \nodata &       \nodata &   & $0.052\pm0.002$ & $3.17\pm0.13$ \\
   NGC\,2623 &   & $78.2\pm4.7$ &   &   & $0.877\pm0.028$ & $68.5\pm4.6$ &   &         \nodata &       \nodata &   & $0.021\pm0.001$ & $1.65\pm0.11$ \\
PG\,0804+761 &   & $3270\pm160$ &   &   &         \nodata &      \nodata &   &   $2.00\pm0.06$ &  $6530\pm327$ &   &         \nodata &       \nodata \\
    Mrk\,463 &   &   $233\pm10$ &   &   & $0.114\pm0.003$ & $26.5\pm0.9$ &   &   $2.00\pm0.06$ &    $466\pm16$ &   &         \nodata &       \nodata \\
   NGC\,6240 &   & $52.4\pm2.1$ &   &   &   $1.43\pm0.02$ & $75.1\pm2.8$ &   & $0.017\pm0.001$ & $0.88\pm0.03$ &   & $0.032\pm0.002$ & $1.65\pm0.11$ \\
   Mrk\,1014 &   &      \nodata &   &   & $0.173\pm0.001$ &      \nodata &   & $0.661\pm0.001$ &       \nodata &   & $0.012\pm0.001$ &       \nodata 
\enddata

\tablenotetext{a}{Ratio of the total integrated dust luminosity to the total
  gas mass.}

\tablenotetext{b}{Ratios of the extinction-corrected \Starburst\ source
  component luminosity to the total dust luminosity and total gas mass.}

\tablenotetext{c}{Ratios of the extinction-corrected \AGN\ source component
  luminosity to the total dust luminosity and total gas mass.}

\tablenotetext{d}{Ratios of the total \PAHs\ component luminosity
  (integrated over all features) to the total dust luminosity and total gas
  mass.}

\tablecomments{Formal statistical uncertainties for the diagnostic ratios
  are given along with their values. Mass-normalized values are not given
  for Mrk\,1014 since no \MIPS\ data is unavailable to constrain its far-IR
  SED (and therefore $M_\gas$).  This absence of data also results in a
  $\sim$20\% uncertainty for the value of $L_\dust$ for Mrk\,1014 (i.e.  the
  difference between its nuclear and global values), and therefore in all
  quantities normalized by this total dust luminosity.}

\end{deluxetable*}
%

In Table~\ref{tab:DiagnosticRatios}, we present several `diagnostic' ratios
for each galaxy derived from the nuclear decomposition masses and
luminosities presented in Tables~\ref{tab:SourceAndDustLuminosities} and
\ref{tab:NuclearDustMasses}.  The first entry in this table gives the ratio
of the total dust luminosity to the total gas mass. The total gas mass is
calculated from the fitted value of the total dust mass and the gas-to-dust
ratio of our adopted dust model: $M_\gas / M_\dust = 1.4\,M_{\rm H} /
M_\dust = 124$
%
\citep{2001ApJ...554..778L}.
The definition of this quantity is motivated by the fact that two sources
with equivalent values of $L_\dust$ or $\LIR$ may have completely different
physical properties (e.g. a very massive but inactive system may have the
same dust luminosity as a relatively small but highly active galaxy).  The
total dust luminosity is therefore a degenerate quantity which does not
alone provide great insight into the nature of a galaxy.  Normalizing the
dust luminosity by the dust mass, however, does provide a means of breaking
this degeneracy. This ratio characterizes the radiative efficiency of a
galaxy, with smaller $L_\dust/M_\gas$ ratios indicating
lower-efficiencies and cooler dust temperatures and larger ratios indicating
higher-efficiencies and warmer dust temperatures.


Also included in Table~\ref{tab:DiagnosticRatios} are the ratios of the
extinction-corrected \Starburst\ and \AGN\ source component luminosities to
the total nuclear dust luminosity. A ratio of unity or greater in these
quantities indicates that the total observed dust emission could be heated
by the fitted source component. A ratio significantly greater than unity
indicates that either the source extinction correction was over-predicted,
or that the dust covering fraction is less than unity. In the latter case,
the fraction of the source which is covered by dust is obtained from the
inverse of the source-to-dust ratio. A ratio significantly less than unity
indicates that either the extinction to the source component is
underpredicted, or that additional unmodeled source emission is present
which has not been accounted for in the decomposition (e.g. the apparent
near-IR luminosity of a deeply embedded source may be small in comparison to
a relatively unextinguished source, even though the two may be intrinsically
similar in strength). In addition to the source-to-dust {\it luminosity}
ratios, we also calculate source-to-dust {\it mass} ratios, where the
extinction-corrected \Starburst\ and \AGN\ luminosities are given as a
fraction of the total gas mass. This quantity indicates the relative
`dustiness' (in terms of mass) of a source compared to others of the same
luminosity.

\subsubsection{PAH Feature Luminosities}
\label{sec:PAHFeatureLuminosities}

%
\begin{deluxetable*}{cccccc}
\tabletypesize{\scriptsize}
\tablewidth{0pc}

\tablecaption{PAH Feature Luminosities\label{tab:PAHFeatureLuminosities}}

\tablehead{
  \colhead{Galaxy} &
  \colhead{$L_{6.2} / L_\pahs$} &
  \colhead{$L_{7.7} / L_\pahs$} &
  \colhead{$L_{11.3} / L_\pahs$} &
  \colhead{$L_{17} / L_\pahs$}
}

\startdata
     NGC\,7714 & $0.114\pm0.002$ & $0.396\pm0.004$ & $0.114\pm0.002$ & $0.067\pm0.004$ \\
     NGC\,2623 & $0.088\pm0.002$ & $0.407\pm0.005$ & $0.074\pm0.001$ & $0.026\pm0.002$ \\
  PG\,0804+761 &         \nodata &         \nodata &         \nodata &         \nodata \\
      Mrk\,463 &         \nodata &         \nodata &         \nodata &         \nodata \\
     NGC\,6240 & $0.115\pm0.005$ & $0.371\pm0.009$ & $0.105\pm0.002$ & $0.069\pm0.004$ \\
     Mrk\,1014 & $0.112\pm0.013$ & $0.211\pm0.014$ & $0.067\pm0.008$ & $0.051\pm0.016$ \\
\hline \\ [-1.8ex]
Mean Starburst &          $0.10$ &          $0.39$ &          $0.11$ &          $0.08$ \\
  Median SINGS &          $0.11$ &          $0.42$ &          $0.12$ &          $0.06$
\enddata

\tablecomments{Ratios of the luminosities of each of the four main PAH
  complexes to the total PAH luminosity.  Formal statistical uncertainties
  for the fitted properties are given along with their values.  Also shown
  are the ratios for the mean starburst galaxy spectrum used to derive our
  PAH template (see Table~\ref{tab:PAHTemplate}) and the median values for
  the SINGS sample of star-forming galaxies presented in
  \citet{2007ApJ...656..770S}.}

\end{deluxetable*}
%
%
\begin{deluxetable}{ccccc}
\tabletypesize{\scriptsize}
\tablewidth{0pc}

\tablecaption{Extinguished Mean Starburst PAH Feature
  Luminosities\label{tab:ExtinguishedPAHs}}

\tablehead{
  \colhead{$\tau_{9.7}$} &
  \colhead{$L_{6.2} / L_\pahs$\tablenotemark{a}} &
  \colhead{$L_{7.7} / L_\pahs$\tablenotemark{a}} &
  \colhead{$L_{11.3} / L_\pahs$\tablenotemark{a}} &
  \colhead{$L_\pahs^\prime / L_\pahs^{}$\tablenotemark{b}}
}

\startdata
$0.25$ & $0.101$ & $0.402$ & $0.106$ & $1.09$ \\
 $0.5$ & $0.104$ & $0.411$ & $0.099$ & $1.19$ \\
   $1$ & $0.111$ & $0.429$ & $0.087$ & $1.40$ \\
   $2$ & $0.123$ & $0.462$ & $0.067$ & $1.91$
\enddata

\tablenotetext{a}{Complex-to-total PAH luminosity ratio for the indicated
  screen $\tau_{9.7}$.}

\tablenotetext{b}{Unextinguished-to-apparent PAH luminosity ratio for the
  indicated screen $\tau_{9.7}$.}

\end{deluxetable}
%

The total apparent PAH luminosity of each source (integrated over all
features), normalized to both the total dust luminosity and the total gas
mass, are presented in Table~\ref{tab:DiagnosticRatios}. Additionally, the
fitted ratios of the luminosities of the four primary PAH complexes (see
Table~\ref{tab:PAHTemplate}) to the total PAH luminosity are presented in
Table~\ref{tab:PAHFeatureLuminosities}. For reference, we also provide the
corresponding ratios for the mean starburst galaxy spectrum from
%
\citet{2006ApJ...653.1129B}
used to create our PAH template, and the median ratios for the star-forming
SINGS galaxies from
%
\citet{2007ApJ...656..770S}.
With the exception of Mrk\,1014, the fitted PAH luminosity ratios of the
four sources in our sample with well-detected PAH emission are fairly
consistent with the values of the mean starburst and SINGS templates (see
\S\ref{sec:DetailedDescriptions} for more on the PAH emission of Mrk\,1014).


The apparent ratios of PAH emission from a galaxy may differ from those of
the low-obscuration templates in Table~\ref{tab:PAHFeatureLuminosities} for
several reasons.  First, the size of a PAH (i.e. the number of carbon atoms)
and its ionization state can alter the relative strengths of its emission
features
%
\citep[see][]{2001ApJ...551..807D}.
Additionally, the apparent ratios may vary as a result of extinction from
intervening dust.  In Table~\ref{tab:ExtinguishedPAHs}, we present the
ratios of the 6.2, 7.7, and $11.3\um$ PAH complexes to the total PAH
luminosity for the mean starburst galaxy from
Table~\ref{tab:PAHFeatureLuminosities} after passing through screens of dust
with various values of $\tau_{9.7}$. In addition to the feature ratios, we
also provide correction factors used to convert the total apparent PAH
luminosity into the actual emitted PAH luminosity for a particular value of
$\tau_{9.7}$ estimated from the feature ratios.

\section{Discussion}
\label{sec:Discussion}

\subsection{Detailed Descriptions of Decompositions}
\label{sec:DetailedDescriptions}

\paragraph{NGC\,7714}

Our low-obscuration template starburst is a massive barred spiral galaxy
that makes up the western component of the interacting system Arp\,284.  The
compact ultraviolet-luminous starburst nucleus of NGC\,7714 has very low
extinction, and there is no evidence at any wavelength for the presence of
an AGN. The \IRS\ spectrum of this galaxy
%
\citep[see][]{2004ApJS..154..188B}
exhibits a rising continuum, strong PAH feature and line emission, and no
perceptible $9.7\um$ silicate absorption.  As suggested by the relatively
flat continuum near $10\um$, all parameters from our fitting are consistent
with this starburst galaxy being relatively unobscured at all wavelengths.
The \Warm\ dust and \Starburst\ source components have fitted values of
$\tau_{9.7} \approx 0.19$ and $\tau_{V} = 0.97$, respectively.
%
\citet{1994MNRAS.266..431P}
estimate a screen extinction to the ionized gas within NGC\,7714 of $A_V
\approx 0.86\pm0.13$ using optical and near-IR recombination lines, a value
which is entirely consistent with our \Starburst\ component optical depth.
For our adopted dust model, $A_V \approx 0.86$ corresponds to $\tau_{9.7}
\approx 0.07$--$0.22$, depending on the influence of scattering in the total
effective opacity (the latter value is for zero scattering contribution, as
expected for spherical symmetry).  This range is again completely consistent
with the extinction we find for the \Warm\ dust component.


The extinction-corrected \Starburst-to-dust ratio given in
Table~\ref{tab:DiagnosticRatios} is $L_\starburst^\prime / L_\dust = 0.91$,
indicating that the fitted source component is capable of powering nearly
all (i.e. 91\%) of the total dust emission. Although we attempt to calculate
the true far-IR dust luminosity of the nuclear point-source, we may
nonetheless overestimate this quantity since our \MIPS\ photometry may still
contain flux from outside the nucleus. Such an overestimate could give rise
to the fitted slight source-luminosity deficit (i.e. there is not enough
source emission to explain the observed dust emission).  Given this possible
explanation, and the fact that the deficit is small ($< 10$\%), we conclude
that the observed source-to-dust ratio of NGC\,7714 is consistent with the
total system being powered by star-formation.


Our decomposition of this galaxy therefore indicates that the warmest dust
in pure low-obscuration starbursts typically has a temperature
$\sim$165$\K$, and that emission from $\sim$1000$\K$ dust is therefore
negligible. Also, the observed PAH luminosity in NGC\,7714 is $\sim$5.2\% of
its total nuclear dust luminosity, providing an estimate of the typical
total PAH strength in low-obscuration starbursts.
%
\citet{2007ApJ...656..770S}
find a median PAH-to-dust luminosity ratio of $\sim$10\% for the `normal'
(i.e. non-starburst) star-forming galaxies in the SINGS sample, with the
ratio ranging between $3.2$ and $16$\% for the sample. Thus, the ratio
measured for NGC\,7714 is consistent with this range, with the nearly factor
of two difference from the median value likely indicating real differences
in the emission from galaxies with different levels of star-formation. Note
also that the PAH ratio $L_{11.3} / L_{7.7} \approx 0.28$ for NGC\,7714 is
very similar to the value of this ratio for the unobscured mean starburst
template ($\sim$0.27).  The PAH emission in NGC\,7714 therefore appears to
be relatively unobscured, consistent with the results from other extinction
indicators.


The total global dust emission in our sample (i.e. obtained using \IRAS\ and
\ISO\ far-IR data) ranges from $\sim$1.0 to 1.4 times the nuclear emission
(i.e. obtained using \MIPS\ data).
%
\citet{2001ApJ...552..150L}
estimate that only $10$--$50$\% of the stellar emission in NGC\,7714 emerges
from the $2\arcsec\times2\arcsec$ ($\sim$0.37$\pc^2$) nucleus, and therefore
infer that $50$--$90$\% of the far-IR \IRAS\ emission likely emerges from
the extended disk of the galaxy. They further reason that the strength of
the extra-nuclear far-IR emission is likely to be towards the lower end of
this range, since the ultraviolet observations used to quantify the stellar
emission strength are prone to missing emission from extinguished
populations. In comparing the results of our global and nuclear spectral
decompositions of NGC\,7714, we calculate that at least $\sim$50\% of the
\Cold\ dust component luminosity (and 30\% of the total dust emission)
emerges from extended regions, consistent with the
\citet{2001ApJ...552..150L} prediction.

\paragraph{NGC\,2623} 

Our extinguished template starburst galaxy is a late stage merger with long
tidal tails and a ${\rm r}^{1/4}$ nuclear stellar light profile
%
\citep{1990Natur.344..417W, 2000AJ....119..991S}.
The radio continuum
%
\citep{1991ApJ...378...65C}
and mid-IR 
%
\citep{2001AJ....122.1213S}
emission is similarly dominated by a compact nuclear source. The \IRS\
spectrum of NGC\,2623 is dominated by deep silicate absorption features at
$9.7$ and $18\um$ and a very steep continuum beyond $20\um$. The fitted
optical depths to the \Warm\ dust and \Starburst\ source components of
NGC\,2623 are $\tau_{9.7} \approx 3.5$ and $\tau_{V} \approx 4.8$. Note that
%
\citet{2006ApJ...653.1129B}
find an apparent $9.7\um$ optical depth of $\sim$1.5, which is more than a
factor two lower than our fitted value. This difference occurs as a result
of the method they employ to determine the {\it apparent} optical
depth---namely interpolating a smooth continuum above the silicate
absorption feature and using the ratio of this `unextinguished' continuum to
the observed value to infer $\tau_{9.7}$. In contrast, the unextinguished
continuum in the dust model used in our decomposition method is not smooth
at $9.7\um$, but instead features a silicate {\it emission} feature (the
presence of which is consistent with observations---e.g. see the fit for
PG\,0804+761), therefore requiring a higher optical depth to obtain the same
extinguished continuum.


We find no evidence for the presence of an AGN contribution to the SED of
NGC\,2623, consistent with the similar conclusion in
%
\citet{2000A&A...357...13R}
based upon hard X-ray observations. The extinction-corrected
\Starburst-to-dust ratio given in Table~\ref{tab:DiagnosticRatios} is
$L_\starburst^\prime / L_\dust = 0.88$, indicating that the fitted source
component is capable of powering most of the total dust emission. As with
NGC\,7714, it is likely that this ratio is indicative of an entirely
star-formation powered system (taking into account the likely small
overprediction of the nuclear dust luminosity due to the \MIPS\ resolution).
Like NGC\,7714, the mid-IR spectrum of NGC\,2623 also exhibits significant
emission from PAHs, albeit a factor $\sim$2.5 weaker. The PAH ratio
$L_{11.3} / L_{7.7} \approx 0.18$ for NGC\,2623 is also lower than the value
for NGC\,7714, with the lower ratio being consistent with PAH emission
obscured by a screen of dust with $\tau_{9.7} \approx 1$ (as determined by
comparison with the obscured mean starburst template properties in
Table~\ref{tab:ExtinguishedPAHs}). For this optical depth, the apparent PAH
luminosity of the mean starburst template is only $\sim$71\% its true
emitted value. Using this correction factor, we estimate that the true
PAH-to-dust luminosity ratio of NGC\,2623 is $L_\pahs^\prime / L_\dust^{}
\approx 0.029$, a factor $\sim$1.8 lower than NGC\,7714. See
\S\ref{sec:ComparisonStarburstSEDs} for a detailed comparison of the
properties of the low obscuration system NGC\,7714 and the obscured
starburst NGC\,2623.

\paragraph{PG\,0804+761} 

Our Type-1 template AGN is a variable source displaying $\sim$40\% near-IR
and $10.6\um$ brightness fluctuations over the course of a decade
%
\citep{1999AJ....118...35N}.
The first deep infrared observations of this source were obtained with \ISO\
%
\citep{2003A&A...402...87H},
which revealed a power-law spectral shape and a very warm infrared SED
peaking at $\sim$30$\um$. Analysis of its \IRS\ spectrum by
%
\citet{2005ApJ...625L..75H}
revealed the presence of strong $9.7$ and $18\um$ amorphous silicate
emission features. In our decomposition, these emission features are
produced by grains in the \Warm\ dust component. If we assume that the
\Warm\ component is directly illuminated by the PG\,0804+761 accretion disk
($L_\agn = 1.26\times10^{12}\LSun$), then equation~(\ref{eq:ThermalHeating})
with $\tilde{U} c \hat{u}_\nu \rightarrow L_\nu / 4 \pi r^2$ indicates that
this dust is located $\sim$19$\pc$ from the nucleus.
%
\citet{2004Natur.429...47J}
analyze $10\um$ VLTI interferometric observations of the nucleus of the
nearby Type-2 AGN NGC\,1068, and conclude that its torus emission must be
confined to a region $\la 2\pc$ in size. Thus, some of the optically-thin
\Warm\ component emission from PG\,0804+761 may emerge from regions beyond
the torus, perhaps from clouds in the narrow-line-region.  Note that this
distance to the \Warm\ dust is actually an upper-limit to the true distance,
since the emission may also originate from indirectly illuminated torus
clouds (see Fig.~\ref{fig:SourceModel}) at $r < 19\pc$.


The unextinguished \AGN-to-dust ratio in the PG\,0804+761 decomposition is
$\sim$2 (see Table~\ref{tab:DiagnosticRatios}), indicating that the \AGN\
source component is more than capable of heating the total observed dust
emission. The inverse of this ratio provides an estimate of the total dust
covering factor (i.e. the fraction of sky covered by the torus as seen from
the nucleus), which is $f_{\rm torus} \approx 0.5$ (note that this value is
at the constraint imposed during the fit---see
\S\ref{sec:DecompositionParametersAndConstraints}). This covering factor is
consistent with a torus opening angle of $\sim$30$^\circ$ (i.e. $\sin
\theta_{\rm torus} = f_{\rm torus}$, where $\theta_{\rm torus}$ is the angle
extending from the mid-plane to the top of the torus as viewed from the
accretion disk). Based upon the statistics of Seyfert galaxies (i.e. the
relative numbers of Type-1 and 2 galaxies),
%
\citet{2001ApJ...555..663S}
find approximately twice as many Type-2 AGN, and hence estimate $\theta_{\rm
  torus} = 45^\circ$. More recently,
%
\citet{2005AJ....129.1795H}
study the AGNs in the Sloan Digital Sky Survey and conclude that
$\theta_{\rm torus} \approx 30^\circ$, consistent with the value obtained
from our decomposition.

%
\citet{2004Natur.429...47J}
find that the $8$--$13\um$ VLTI spectrum of NGC\,1068 can be decomposed into
a $\bar{T} > 800\K$ dust component obscured by a $\tau_{9.7} = 2.1\pm0.5$
screen and a $\bar{T} = 320\pm30\K$ component behind a $\tau_{9.7} =
0.3\pm0.2$ screen. The temperatures and screen optical depths of these
components are comparable to those obtained in our spectral decomposition
(see also the properties of Mrk\,463). PG\,0804+761 has the worst reduced
$\chi^2$ of the galaxies in our sample, due primarily to the poor fit to its
\IRS\ spectrum between $7.5$--$10\um$ (although the fit to the rest of the
SED is quite good).
%
\citet{2004Natur.429...47J}
report a similar problem in this wavelength range in their fit to the
spectrum of NGC\,1068, and conclude that the fit is improved using
high-temperature calcium aluminium silicate dust instead of the standard
olivine-type silicates (the opacity of the $9.7\um$ feature of the former
species begins near $9\um$ as opposed to the $\sim$8$\um$ onset for
olivine). We finally note that our decomposition of the ultraviolet and
optical emission is consistent with a largely uncovered \AGN\ accretion
disk---with only $\sim$25\% covered---consistent with a Type-1 source.

\paragraph{Mrk\,463} 

Our Type-2 template AGN is a merging system with two nuclei separated by
$\sim$4$\arcsec$
%
\citep{1991AJ....102.1241M}.
The portion of the \IRS\ spectrum obtained with the short-low module is
likely dominated by Mrk\,463e (since the slit of this module is only
$\sim$3.6$\arcsec$ wide and the observation targeted the eastern component),
while the portion obtained with the wider-slit of the long-low module likely
contains emission from both nuclei (although the optically-bright western
component may not contribute significantly at these wavelengths). Mrk\,463
has a luminous steep-spectrum radio core, and broad lines are seen in
scattered optical
%
\citep{1990ApJ...355..456M}
and direct near-IR light
%
\citep{1994ApJ...422..521G, 1997ApJ...477..631V}.
Like the quasar PG\,0804+761, the mid-IR spectrum of this galaxy shows no
discernible PAH emission features
%
\citep{2004ApJS..154..178A}.
Silicate absorption at $9.7\um$ is clearly seen in the spectrum of Mrk\,463,
producing a mid-IR continuum dominated by a broad emission-like feature at
$\sim$8$\um$, characteristic of self-absorbed silicate dust. In our
decomposition, this feature is primarily produced by the \Warm\ component
which has a screen optical depth $\tau_{9.7} \approx 1.3$.  Like the quasar
PG\,0804+761, Mrk\,463 has a moderately obscured ($\tau_{9.7} \approx 2.1$)
\Hot\ dust component. The \Hot\ component optical depth and the deviations
of the fit to the \IRS\ spectrum from $7.5$--$9\um$ are both similar to the
reported properties of PG\,0804+761 and NGC\,1068 described above. Unlike
the fit to the quasar PG\,0804+761, the \AGN\ component of Mrk\,463 is
completely covered by obscuring clouds---i.e. $\tilde{\epsilon}_\agn =
1$---as is expected for a Type-2 source.


The \AGN-to-dust ratio for Mrk\,463 is $1.93$ (similar to the value for
PG\,0804+761), consistent with the \AGN\ source component powering the total
observed dust emission, and implying a torus opening angle $\theta_{\rm
  torus} \approx 31^\circ$. There is evidence for \ISRF\ and \Starburst\
contributions in the decomposition, although these are bolometrically weak
compared to the power of the AGN ($< 5$\%). Although Mrk\,463 and
PG\,0804+761 have similar \AGN-to-dust ratios, they have very different
quantities of cooler dust. Mrk\,463 has 16 times more \Cold\ dust by mass
(see Table~\ref{tab:NuclearDustMasses}), and it also has a factor 14 times
smaller dust luminosity-to-mass ratio (also indicative of cooler mean dust
temperatures---see Table~\ref{tab:DiagnosticRatios}).  Similarly, the ratio
of the unextinguished \AGN\ component luminosity to the total gas mass is a
factor 14 smaller for Mrk\,463 (i.e. it has 14 times more gas mass for the
same accretion disk luminosity). In the standard AGN unification scenario,
the only difference between a Type-1 source (such as PG\,0804+761) and a
Type-2 source (such as Mrk\,463) is the orientation of the obscuring torus
with respect to the observer. Thus, within this picture, Type-1 and Type-2
AGNs should have similar \Cold\ dust masses relative to their total masses,
and they should therefore have similar \AGN\ luminosity-to-mass ratios
(since the accretion disk luminosity and the total mass of dust do not
depend on orientation). In \S\ref{sec:FarIRInAGNs}, we suggest that the
relative orientations of the accretion and host galaxy disks may explain why
this simple consequence of the unification scenario does not hold for some
galaxies. We therefore emphasize that the SEDs of individual AGNs may appear
falsely inconsistent with the unification scenario due to local geometrical
factors, even if unification holds for AGNs as a class.

\paragraph{NGC\,6240} 

The first of our two composite sources is an interacting system containing
two nuclei separated by $\sim$1$\arcsec$ (so that both nuclei are contained
within all \IRS\ slits). With an $\LIR \approx 6.5 \times 10^{11}\LSun$
NGC\,6240 is technically a LIRG, although it is often treated as a ULIRG
since it shares most properties with other members of the class.  Based upon
its optical nuclear spectrum, NGC\,6240 is classified as a LINER
%
\citep{1989ApJ...347..727A},
and X-ray observations
%
\citep{2003ApJ...582L..15K}
provide clear evidence for the presence of a pair of AGNs located behind
significant columns of absorbing neutral gas ($\NH = 1$--$2\times10^{24}
{\rm cm}^{-2}$). The mid-IR spectrum of NGC\,6240 is extremely rich,
displaying strong PAH features, both low and high ionization lines (e.g.
[\ion{Ne}{2}]~$12.8\um$ and [\ion{Ne}{5}]~$14.3\um$) as well as strong
emission from molecular hydrogen
%
\citep[see][]{2006ApJ...640..204A}.  
Like NGC\,2623, the mid-IR continuum of this galaxy is shaped by strong
absorption from silicate grains at $9.7$ and $18\um$, as indicated by the
\Warm\ component screen optical depth of $\tau_{9.7} \approx 3.6$.  The
near-IR continuum of NGC\,6240 is well-fitted by a combination of apparently
unobscured weak \Hot\ dust and extinguished ($\tau_{V} = 4.7$) \Starburst\
component emission. The detection of [\ion{Ne}{5}]~$14.3\um$ in the
high-resolution \IRS\ spectrum of NGC\,6240 presented in
%
\citet{2006ApJ...640..204A}
and the evidence for X-rays are both consistent with the presence of \Hot\
dust in our decomposition---indicating that a (perhaps small) fraction of
the near-IR emission is powered by an AGN.


Based upon {\it Chandra} X-ray observations,
%
\citet{2002MNRAS.333..709L} 
conclude that the AGN in NGC\,6240 likely contributes between $30$--$50$\%
of the bolometric luminosity. However, as reported in 
%
\citet{2006ApJ...640..204A},
the small [\ion{Ne}{5}$]/[$\ion{Ne}{2}] and [\ion{Ne}{5}$]/\LIR$ flux ratios
are both consistent with an AGN contribution of only $3$--$5$\% of the
bolometric luminosity. Similarly, the apparent \Hot\ dust emission in our
decomposition makes up just $\sim$2\% of the total dust luminosity (compared
to $> 36$ and $30$\% for PG\,0804+761 and Mrk\,463, respectively), so that
any contribution to the observed near- and mid-IR from an AGN is very small.
On the other hand, the \Starburst-to-dust ratio for NGC\,6240 is 1.43,
indicating that the total observed dust emission could be powered by the
observed \Starburst\ component. This galaxy therefore presents quite a
puzzle, whereby its appearance changes radically depending upon where you
look---although X-rays suggest powerful AGNs, data at optical and longer
wavelengths do not require such a presence.  We note that even though the
\Hot\ component in the decomposition is unobscured, it could still be
associated with a deeply obscured AGN in a clumpy geometry where most of the
hot dust is too obscured to be seen, and we view only the most unobscured
portions (see \S\ref{sec:HotDustInAGNTori}).  If this scenario is true, the
actual \Hot\ dust contribution to the total bolometric luminosity could
easily be pushed higher.

\paragraph{Mrk\,1014} 

Our second composite source is a radio-quiet infrared-luminous QSO with
broad optical emission lines (FWHM H$\beta > 4000$ km s$^{-1}$). Mrk\,1014
displays twin tidal tails, indicative of a recent interaction and merger
%
\citep{1984ApJ...283...64M}.
It is a relatively warm far-IR source, with the peak of its SED located
around 70$\um$. Its mid-IR spectrum is characterized by a nearly power-law
continuum, with no obvious silicate emission or absorption features, and
weak PAH emission
%
\citep[see][]{2004ApJS..154..178A}.
This power-law continuum is rather remarkable (assuming the continuum is
thermal in origin) since a nearly featureless SED must be constructed from
emission components which are known to have significant features (i.e.
silicates). The decomposition which provides a solution to this puzzle is
consistent with Mrk\,1014 being a composite source, containing a combination
of dust components seen in both our template starbursts and AGNs. Like the
AGNs, Mrk\,1014 contains a moderately obscured ($\tau_{9.7} \approx 2.7$)
\Hot\ dust component and a much less obscured \Warm\ component ($\tau_{9.7}
\approx 0.14$). Like the starbursts (and unlike the quasar), Mrk\,1014 also
contains bolometrically significant emission from cooler dust, comprising
$\sim$60\% of its total dust luminosity (we argue in \S\ref{sec:FarIRInAGNs}
that some of this far-IR emission is likely powered by the AGN).


The extinction-corrected \AGN-to-dust and \Starburst-to-dust luminosity
ratios of Mrk\,1014 are 0.66 and 0.17, respectively, indicating that the AGN
is capable of powering $\sim$66\% of the observed dust emission from the
galaxy. As fitted, the \AGN\ and \Starburst\ components are together capable
of powering 83\% of the observed dust emission, thereby requiring the
presence of at least an additional 17\% source luminosity to power the
remaining dust emission (note also that the total nuclear dust luminosity
from Mrk\,1014 is uncertain by $\sim$20\% since no \MIPS\ data is available
to constrain its far-IR SED). If all of this undetected source luminosity
emerges from the AGN accretion disk (e.g. if we underpredict the \AGN\
component luminosity as a result of too rigid a constraint on the
\Hot-to-\AGN\ luminosity ratio), the AGN would account for 83\% of the
bolometric luminosity of Mrk\,1014. If, on the other hand, the remaining
undetected source luminosity is entirely produced by a starburst (e.g. if we
underpredict the \Starburst\ luminosity as a result of missing emission from
highly embedded regions), then the AGN would account for 66\% of the
bolometric luminosity.

%
\citet{2007ApJ...656..148A}
create diagnostic diagrams based upon mid-IR spectral lines (see their Figs.
5--8) which are all consistent with an AGN fraction between 50--90\% for
Mrk\,1014.  Additionally,
%
\citet{2002MNRAS.336.1143B}
conclude that Mrk\,1014 is dominated by an AGN and not star-formation based
upon X-ray observations.  If we take the \PAHs-to-dust luminosity fractions
of the galaxies NGC\,2623 and NGC\,6240 to be representative of the range of
values in obscured starbursts, then the \PAHs-to-dust ratio of Mrk\,1014
suggests that between $\sim$38 and 57\% of its bolometric luminosity is
powered by obscured star-formation (assuming all PAH emission comes from
star-formation). Given our conclusion above that the starburst contribution
is between 17 and 34\%, it is therefore likely either that we are
underestimating the total starburst contribution based upon the source
components (which seems improbable given the corroborating X-ray and
emission line evidence), or that some of the observed PAH emission may
actually be powered by the AGN (see \S\ref{sec:FarIRInAGNs}). We note as
well that the PAH feature luminosity ratios for Mrk\,1014 (see
Table~\ref{tab:PAHFeatureLuminosities})---in particular the $7.7\um$
complex---differ significantly from the template starbursts, perhaps
indicative of an AGN contribution giving rise to non-standard PAH emission.

\subsection{Comparison of Starburst SEDs}
\label{sec:ComparisonStarburstSEDs}

From our modest sample of two starburst galaxy decompositions, it is
apparent that starburst SEDs do not form a homogeneous group.  Comparing the
fits of NGC\,2623 and NGC\,7714, noticeable differences include a 30\%
higher maximum dust temperature, a factor 20 times higher maximum optical
depth, a factor 2.5 times weaker \PAHs-to-dust luminosity ratio, and
comparatively weaker [\ion{Ne}{3}]~$15.56\um$, [\ion{S}{3}]~$18.71\um$,
[\ion{S}{3}]~$33.48\um$, and [\ion{Si}{2}]~$34.82\um$ line emission. If the
nuclear emission from each starburst emerges from distinct star-forming
clouds consisting of embedded stars surrounded by cocoons of dust (see
Fig.~\ref{fig:GalaxyModel}), then many of the observed variations in the
SEDs of NGC\,7714 and NGC\,2623 can be understood as resulting from: (1)
differences in the total number of star-forming clouds (i.e. if individual
cloud complexes in each galaxy have similar intrinsic luminosities, then the
total luminosity of each galaxy is determined by the total number of clouds
it contains), and (2) the mean optical depth through an individual cloud.
There are, of course, many other factors shaping the SEDs of starbursts as
well---e.g. age and metallicity.  The simple model described in this section
is therefore not intended to provide a comprehensive explanation of all
starburst properties, but rather a structure within which to understand
several general properties.


If we assume for simplicity that each cloud in a starburst has constant gas
density, then the optical depth through a cloud is $\tau(r) = \tau_{\rm
  cloud}(r / r_{\rm cloud})$, where $\tau_{\rm cloud}$ and $r_{\rm cloud}$
are the total optical depth and radius of the cloud. As described in
\S\ref{sec:OpticallyThinEmission}, direct emission from stars heats a shell
of dust out to a radius $r_{\rm shell}$, corresponding to the optical depth
$\tau_{9.7}^{\rm shell}$, where the cloud becomes optically thick to the
heating stellar photons (i.e. essentially the outer edge of the
photodissociation region). Using the equation for the optical depth above,
we obtain $r_{\rm shell}^{\rm 2623} / r_{\rm shell}^{\rm 7714} = (\tau_{\rm
  cloud}^{\rm 7714} / \tau_{\rm cloud}^{\rm 2623}) (r_{\rm cloud}^{\rm 2623}
/ r_{\rm cloud}^{\rm 7714})$. Here, we assume that $\tau_{9.7}^{\rm shell}$
is the same for all clouds since it depends only on the properties of dust
and the illuminating radiation field.  With $\tau_{\rm
  cloud}\rightarrow\tau_\warm$ (i.e. using the \Warm\ component optical
depth as a proxy for the total cloud optical depth), this gives $r_{\rm
  shell}^{\rm 2623} / r_{\rm shell}^{\rm 7714} \approx (0.05)(r_{\rm
  cloud}^{\rm 2623} / r_{\rm cloud}^{\rm 7714})$.  If the clouds in each
galaxy are similar in size, this suggests that $r_{\rm shell}^{\rm 2623} /
r_{\rm shell}^{\rm 7714} < 1$ (i.e. the \Warm\ dust shell from
Fig.~\ref{fig:SourceModel} is geometrically more extended in NGC\,7714 than
NGC\,2623), so that stellar photons penetrate to much larger radii in the
star-forming clouds of NGC\,7714.  Furthermore, the fractional cloud volume
containing dust at the temperature of the \Warm\ component should be smaller
in NGC\,2623 than NGC\,7714, and consequently the mass of \Warm\ dust
relative to the total mass should be smaller as well.  Indeed, from
Table~\ref{tab:NuclearDustMasses} we find that $(M_\warm / M_\dust)_{\rm
  7714} \approx 1.3\times10^{-4}$ and $(M_\warm / M_\dust)_{\rm 2623}
\approx 4\times10^{-5}$, consistent with this prediction.


For $r < r_{\rm shell}$, energy conservation dictates that the temperatures
of grains within a cloud scale roughly as $T \propto r^{-1/2}$.  This,
coupled with the fact that $r_{\rm shell}^{\rm 2623} < r_{\rm shell}^{\rm
  7714}$, implies that the mean temperature of such grains should be higher
in NGC\,2623 than NGC\,7714---a prediction which is consistent with its 25\%
higher \Warm\ component temperature.  The luminosities per unit mass of dust
at the temperatures of the \Warm\ components in the two galaxies scale
approximately as $(\Tavg_\warm^{\rm 2623} / \Tavg_\warm^{\rm 7714})^4
\approx (1.25)^4 \approx 2.4$ (i.e. a mass of \Warm\ dust from NGC\,2623
will be $\sim$2.4 times more luminous than an equivalent mass from
NGC\,7714). This ratio, along with the absolute \Warm\ component dust masses
for the two galaxies, suggests that the ratio of the \Warm\ component
luminosities should be $L_\warm^{\rm 2623} / L_\warm^{\rm 7714} \approx
(\Tavg_\warm^{\rm 2623} / \Tavg_\warm^{\rm 7714})^4 (M_\warm^{\rm 2623} /
M_\warm^{\rm 7714}) \approx 4$. The actual fitted ratio is 4.7, in
reasonable agreement with this prediction.  Furthermore, the $L_\dust /
M_{\rm H}$ ratio from Table~\ref{tab:DiagnosticRatios} is $\sim$30\% higher
in NGC\,2623 than NGC\,7714. If the \Warm\ dust in NGC\,7714 were at the
temperature of the \Warm\ dust in NGC\,2623, the total dust luminosity of
NGC\,7714 would increase by $\sim$40\%, and the resulting dust
luminosity-to-mass ratios of the two sources would differ by only $\sim$7\%.
Thus, the different values of $L_\dust / M_\gas$ are directly related to the
higher \Warm\ dust temperature in NGC\,2623, which itself is likely directly
related to the higher optical depth through the star-forming clouds in that
galaxy.


The apparent \PAHs-to-dust luminosity ratio is a factor $\sim$2.5 times
lower in NGC\,2623 than NGC\,7714. We estimate that extinction to the \PAHs\
component in NGC\,2623 reduces its luminosity by a factor 1.4 (see
\S\ref{sec:DetailedDescriptions}), so that the unextinguished \PAHs-to-dust
luminosity ratios for the two galaxies differ by a factor 1.8.  With this
correction, we also find that $L_\pahs / M_\gas$ is only a factor 1.4
greater in NGC\,7714, as opposed to the observed factor 1.9.  As described
above, the \Warm-to-total dust mass ratio is less in NGC\,2623, so that the
volume within which PAH molecules are heated is smaller for this source.
Therefore, some portion of the reduced \PAHs\ luminosity giving rise to the
differences in ratios described above likely results from such a difference
in heating geometries.  Using the ratio of the extinction-corrected $L_\pahs
/ M_\gas$ values derived above, we estimate that this geometrical effect
must decrease the observed \PAHs\ luminosity of NGC\,2623 by a factor 1.4
(i.e. assuming that the full difference is caused by this geometrical
effect).  Note that this factor is similar to the value $[(M_\warm /
M_\dust)_{\rm 2623} / (M_\warm / M_\dust)_{\rm 7714}] (L_\warm^{\rm 2623} /
L_\warm^{\rm 7714}) \approx 1.4$, where the first term accounts for the
difference in volume within which PAHs are heated, and the second term
accounts for the difference in heating intensity. With these two correction
factors of 1.4, the \PAHs-to-dust luminosity ratio of NGC\,2623 differs by
only $\sim$30\% from that of NGC\,7714.  Recalling that the the dust
luminosity of NGC\,7714 would increase by $\sim$40\% if its \Warm\ dust were
at the same temperature of the dust in NGC\,2623, the \PAHs-to-dust
luminosity ratios of the two galaxies differ by less than 10\%.


The differences in the apparent PAH emission from our two starburst
templates may therefore be understood as resulting from a combination of:
(1) extinction to the PAHs, (2) the geometry of the PAH emitting regions,
and (3) the method of normalizing the \PAHs\ component luminosity.  We
suggest that the ratio $L_\pahs / M_\gas$ is a somewhat better tracer of
the true strength of PAH emission than $L_\pahs / L_\dust$, since it is less
affected by temperature effects.  This latter ratio, or a related quantity
such as $L_{6.2} / L_{\rm FIR}$, is often used to quantify the amount of
star-formation in galaxies
%
\citep[e.g.][]{2004ApJ...613..986P}.
Given that both our template starbursts are believed to be entirely powered
by star-formation, it is clear that there is significant variation in the
relative strength of PAH emission in pure starbursts. We therefore caution
that great care must be taken when using any of these metrics to derive
absolute quantities of star-formation, since both extinction and geometrical
effects may result in different ratios for pure starburst galaxies.

\subsection{Hot Dust Emission from Clumpy AGN Tori}
\label{sec:HotDustInAGNTori}

%
\begin{figure}
  \epsscale{1.1}
  \plotone{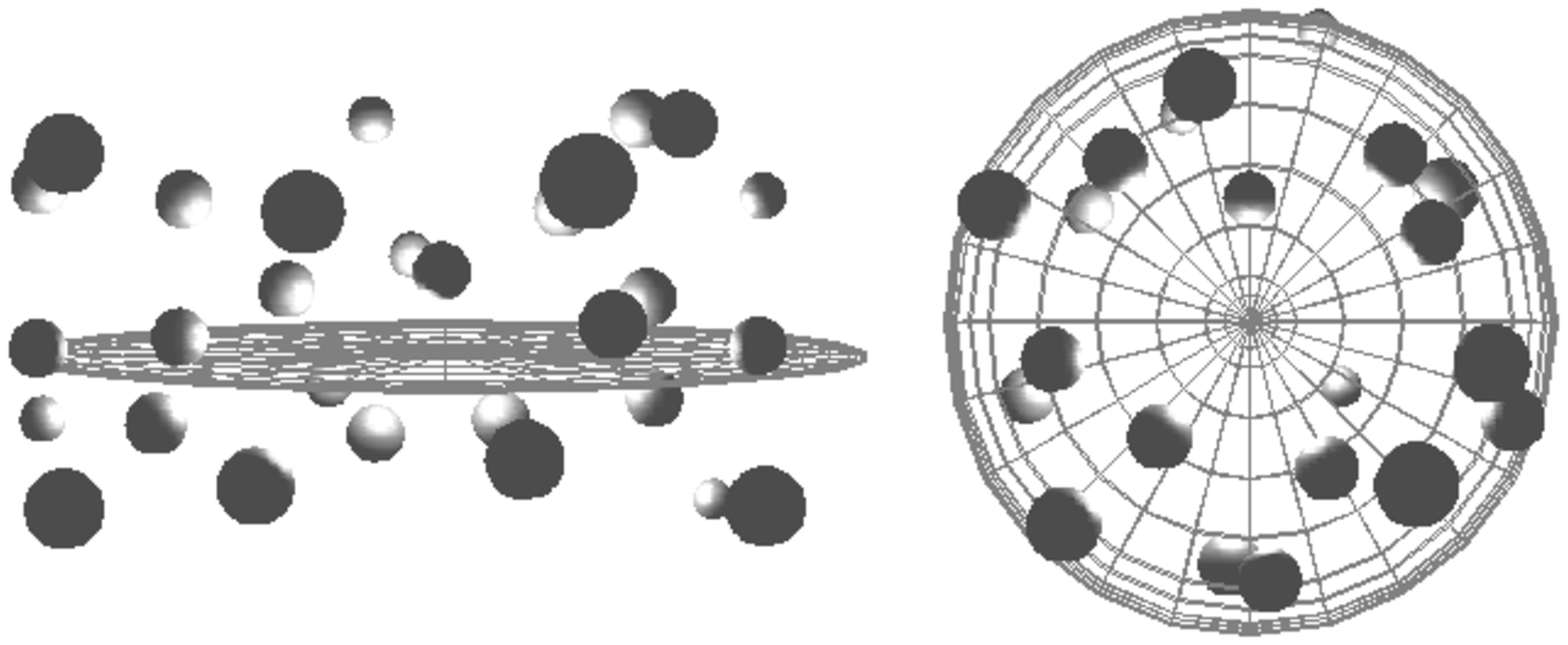}
  \plotone{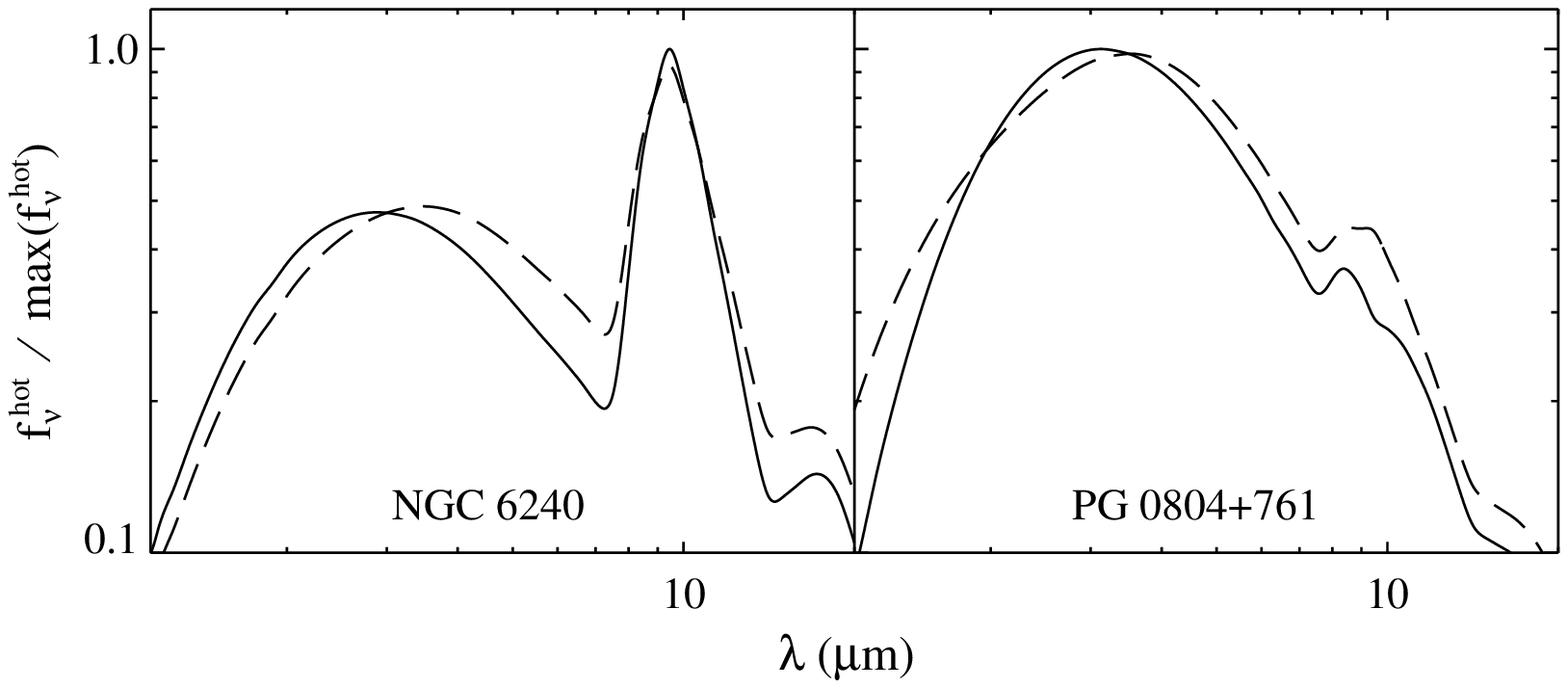}
  \caption{Edge-on ({\it upper-left}) and face-on ({\it upper-right}) views
    of a clumpy torus comprised of a toroidal distribution of discrete
    clouds.  Directly illuminated cloud faces (i.e. the sources of \Hot\
    emission) are obscured from most viewpoints by dust on the shaded side
    of each cloud. SEDs of the fitted NGC\,6240 and PG\,0804+761 \Hot\
    components are shown beneath the schematic torus model representing
    their geometry ({\it solid lines}), along with model SEDs derived
    assuming that a fraction of the \Hot\ component emission (50\% for
    NGC\,6240 and 75\% for PG\,0804+761) is obscured by a $\tau_V = 25$
    screen, with the remainder emerging unobscured ({\it dashed lines}).}
  \label{fig:TorusModel}
  \epsscale{1.0}
\end{figure}
%

If the obscuring dust surrounding an AGN accretion disk is arranged in a
smooth (i.e. non-clumpy) toroidal distribution, then the $\sim$3--$8\um$
SEDs of Type-1 sources should be dominated by emission from relatively
unobscured \Hot\ dust (since this \Hot\ dust is viewed directly by the
observer, with no intervening material from the torus along the
line-of-sight to obscure it).  If, however, the obscuring structure is
instead composed of discrete clouds distributed in a roughly toroidal shape
(i.e. a clumpy torus), then \Hot\ dust emission is not expected to emerge
unobscured.  As illustrated in Figure~\ref{fig:TorusModel}, the directly
illuminated (and therefore hotter) faces of clouds in a clumpy torus always
point radially towards the nucleus, and therefore away from the observer.
Thus, emission from these hot cloud faces, and therefore \Hot\ component
emission from such sources, is obscured by cooler dust on the shaded sides
of clouds.  Our decomposition of the quasar PG\,0804+761 includes a
moderately obscured ($\tau_{9.7} \approx 1.2$) \Hot\ dust component,
consistent with such a clumpy obscuring torus.


The SED of the \Hot\ dust component from the PG\,0804+761 decomposition is
shown in the lower-right panel of Figure~\ref{fig:TorusModel}. Also shown is
a curve representing emission from dust at the temperature of the
PG\,0804+761 \Hot\ component, with 75\% obscured by a $\tau_{V} = 25$ screen
(i.e. equivalent to the screen obscuring the \AGN\ component), and the
remaining 25\% emerging unobscured.  While this model with non-uniform
coverage does not exactly reproduce the SED from the fully covered model (in
particular at shorter wavelengths), it nonetheless produces a very similar
spectrum which would also provide an adequate fit to the PG\,0804+761 SED
(with suitable small adjustments to the \AGN\ and \Warm\ components). Thus,
the SED of \Hot\ component emission from a highly obscured geometry with a
few clear lines-of-sight may appear similar to the emission from a fully
covered but less obscured geometry.  For a traditional torus made from a
smooth dust distribution, \Hot\ dust emission from a face-on source would
emerge unobscured with strong silicate emission features (since there are no
shaded sides of clouds to obscure the \Hot\ emission).  Our decomposition of
PG\,0804+761 does not show such strong features from the \Hot\ component,
suggesting that its \Hot\ dust, and possibly that of most Type-1 sources,
may therefore be significantly obscured by the shadowed sides of clumpy
torus clouds.


The lower-left panel of Figure~\ref{fig:TorusModel} depicts the fitted
NGC\,6240 \Hot\ dust component SED (unobscured in the decomposition), along
with a model in which $50$\% of its \Hot\ dust is covered by a screen with
$\tau_{V} = 25$ and the remainder emerges unobscured.  As with PG\,0804+761,
this partially covered model does not precisely reproduce the fitted \Hot\
dust component.  It is, however, similar enough to the fitted component to
provide a reasonably good fit (i.e.  since the \Hot\ component of NGC\,6240
contributes in a region with strong contributions from several other
components, the small deviations between the two models would not affect the
decomposition greatly). The near equivalence of these two models
demonstrates that emission from a highly obscured, but only partially
covered, geometry may appear quite similar to emission from a fully covered
geometry at much lower levels of obscuration.  Therefore, despite the small
apparent optical depth derived from the fit to NGC\,6240, it is possible
that the \Hot\ component in this source is actually much more significantly
obscured (i.e. we may only see direct emission from a few clear
lines-of-sight, while the majority of the \Hot\ dust is completely
obscured).

\subsection{Origin of the Far-IR Emission in AGNs}
\label{sec:FarIRInAGNs}

%
\begin{figure}
  \epsscale{1.1}
  \plotone{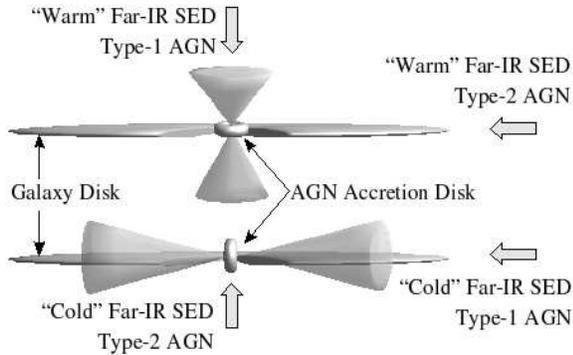}
  \caption{If the disk of the host galaxy and the AGN accretion disk are
    co-planar ({\it top}), very little dust in the disk of the galaxy is
    illuminated by the AGN (i.e. the radiation cones from the accretion disk
    do not intersect the host galaxy disk). In contrast, if the host galaxy
    and AGN accretion disks are orthogonal ({\it bottom}), the AGN heats
    dust over a large area of the galaxy disk (i.e. the radiation cones {\it
      do} intersect the host galaxy disk), resulting in far-IR and possibly
    PAH emission. Broad arrows label the line-of-sights giving rise to the
    indicated properties of the emergent SEDs for the various geometries.}
  \label{fig:AGNOrientation}
  \epsscale{1.0}
\end{figure}
%

Since emission from dust at far-IR ($\lambda > 50\um$) wavelengths is likely
to emerge unobscured from galaxies, the average properties of the far-IR
SEDs of AGNs (e.g. their relative cold dust luminosities and masses) should
be similar for all sources (assuming their nuclei are similarly structured),
independent of their orientation-based (i.e. Type-1 versus Type-2)
classification. As described in \S\ref{sec:DetailedDescriptions}, the two
template AGNs in our sample have very different far-IR properties (e.g.
Mrk\,463 has a dust luminosity-to-mass ratio $\sim$15 times that of
PG\,0804+761), in sharp contrast to these expectations.
%
\citet{2003MNRAS.343..585F}
use a two-component (starburst+AGN) template decomposition to conclude that
some optically classified AGNs with strong far-IR emission likely have
significant contributions from starbursts. While this is likely true for
many AGNs with cool far-IR SEDs, it does not appear to hold for Mrk\,463
(see \S\ref{sec:DetailedDescriptions}). As illustrated in
Figure~\ref{fig:AGNOrientation}, we instead suggest that the far-IR emission
from {\it some} AGNs (including Mrk\,463) is powered by the AGN itself and
originates from cool dust in the disk of the host galaxy. We note here that
this suggestion is based upon the SEDs of only two AGNs, so further study
using a much larger sample (to be presented in a subsequent paper) is needed
before conclusions are drawn about the origin of far-IR emission from the
class of AGNs.


While it is often assumed that an AGN accretion disk and the disk of the
galaxy hosting the AGN are co-planar (as in the top of
Fig.~\ref{fig:AGNOrientation}), this geometrical assumption is not
necessarily always true.  After the merger of two galaxies, the resulting
accretion and host galaxy disks may eventually come to rest in the same
plane (e.g.  due to torques), but there is a period of time after the merger
during which their relative orientation is essentially random. If the two
disks are tilted with respect to each other, a significant quantity of dust
in the disk of the host galaxy may be heated by the AGN out to large radial
distances (see the bottom of Fig.~\ref{fig:AGNOrientation}).  Assuming that
the galaxy disk is not a significant source of opacity, a galaxy therefore
appears as a Type-1 or Type-2 AGN depending upon the relative orientation of
the obscuring torus with respect to the observer (i.e. the standard AGN
unification scenario).  Additionally, depending on the relative orientation
of the host galaxy and AGN accretion disks (but independent of the
orientation of the observer), the source may exhibit weak or strong far-IR
emission (i.e.  ``Warm'' or ``Cold'' far-IR SEDs as labeled in
Fig.~\ref{fig:AGNOrientation}).  In this model, Mrk\,463 and PG\,0804+761
are therefore ``Cold'' and ``Warm'' far-IR sources, respectively, with
emission arising from geometries similar to those in the bottom and top of
Figure~\ref{fig:AGNOrientation}, respectively.  We note that the geometry
suggested here has a similar impact on the SEDs of galaxies as the
`warped-disk' geometry from
%
\citet{1989ApJ...347...29S}.
%


As described in \S\ref{sec:DetailedDescriptions}, we estimate that
$\sim$35\% of the bolometric luminosity of Mrk\,1014 is powered by
star-formation. If all of this luminosity emerges in the form of \Cool\
dust, this still leaves $\sim$30\% of its $70\K$ emission component to be
powered by its AGN. We therefore suggest that the geometry of Mrk\,1014 is
likely to be similar to that of Mrk\,463 (i.e.  the bottom of
Figure~\ref{fig:AGNOrientation}).  Additionally, the luminosity of the
Mrk\,1014 \PAHs\ component compared to its inferred starburst fraction
suggests that between $\sim$20--50\% of its PAH emission is powered by its
AGN (i.e.  derived by comparison with the properties of the obscured
starbursts NGC\,2623 and NGC\,6240---see \S\ref{sec:DetailedDescriptions}).
%
\citet{2006ApJ...649...79S}
argue against AGN powered far-IR emission in a sample of QSOs observed with
\Spitzer\ based largely on the fact that PAHs are destroyed in a hard AGN
radiation field. However, in the geometry of
Figure~\ref{fig:AGNOrientation}, the radiation field gradually softens as
the photons propagate through the galaxy disk, so that at some radius PAH
molecules illuminated by the AGN may survive. Clearly, more work is required
(e.g. radiative transfer calculations with realistic PAH destruction models)
to verify if this scenario is indeed realistic, and we do not dispute that
some far-IR emission in AGNs is powered by starbursts, but the suggested
mechanism to generation far-IR and PAH emission in AGNs as a result of the
interaction of the accretion disk with its host galaxy should be considered
as well.  Furthermore, we warn against the blanket assumption that all PAH
and far-IR emission is directly associated with star-formation, as this
could result in an overestimation of the true level of star-formation
activity in galaxies.

\section{Summary}
\label{sec:Summary}

We have presented a new multi-component spectral energy distribution
decomposition method. Our principal goal for this paper is to demonstrate
the effectiveness of the method for studying composite dusty galaxies,
although we note that the method is quite general and may be applicable to
much broader selections of galaxies. To demonstrate the efficacy of the
method, we have applied it to the ultraviolet through millimeter wavelength
SEDs of the nuclei of the unobscured and obscured template starbursts
NGC\,7714 and NGC\,2623, as well as the template Type-1 and Type-2 AGNs
PG\,0804+761 and Mrk\,463. We find that the pan-spectral SEDs of these
sources are all consistent with being entirely powered by star-formation and
accretion, respectively. We compare the SEDs of the two starburst galaxies
and demonstrate that the differences in their properties (including the
luminosities and masses of their dust components and the strength of their
PAH emission) can largely be explained in terms of the mean optical depth
through the dusty cocoons surrounding their newly formed stars.  In
comparing the template AGNs, we find that the far-IR SEDs of the two sources
differ significantly, which we suggest may result from variations in the
relative orientation of their host galaxy and AGN accretion disks.


We also decompose the nuclear SEDs of the composite sources NGC\,6240 and
Mrk\,1014. Our decomposition finds very little evidence for an AGN
contribution in NGC\,6240 ($\sim$2\%), with the total dust emission in that
galaxy being powered by a starburst with mean optical depth $\tau_V \approx
5$. We note, however, that we cannot rule out a larger contribution from a
deeply embedded AGN visible only in X-rays.  The SED of Mrk\,1014 is
consistent with a galaxy which is $\sim$65\% AGN powered, with a $\sim$35\%
contribution from star-formation. Like Mrk\,463, we find that Mrk\,1014 has
a far-IR excess (compared to the Type-1 AGN PG\,0804+761), and attribute
this to cold dust in the disk of its host galaxy which is heated directly by
its AGN. We also estimate that up to 50\% of the detected PAH emission in
Mrk\,1014 may be powered by radiation from the AGN that has softened as it
propagates through the galaxy disk (and therefore does not destroy the
fragile PAH molecules). In a future paper, we will apply this decomposition
method to a much larger sample of starbursts, AGNs, LIRGs, and ULIRGs
observed as part of the \Spitzer\ GTO program in order to provide a suite of
templates to aid in understanding both the local dusty galaxy population and
the high-redshift LIRGs and ULIRGs now being uncovered by \Spitzer.
%

\acknowledgements We would like to thank Moshe Elitzur, Vandana Desai, and
Kieran Cleary for many helpful discussions. We also wish to acknowledge the
assistance of the anonymous referee, whose review has resulted in a much
improved paper. This work is based [in part] on observations made with the
Spitzer Space Telescope, which is operated by the Jet Propulsion Laboratory,
California Institute of Technology under NASA contract 1407.  Support for
this work was provided by NASA through Contract Number 1257184 issued by
JPL/Caltech. This research has made use of the NASA/IPAC Extragalactic
Database (NED) which is operated by the Jet Propulsion Laboratory,
California Institute of Technology, under contract with the National
Aeronautics and Space Administration.

\begin{appendix}

\section{Weighting of Non-Uniformly Sampled Data}
\label{app:DataWeighting}

The sampling term, $\hat{\Lambda}$, in equation~(\ref{eq:Weights}), is
required to produce sensible and fair fits of non-uniformly sampled data, as
is the case when coarsely sampled photometry is joined together with
spectral data. If such a term is excluded (i.e. $\hat{\Lambda}_k = 1$ for
all $\lambda_k$), then $\chi^2$ will be dominated by the more finely sampled
spectral data, simply because it contains many more data points. To
establish a weighting scheme for non-uniformly sampled data, we define the
quantity
\begin{equation}
  n_\IRS \equiv N_\IRS / \log(\lambda_\Max^\IRS / \lambda_\Min^\IRS),
\end{equation}
which is the number of spectral samples per logarithmic wavelength interval
in an \IRS\ spectrum. Here, $N_\IRS$ is the total number of spectral samples
in the IRS spectrum, and $\lambda_\Min^\IRS \equiv \min(\lambda_\IRS)$ and
$\lambda_\Max^\IRS \equiv \max(\lambda_\IRS)$, where $\lambda_\IRS$ is the
array of wavelength points in the \IRS\ spectrum. We then divide the full
SED into $N_{\rm bins}$ wavelength bins (typically 10), distributed in even
logarithmic wavelength intervals, with each bin $l \in [0,1,\ldots,N_{\rm
  bins}-1]$ containing minimum and maximum wavelengths
\begin{equation}
  \lambda_\Min^l \equiv \min(\lambda_l) = \lambda_\Min
  \left(\frac{\lambda_\Max}{\lambda_\Min}\right)^{l / N_{\rm bins}}
  \textnormal{\quad and \quad}
  \lambda_\Max^l \equiv \max(\lambda_l) = \lambda_\Min
  \left(\frac{\lambda_\Max}{\lambda_\Min}\right)^{(l + 1) / N_{\rm bins}},
\end{equation}
where $\lambda_\Min \equiv \min(\lambda)$ and $\lambda_\Max \equiv
\max(\lambda)$ are the minimum and maximum wavelength in the SED having
wavelength array $\lambda$. Each bin is then assigned a total weight based
upon the sampling of the \IRS\ spectrum
\begin{equation}
  \Lambda_{\rm bins}^l = 
  n_\IRS \log\left(\frac{\lambda_\Max^l}{\lambda_\Min^l}\right) 
  = \frac{N_\IRS}{N_{\rm bins}} \frac{\log(\lambda_\Max /
    \lambda_\Min)}{\log(\lambda_\Max^\IRS / \lambda_\Min^\IRS)}.
\end{equation}
The weight of bins containing no data points is redistributed to the nearest
bins containing data above and below, with the weight going to each bin
being a linear function of the proximity to the bin with no data; i.e. for a
bin $l^\prime$ containing no data, the weight distributed to the bins
$l^\prime_{+}$ and $l^\prime_{-}$ above and below is
\begin{equation}
  \delta\Lambda_{\rm bins}^{l^\prime \rightarrow l^\prime_{+}} = 
  \Lambda_{\rm bins}^{l^\prime}
  \frac{l^\prime_{+} - l^\prime}{l^\prime_{+} - l^\prime_{-}}
  \textnormal{\quad and \quad}
  \delta\Lambda_{\rm bins}^{l^\prime \rightarrow l^\prime_{-}} = 
  \Lambda_{\rm bins}^{l^\prime} 
  \frac{l^\prime - l^\prime_{-}}{l^\prime_{+} - l^\prime_{-}}.
\end{equation}
The weighting in each bin is then distributed to the data points within that
bin, so that for a data point at wavelength $k$ which is in bin $l$
\begin{equation}
  \Lambda_{k \in l} = \frac{\Lambda_{\rm bins}^l}{N_l},
\end{equation}
where $N_l$ is the number of data points in bin $l$. Thus, the $\Lambda_k$
represent the number of effective spectral samples that each data point
represents so that the sampling over the entire SED is equivalent to the
\IRS\ sampling. Finally, we normalize and scale the weighting function
\begin{equation}
  \hat{\Lambda}_k = N_{\rm tot} \frac{\Lambda_k}{\sum_k \Lambda_k},
\end{equation}
so that $\sum_k \hat{\Lambda}_k = N_{\rm tot} = N_\phot + N_\IRS$, the total
number of data points in the spectral energy distribution, as would be the
case if the weighting function were taken to be unity at all wavelengths
(i.e. $\hat{\Lambda}_k = 1$ for all $\lambda_k$).

\section{IRS Statistical Flux Density Uncertainties}
\label{app:IRSUncertainties}

We derive the statistical flux density uncertainty at each wavelength
element of an \IRS\ spectrum by differencing two spectra obtained at
different nod positions of each slit. The initial estimate of the
uncertainty at each wavelength is
\begin{equation}
  \delta f_k = \frac{f_1(\lambda_k) - f_2(\lambda_k)}{(\sqrt{2})^2},
\end{equation}
where $f_1$ and $f_2$ are the observed flux densities in nod positions 1 and
2, and the two factors of $\sqrt{2}$ both correct for the introduction of
noise resulting from the subtraction, and account for the reduction in noise
of the final flux densities resulting from averaging the two nods. At each
wavelength, this array contains a single sample of the true uncertainty
drawn from the probability distribution function having standard deviation
$\sigma(\lambda_k)$. We estimate the true value of the statistical
uncertainty at each wavelength by calculating the standard deviation of the
points in a window of width $W$ (typically around 20 wavelength bins, with
the window size decreasing near the edges of the spectrum), and assigning
this value as the standard deviation of the central wavelength, i.e.
\begin{equation}
  \sigma^2(\lambda_k) = \frac{1}{W + 1} \sum_{i = k - W/2}^{k + W/2} 
  \left[\delta f_i - \langle \delta f \rangle \right]^2,
\end{equation}
where $\langle \delta f \rangle \approx 0$ is the mean value of $\delta f$
for the points within the window. Since we are approximating a function
which in principle should be sampled on an infinitely fine grid, our
estimated $\sigma$ array will always show some residuals from our finite
grid size. As a result, the $\sigma$ array does not always have the expected
property that $\sim$68\% of the $\delta f_k$ are enveloped by the
$1$--$\sigma$ contour. We therefore perform one final step, smoothing our
estimated $\sigma$ array with a window of width $W$, and scaling the entire
array until it has the expected behavior that the $1$--$\sigma$ error
contour envelops 68\% of the $\delta f_k$.

\end{appendix}

\end{document}